\documentclass[letterpaper,11pt]{article}

\pdfoutput=1

\PassOptionsToPackage{nosort}{cite}

\usepackage{amssymb}
\usepackage{amsfonts}
\usepackage{dsfont}
\usepackage{mathrsfs}
\usepackage{enumitem}
\usepackage[T1]{fontenc}
\usepackage[british]{chet}
\usepackage[font=small,format=hang,labelfont={sf,bf}]{caption}
\usepackage{float}
\usepackage{pgfplots}
\usepackage{multirow}
\usepackage{makecell}
\usepackage{tikz}
\usepackage{slashed}

\usepackage{soul}

\renewcommand{\arraystretch}{1.1}

\usetikzlibrary{intersections,arrows.meta,patterns,external}

\pgfplotsset{width=10cm, compat=1.10,every non boxed y axis/.style={}}
\usepgfplotslibrary{colorbrewer}
\usepgfplotslibrary{fillbetween}
\usepgfplotslibrary{colormaps}

\DeclareMathOperator*{\SumInt}{%
	\mathchoice%
	{\ooalign{$\displaystyle\sum$\cr\hidewidth$\displaystyle\int$\hidewidth\cr}}
	{\ooalign{\raisebox{.14\height}{\scalebox{.7}{$\textstyle\sum$}}\cr\hidewidth$\textstyle\int$\hidewidth\cr}}
	{\ooalign{\raisebox{.2\height}{\scalebox{.6}{$\scriptstyle\sum$}}\cr$\scriptstyle\int$\cr}}
	{\ooalign{\raisebox{.2\height}{\scalebox{.6}{$\scriptstyle\sum$}}\cr$\scriptstyle\int$\cr}}
}

\newcommand{\lsp}{\hspace{1pt}}
\newcommand{\llsp}{\hspace{0.5pt}}
\newcommand{\lnsp}{\hspace{-1pt}}

\renewcommand{\geq}{\geqslant}
\renewcommand{\leq}{\leqslant}
\def\nn{\nonumber}
\newcommand{\p}{\partial}
\newcommand{\f}{\frac}

\newcommand{\cA}{\mathcal{A}}
\newcommand{\cF}{\mathcal{F}}
\newcommand{\bF}{\mathbb{F}}
\newcommand{\wt}{\widetilde}

\DeclareMathOperator{\tr}{tr}

\allowdisplaybreaks

\definecolor{forestgreen}{rgb}{0.0, 0.27, 0.13}
\definecolor{darkblue}{rgb}{0.0, 0.0, 0.55}
\definecolor{darkred}{rgb}{0.55, 0.0, 0.0}
\hypersetup{colorlinks,
           linkcolor={forestgreen},
           citecolor={darkblue},
           urlcolor={darkred},
           pdftitle={Magnetised Bounds for Conformal Field Theories},
           pdfdisplaydoctitle
}

\tikzset{cross/.style={path picture={
      \draw[black]
            (path picture bounding box.south east) --
            (path picture bounding box.north west)
            (path picture bounding box.south west) --
            (path picture bounding box.north east);}}}

\title{Magnetised Bounds for Conformal Field Theories}

\author{Christopher P.\ Herzog, William H.\ Pannell, Biswajit Sahoo, and Andreas Stergiou}

\affiliation{Department of Mathematics, King's College London, Strand, London WC2R 2LS, United Kingdom}

\abstract{Aspects of parity-preserving, charge-conjugation-invariant, three-dimensional conformal field theories (CFTs) with a global $U(1)$ symmetry in the presence of a background magnetic field are investigated. A local effective action is constructed to four-derivative order, based on an assumption that the magnetic field drives the theory into a gapped phase. This action is evaluated in a variety of backgrounds, and is used to obtain one- and two-point functions of the conserved current and stress-energy tensor. Dispersive arguments are developed and shown to impose powerful constraints on the Wilson coefficients of the effective action, leading to universal predictions for the CFT response at large magnetic field and the scaling dimensions of background monopole operators. These general results are further examined through explicit calculations in the free complex scalar, free Dirac fermion, and a holographic Einstein--Hilbert--Maxwell model.}

\date{May 2025}

\begin{document}

\maketitle

\toc

\section{Introduction}
Deforming conformal field theories (CFTs) with external probes reveals novel observables that offer fresh insights into their dynamics, often with direct experimental relevance. There is already a significant body of work introducing background gravity, finite temperature, and chemical potential to illuminate key aspects of CFTs and the renormalisation group (RG) flows that connect them. In similar spirit, we explore here a particularly compelling and physically motivated deformation, that of a background magnetic field. CFTs with global $U(1)$ symmetry and associated conserved current $J_\mu$ can be coupled through the interaction term $\int\lnsp J^\mu\mathcal{A}_\mu$ in the action, where $\mathcal{A}_\mu$ is a suitable background gauge field. This setup uncovers new facets of CFT behaviour, which we analyse in specific models and constrain more generally across broad classes of theories.

We confine our attention to three-dimensional CFTs that preserve parity and charge conjugation. This offers an arena where general calculations and constraints can be examined in detail in specific models. When deformed by a magnetic field, CFTs will typically develop a mass gap. While not universal, this feature is wide-ranging enough for us to use it as one of our basic assumptions. The presence of this mass gap allows the construction of a local effective action in a derivative expansion, in which the magnetic field provides the suppression scale. Parity invariance guarantees that only operators with an even number of derivatives can appear, and charge-conjugation invariance that only operators even under a sign reversal of the magnetic field can appear. These assumptions enable the derivation of the results that follow, and the extent to which they can be relaxed in a controllable way is of significant interest and will be briefly discussed in the conclusion.

The effective action we obtain arises if we imagine that the original CFT is ``integrated out'' after coupling to the background. Besides coupling to a background gauge field $\mathcal{A}_\mu$, we also couple to a background metric $g_{\mu\nu}$ to find the effective action
\begin{equation}
    {\mathcal W}[{\mathcal A}, g] = c_0 S_0 + c_{2,1} S_{2,1} + c_{2,2} S_{2,2} + c_{2,3} S_{2,3} + c_{2,4} S_{2,4} + \cdots\,,
\end{equation}
where we have displayed, schematically, terms with zero and two derivatives. There is one and four distinct terms of each type, respectively, whose specific form is discussed in detail in Section \ref{S:EFT_construction} below. The form of $\mathcal{W}$ is theory-independent, within the assumptions described above, while the numerical Wilson coefficients $c_0,c_{2,1},c_{2,2},c_{2,3},c_{2,4},\ldots$ are theory-dependent. $\mathcal{W}$ can be determined from Weyl invariance, as was recently described in \cite{Boyack:2023uml}.

The effective action $\mathcal{W}$ can be computed in specific backgrounds. The Wilson coefficients that parametrise it, then, can be determined for different theories using the partition functions of these theories computed in these backgrounds. The free complex boson and free Dirac fermion have already been discussed in this light \cite{Cangemi:1994by, Gusynin:1998bt}, in an expansion in the magnetic field $B$ in flat space. It can be shown that only $c_0$ and $2c_{2,1}-c_{2,3}+c_{2,4}$ can be determined from these results. Here we consider further backgrounds, involving a non-trivial background metric, to determine additional linear combinations of Wilson coefficients. Furthermore, we perform explicit computations of two-point functions of the current. The linear combinations of Wilson coefficients we are able to determine are found in Table \ref{table:Wilsoncoeffs}.
\begin{table}[ht]
    \begin{center}
        \begin{tabular}{|l|c|}
            \hline
            Method & Wilson coefficients \\
            \hline
            Expansion in $B$ & $c_0,\, 2 c_{2,1} - c_{2,3}+c_{2,4}$ \\
            $S^3$ with magnetic flux & $c_0,\, 6 c_{2,1} - 2 c_{2,2} - 3 c_{2,3}$\\
            Spinning $S^2$ with magnetic flux & $c_0,\, c_{2,1},\, 2 c_{2,2} + 3 c_{2,3}$\\
            $\langle J^\mu J^\nu \rangle$ two-point function & $c_0,\, c_{2,3},\, 2 c_{2,1} + 2 c_{2,2} - 5c _{2,3}+4c_{2,4}$ \\
            \hline
        \end{tabular}
    \end{center}
    \caption{Different methods of computing second order Wilson coefficients in the effective action.}
    \label{table:Wilsoncoeffs}
\end{table}
Our computations of partition functions and correlation functions allow a full determination of the zero- and two-derivative Wilson coefficients for the free complex boson and free Dirac fermion for the first time. For a holographic example of the Einstein--Hilbert--Maxwell action with a negative cosmological constant we determine three out of four $c_{2,n}$'s. Our results are found in Table \ref{tab:results_Wilson_coeffs}.
\begin{table}[ht]
    \begin{center}
        \begin{equation}
            \renewcommand{\arraystretch}{1.5}
            \begin{array}{|l|c|c|c|c|}
            \hline
            & \mbox{Free scalar} & \mbox{Free fermion} & \mbox{Holographic example} \\
            \hline
            c_0 & \frac{(\sqrt{2}-1) \lsp\zeta \left( - \frac{1}{2} \right)}{2 \pi} & 
            \frac{2 \lsp\zeta \left(-\frac{1}{2}\right) }{\sqrt{2} \pi}& 
            - \frac{2 \cdot 2^{\frac{1}{4}}}{3^{\frac{3}{4}}} \sqrt{\frac{L}{g^3 \kappa}} \\
            c_{2,1}& \frac{5 ( 2 \sqrt{2}-1) \lsp\zeta \left(-\frac{3}{2}\right)}{32 \pi} &  
            -\frac{15 \lsp\zeta \left(\frac{5}{2} \right)}{64 \sqrt{2} \pi^3} &
            - \frac{1}{2 \cdot 6^{\frac{1}{4}}} \sqrt{\frac{L^3}{g \kappa^3}}
            \\
            c_{2,2} & \frac{150(-1+2\sqrt{2})\lsp\zeta\left(-\frac{3}{2}\right)+5(-2+\sqrt{2})\lsp\zeta\left(\frac{1}{2}\right)}{512\pi} &
           \frac{8\pi^{2}\zeta\left(\frac{1}{2}\right)-225\,\zeta\left(\frac{5}{2}\right)}{512\sqrt{2}\pi^{3}} &
            \left( \frac{7}{2 \cdot 6^\frac{1}{4}} + \frac{3}{2} s  \right) \sqrt{\frac{L^3}{g \kappa^3}}-\frac{3}{2} c_{2,4}
            \\
            c_{2,3} & \frac{30(-1+2\sqrt{2})\lsp\zeta\left(-\frac{3}{2}\right)+(-2+\sqrt{2})\lsp\zeta\left(\frac{1}{2}\right)}{256\pi}&
           \frac{8\pi^{2}\zeta\left(\frac{1}{2}\right)-45\,\zeta\left(\frac{5}{2}\right)}{256\sqrt{2}\pi^{3}} &
            - s \sqrt{\frac{L^3}{g \kappa^3}} + c_{2,4}
            \\
            c_{2,4} & -\frac{130(-1+2\sqrt{2})\lsp\zeta\left(-\frac{3}{2}\right)+7(-2+\sqrt{2})\lsp\zeta\left(\frac{1}{2}\right)}{256\pi}& \frac{8\pi^{2}\zeta\left(\frac{1}{2}\right)+195\,\zeta\left(\frac{5}{2}\right)}{256\sqrt{2}\pi^{3}}& \text{undetermined}
            \\
            \hline
            \end{array}
        \end{equation}
    \end{center}
    \caption{Wilson coefficients up to and including second order in derivatives for the three examples we consider in this work: the free complex scalar, the free four-component fermion, and the holographic example. We define $s = \left(\frac{3}{2}\right)^{\frac{3}{4}} \cot^{-1} \sqrt{2}$. Only three combinations of the $c_{2,n}$ are determined in the holographic example.}\label{tab:results_Wilson_coeffs}
\end{table}

While results in specific theories are important, constraints that can be potentially obeyed by Wilson coefficients across theories are much more significant due to their universal character. In situations with Lorentz invariance and dynamical low-energy degrees of freedom, it is widely understood how to constrain Wilson coefficients using dispersive arguments on the amplitudes of low-energy fields, following the pioneering work \cite{Adams:2006sv}. In our case, Lorentz invariance is broken due to the background magnetic field, and there are no low-energy dynamical degrees of freedom due to the mass gap assumption. Having an effective action of the background gauge field and metric, however, enables the calculation of correlation functions of the current and stress-energy tensor. These essential operators are analysed here at the level of two-point functions, first in position and then in momentum space. Tensorial expressions of general form consistent with the symmetries are constructed, and the Ward identities that follow from gauge invariance, Weyl invariance and diffeomorphism invariance are discussed in detail.

Two-point functions of the current and stress-energy tensor have been shown to constitute prime candidates with which to build dispersive arguments in situations like ours \cite{Creminelli:2022onn}. The upshot of a careful analysis of analyticity of momentum-space two-point functions of $J^\mu$ and $T^{\mu\nu}$ is found to be that certain matrices that arise in the frequency expansion of these two-point functions are positive semi-definite. These matrix conditions are examined here for the zero-, two- and four-derivative terms in the effective action $\mathcal{W}$. They are shown to imply
\begin{equation}\label{eq:c0_constraint}
    c_0\leq0\,,
\end{equation}
which is verified to be the case in all examples of Table \ref{table:Wilsoncoeffs}. Constraints on combinations of $c_{2,1}, c_{2,2}, c_{2,3}$, and $c_{2,4}$ are most easily captured pictorially as in Fig.\ \ref{fig:genericquadboundsIntro}.
\begin{figure}[ht]
    \centering
    \begin{tikzpicture}
        \begin{axis}[
            xmin=-10, xmax=10,
            ymin=-10, ymax=10,
            xlabel= $A$,
            ylabel= $B$,
            ylabel style={rotate=-90}
        ]
        \addplot[name path=tr,domain=-10:-2,blue] {x^2/4};
        \addplot[name path=tl,domain=-2:10,blue] {-x-1};
        \path[name path=br] (axis cs:-10,10) -- (axis cs:-2,10);
        \path[name path=bl] (axis cs:-2,10) -- (axis cs:10,10);
        \addplot [
                thick,
                color=blue,
                fill=blue,
                fill opacity=0.1
            ]
            fill between[
                of=br and tr
            ];
        \addplot [
                thick,
                color=blue,
                fill=blue,
                fill opacity=0.1
            ]
            fill between[
                of=bl and tl
            ];
        \end{axis}
    \end{tikzpicture}
    \caption{The allowed region, in blue, of Wilson coefficients in the $A$-$B$ plane, where $A$, $B$ are rational functions of the $c_{2,n}$'s; see e.g.\ \eqref{eq:ABtwoderivative} below.}
    \label{fig:genericquadboundsIntro}
\end{figure}
We find that they are obeyed by the free complex scalar and fermion coefficients. That the free fermion satisfies the bound is perhaps surprising, because the zeroth Landau level is gapless. For the holographic example the verdict rests on the single Wilson coefficient we are unable to determine there: the bound is violated unless $c_{2,4}\geq2c_{2,2}$, that is, unless $c_{2,4}$ is at least half of the combination $c_{2,2}+\tfrac32 c_{2,4}$ that we do determine. The holographic example is in any case gapless, and we will see later that the structure of the effective field theory expansion would need to be completely modified to capture the behaviour of the current-current two-point functions. In this sense, the associated column in Table \ref{tab:results_Wilson_coeffs} is at best misleading. In general, our bounds on Wilson coefficients will be valid in interacting theories that develop a mass gap when coupled to a magnetic field.

Our results also imply constraints on the dimension of background monopole operators. In contrast to the situation at non-zero chemical potential \cite{Hellerman:2015nra, Monin:2016jmo}, these operators are not bona fide operators of the underlying CFT. They were first considered in \cite{Kapustin:2011jm} as a generalisation of the superconformal index that could be used to test dualities between three-dimensional superconformal theories. They also have applications in compressible quantum phases, which arise when the global $U(1)$ charge is given a non-zero chemical potential \cite{Sachdev:2012tj}. Background monopole operators can be assigned a scaling dimension, whose computation in specific theories was considered in detail in \cite{Pufu:2013eda} for the free complex boson, free Dirac fermion as well as the interacting $O(2N)$ model in the $1/N$ expansion. The dimension of background monopole operators was discussed extensively in \cite{Boyack:2023uml} in the effective field theory (EFT) context. Our constraint \eqref{eq:c0_constraint} implies positivity of scaling dimensions of background monopole operators at large magnetic field, yielding a connection between analyticity properties of two-point functions of $J^\mu$ and $T^{\mu\nu}$ and unitarity in the background monopole sector.

The response of a CFT to the introduction of a magnetic field can also be characterised by the behaviour of a Euclidean free energy $\mathcal{E}$ that can be computed as the logarithm of the CFT partition function on the sphere associated with the cylinder of radial quantisation. Our bound \eqref{eq:c0_constraint} implies that at large magnetic field this free energy grows with the magnetic field, which results in a universal diamagnetic behaviour for parity preserving three-dimensional CFTs at large magnetic field.

In the next section we describe the construction of the EFT associated with CFTs in magnetic field. In Section \ref{sec:EFT_backgrounds} we present computations of the effective action in specific backgrounds with curvature and magnetic field. Section \ref{sec:2pt_fun_Ward_identities} discusses two-point functions of the conserved current and stress-energy tensor, and the constraints imposed on them by gauge, Weyl and diffeomorphism invariance. In Section \ref{sec:dispersion} we present our dispersive constraints on these two-point functions. Explicit computations of Wilson coefficients with a variety of methods for the free complex scalar, free Dirac fermion, and our holographic example are presented in Sections \ref{sec:complex_scalar}, \ref{sec:fermion}, and \ref{sec:holo}, respectively. We conclude in \ref{sec:conclusion}, while Appendix \ref{app:squashed} contains technical details relevant for the calculation of the partition function of a massless scalar and a massless fermion on a squashed three-sphere in the presence of magnetic flux.

\section{EFT construction}\label{S:EFT_construction}
In this section we construct an effective field theory for a parity- and charge-conjugation-invariant CFT with a global $U(1)$ symmetry in three-dimensional spacetime. The theory is expected to have a stress-energy tensor $T^{\mu\nu}$ and a conserved current $J^\mu$.  We can couple such a theory to an external gauge field ${\mathcal A}_\mu$ and place it on a curved manifold with metric $g_{\mu\nu}$. The partition function for the theory can be written formally as
\begin{equation}
    {\mathcal Z}[{\mathcal A}, g]=\int [D \phi]\, e^{i S[\phi,{\mathcal A}, g]} \,.
\end{equation}
Defining the functional ${\mathcal W}[{\mathcal A}, g]$ via $e^{i{\mathcal W}[{\mathcal A}, g]} = {\mathcal Z}[{\mathcal A}, g]$, the current and stress-energy tensor one-point functions can be recovered from the functional derivatives
\begin{equation}\label{eq:JTdefns}
    \langle J^\mu(x)\rangle=\frac{1}{\sqrt{-g(x)}} \frac{\delta \mathcal{W}[\mathcal{A},g]}{\delta A_\mu(x)}\,,\qquad \langle T^{\mu\nu}(x)\rangle=\frac{2}{\sqrt{-g(x)}} \frac{\delta \mathcal{W}[\mathcal{A},g]}{\delta g_{\mu\nu}(x)}\,,
\end{equation}
in the usual way, where $\mathcal{A}_\mu=\mathbb{A}_\mu+A_\mu$, with $\mathbb{A}_\mu$ the background gauge field and $A_\mu$ the non-dynamical fluctuation.

In order to preserve a notion of conformal invariance in a general spacetime, we further assume that 
the connected functional ${\mathcal W}[{\mathcal A}, g]$ is invariant under the Weyl transformations
\begin{equation}
    g_{\mu\nu}(x)\rightarrow e^{2\sigma(x)}g_{\mu\nu}(x)\,,\qquad
    \mathcal{A}_\mu(x)\rightarrow \mathcal{A}_\mu(x)\,.
\end{equation}
As we primarily discuss the case $d=3$, we will ignore any issues associated with Weyl anomalies, which could be important in even dimensional spacetimes. 

We are interested in developing an effective field theory description in the limit of a large, constant, background magnetic field. The field strength $\mathcal{F}_{\mu\nu}(x)= \partial_\mu \mathcal{A}_\nu(x) -\partial_\nu \mathcal{A}_\mu(x)$ can be written as a sum of background field $\mathbb{F}_{\mu\nu}$ and a local non-dynamical fluctuation $F_{\mu\nu}(x)$ as
\begin{equation}\label{eq:backgroud_F_choice}
    \mathcal{F}_{\mu\nu}(x)=\mathbb{F}_{\mu\nu}+F_{\mu\nu}(x)\,,
\end{equation}
where $F_{\mu\nu}= \partial_\mu A_\nu -\partial_\nu A_\mu$. The field strength $\mathbb{F}_{\mu\nu}=\partial_\mu \mathbb{A}_\nu -\partial_\nu \mathbb{A}_\mu$ arises due to a background gauge field $\mathbb{A}_\mu(x)$.  

A critical assumption in what follows is that we can express the low energy effective field theory purely in terms of the external fields ${\mathcal A}_\mu$ and $g_{\mu\nu}$.  In particular, this assumption precludes the existence of any massless propagating modes that would also need to be represented in our description. A simple way to guarantee the absence of such modes is to assume a mass gap. For background magnetic field, such a gap forms for a free scalar theory because of Landau levels. The lowest Landau level for a free fermion actually has zero energy, making it an exception. The expectation however for any weakly interacting CFT in a strong magnetic field background, due to the absence of any other scale, is that there should be a gap that scales with the size of the magnetic field. This has been argued to be the case in \cite{Boyack:2023uml}.

The next step is to develop some notation to write the effective field theory as compactly as possible. Our three-dimensional setting allows the use of the Levi--Civita tensor to define the parity-odd vector
\begin{equation}\label{eq:F_vector_def}
    \mathcal{F}^\mu=\frac{1}{2\sqrt{-g}} \epsilon^{\mu\nu\rho} \mathcal{F}_{\nu\rho}\,,\qquad \mathcal{F}_\mu=\frac{\sqrt{-g}}{2}\epsilon_{\mu\nu\rho}\mathcal{F}^{\nu\rho}\,,
\end{equation}
which will considerably simplify the construction of the effective action below. We use conventions in which $\epsilon_{012}=-\epsilon^{012}=1$, and $\frac{1}{\sqrt{-g}}\lsp\epsilon^{\mu\nu\rho}$ and $\sqrt{-g}\lsp\epsilon_{\mu\nu\rho}$ transform like contravariant and covariant tensors, respectively. We also define a scalar $\mathcal{F}$ under general coordinate transformations out of the field strength as
\begin{equation}\label{eq:F_scalar}
    \mathcal{F}=\sqrt{-\mathcal{F}^\mu \mathcal{F}_\mu}=\sqrt{\tfrac12 \mathcal{F}^{\mu\nu}\mathcal{F}_{\mu\nu}}\,.
\end{equation}
The Bianchi identity is simply given by
\begin{equation}\label{eq:bianchi_id}
    \nabla^\mu\lnsp\mathcal{F}_\mu=0\,.
\end{equation}

The connected functional $\mathcal{W}$ can be interpreted as the EFT action below the energy scale $\sqrt{\mathbb{F}}=\left(\f{1}{2}\mathbb{F}^{\mu\nu}\mathbb{F}_{\mu\nu}\right)^{1/4}$, which represents the order of the mass gap. We choose the following configuration for constant background field strength in three-dimensional Minkowski background:
\begin{equation}\label{eq:B_background}
    \mathbb{F}^\mu=\begin{pmatrix}
        -B\\ 
        0\\ 
        0
    \end{pmatrix}\,,\qquad 
    \mathbb{F}_{\mu\nu}= \begin{pmatrix}
        0 & 0 & 0\\
        0 & 0 & B\\
        0 & -B & 0
    \end{pmatrix}.
\end{equation}

The effective field theory up to two derivative orders on the electromagnetic field strength has been constructed in \cite{Boyack:2023uml}. In this section, we focus on parity and charge-conjugation preserving theories and extend this construction to include terms up to four-derivative order.

A Weyl invariant metric can be defined in three dimensions as
\begin{equation}\label{eq:g_hat_definition}
    \hat{g}_{\mu\nu}=g_{\mu\nu}\mathcal{F}\,.
\end{equation}
Note that one may attempt to construct various Weyl invariant metrics of the type $g_{\mu\nu}\mathcal{F}_{(n)}$, with $\mathcal{F}_{(n)}=\big(g^{\mu_1\mu_{2n}}g^{\mu_2\mu_3}\cdots g^{\mu_{2n-2}\mu_{2n-1}}\mathcal{F}_{\mu_1\mu_2}\mathcal{F}_{\mu_3\mu_4}\cdots \mathcal{F}_{\mu_{2n-1}\mu_{2n}}\big)^{1/n}$ for $n>2$. However, $\mathcal{F}_{(n)}$ vanishes for odd values of $n$ due to the anti-symmetry of $\mathcal{F}_{\mu\nu}$, while for even $n\geq 4$ (in three dimensions) the Cayley--Hamilton theorem\footnote{Note that the Cayley--Hamilton theorem implies that in $d$ spacetime dimensions all higher order contracted terms of the form ${[\mathcal{F}^n]_{\mu_1}}^{\mu_{n+1}}= {\mathcal{F}_{\mu_1}}^{\mu_2}{\mathcal{F}_{\mu_2}}^{\mu_3}\cdots {\mathcal{F}_{\mu_n}}^{\mu_{n+1}}$ for $n\geq d$ are fully determined in terms of $[\mathcal{F}^m]_{\mu_1}^{\ \mu_{m+1}}$ with $m=0,1,\ldots,d-1$.} relates such Weyl invariant metrics to $\hat{g}$ given in \eqref{eq:g_hat_definition}.

Now, the Weyl invariance of $\mathcal{W}[\mathcal{A}, g]$ requires that it be expressed in terms of general coordinate invariant terms constructed out of $\hat{g}_{\mu\nu}$ and
\begin{equation}\label{eq:hat_defns}
    \widehat{\mathcal{F}}^\mu=\frac{1}{2\sqrt{-\hat{g}}} \epsilon^{\mu\nu\rho} \mathcal{F}_{\nu\rho}\,,\qquad
    \widehat{\mathcal{F}}_{\mu}=\hat{g}_{\mu\nu}\widehat{\mathcal{F}}^{\nu}\,,\qquad \widehat{\nabla}^\mu=\hat{g}^{\mu\nu}\widehat{\nabla}_{\!\nu}\,.
\end{equation}
Contractions of indices involving hatted tensors should be performed with the hatted metric. Integrations by parts can be used to relate various terms. In $d=3$ the Riemann tensor satisfies
\begin{equation}
    R_{\mu\nu\rho\sigma}=g_{\mu\rho}R_{\nu\sigma} -g_{\mu\sigma}R_{\nu\rho}-g_{\nu\rho}R_{\mu\sigma} +g_{\nu\sigma}R_{\mu\rho}+\tfrac12 R(g_{\mu\sigma}g_{\nu\rho}- g_{\nu\sigma}g_{\mu\rho})\,,
\end{equation}
and so does not need to be involved in the construction of the effective action. Additionally, our definitions imply that
\begin{equation}
    \widehat{\mathcal{F}}^2=-\widehat{\mathcal{F}}^\mu\widehat{\mathcal{F}}_\mu=1\,.
\end{equation}
This equation and its derivatives can be used to relate potential contributions to the effective action, to obtain a basis that is complete but not over-complete. Finally, since
\begin{equation}
    \widehat{\nabla}^\mu\lnsp\widehat{\mathcal{F}}_\mu=\mathcal{F}^{-3/2}\lsp\nabla^\mu\lnsp\mathcal{F}_\mu\,,
\end{equation}
the Bianchi identity \eqref{eq:bianchi_id} implies $\widehat{\nabla}^\mu\lnsp\widehat{\mathcal{F}}_\mu=0$, which needs to also be used as a constraint in the construction of the effective action.
\begin{table}
    \begin{center}
        \begin{equation}
            \begin{array}{|c|cccccccc|}
                \hline
                {\rm Operator} & \widehat{\mathcal F}^\mu & \widehat{\mathcal F}_\mu & 
                \widehat \nabla^\mu & 
                \widehat \nabla_\mu & \widehat R & \widehat R^{\mu\nu} & \widehat R_{\mu\nu} & \sqrt{-\hat{g}}\\
                \mbox{Scale weight} & -1 & 1 & -1 & 1 & 0 & -2 & 2 & 3 \\
                \hline
            \end{array}
        \end{equation}
        \caption{The scaling weights of the quantities appearing in the effective field theory description with respect to the mass scale set by $\sqrt{B}$. These values also represent the mass dimensions of the corresponding quantities.}
        \label{table:scalingweights}
    \end{center}
\end{table}
While $\mathcal{F}^\mu$ and $\mathcal{F}_\mu$ both scale as a power of the magnetic field $B$, after the rescaling procedure, the weights are altered in the way summarised by Table \ref{table:scalingweights}.

We organise the effective field theory in a derivative expansion. Each term in the expansion will be Weyl invariant because it will be given by quantities constructed from the building blocks $\hat g_{\mu\nu}$, $\widehat{\mathcal F}^\mu$.  Each term must have mass dimension three to compensate the $d^{\lsp 3} x$ in the integration measure for the action. The parity- and charge-conjugation-invariant effective action takes the form
\begin{equation}\label{eq:curved_space_W}
    {\mathcal W}[{\mathcal A}, g] = \sum_{a=0}^\infty \sum_{n=1}^{N_a} c_{2a,n} S_{2a,n} = c_{0,1} S_{0,1} + c_{2,1} S_{2,1} + c_{2,2} S_{2,2} + c_{2,3} S_{2,3} + c_{2,4} S_{2,4} + \cdots\, ,
\end{equation}
where the $c_{2a,n}$ are dimensionless Wilson coefficients that depend on the microscopic details of the theory and the index $a$ counts the derivative order in the expansion. Below we define $c_0=c_{0,1}$ and $S_0=S_{0,1}$. Because we assume parity, the expansion is in even numbers of derivatives on the parity-odd $\widehat{\mathcal{F}}$. Furthermore, charge conjugation flips the sign of the magnetic field, which implies that we only have terms with an even number of $\widehat{\mathcal{F}}$'s. This excludes, for example, the parity-even two-derivative structure $\frac{1}{\sqrt{-\hat{g}}}\lsp\epsilon^{\mu\nu\rho}\lsp\widehat{\mathcal{F}}_\mu\widehat{\nabla}_{\!\nu}\widehat{\mathcal{F}}^\sigma\widehat{\nabla}_{\!\sigma}\widehat{\mathcal{F}}_\rho$, which contains three $\widehat{\mathcal{F}}$'s. At the first two orders in the derivative expansion, there are only five terms:
\begin{align}
    S_{0} &= \int d^{\llsp 3}x\sqrt{-\hat{g}} \,, \nonumber \\
    S_{2,1} &=\int d^{\llsp 3}x\sqrt{-\hat{g}}\, \widehat{R} \,, \qquad
    S_{2,2} =\int d^{\llsp 3}x\sqrt{-\hat{g}}\, \widehat{R}^{\mu\nu}\widehat{\mathcal{F}}_\mu\widehat{\mathcal{F}}_\nu \,,\\
    S_{2,3} &= \int d^{\llsp 3}x \sqrt{-\hat{g}}\,\big(\widehat{\nabla}^\mu\lnsp \widehat{\mathcal{F}}^\nu\,\widehat{\nabla}_{\!\mu}\widehat{\mathcal{F}}_\nu -\tfrac12 \widehat{\nabla}^\mu\lnsp\widehat{\mathcal{F}}^\nu\,\widehat{\nabla}_{\!\nu}\widehat{\mathcal{F}}_\mu\big)\,,\qquad S_{2,4} = \int d^{\llsp 3}x \sqrt{-\hat{g}}\,\widehat{\mathcal{F}}^\mu\widehat{\mathcal{F}}^\nu\widehat{\nabla}_{\!\mu}\widehat{\mathcal{F}}^\rho\widehat{\nabla}_\nu\widehat{\mathcal{F}}_\rho\,.\nonumber
\end{align}
At order four in derivatives there are 29 terms, which we list in Table \ref{tab:four-der-effact-terms}. (We give the Lagrangian densities $\hat{\mathscr{L}}_{4,n}$ as opposed to the lengthier $S_{4,n}=\int d^{\llsp 3}x \sqrt{-\hat{g}}\, \hat{\mathscr{L}}_{4,n}$.)

We may expand these hatted terms into unhatted notation. The first four terms expand into
\begin{align}
    \label{eq:0_der_term}
    S_{0} &=\int d^{\llsp 3}x\sqrt{-g}\, \mathcal{F}^{3/2}\,, \\
    \label{eq:2_der_term1}
    S_{2,1}&=\int d^{\llsp 3}x\sqrt{-g}\,\mathcal{F}^{1/2}\bigg(R +\frac{\partial^\mu\lnsp\mathcal{F}\,\partial_\mu\mathcal{F}}{2\mathcal{F}^2}\bigg)\,, \\
    \label{eq:2_der_term2}
    S_{2,2}&=\int d^{\llsp 3}x\sqrt{-g}\,\mathcal{F}^{-3/2}\bigg(R^{\mu\nu}\lnsp \mathcal{F}_\mu \mathcal{F}_\nu-\frac{\mathcal{F}^\mu\mathcal{F}^\nu\,\nabla_{\!\mu}\partial_\nu\mathcal{F}}{2\mathcal{F}}+\frac{3\big(\mathcal{F}^\mu\partial_\mu\mathcal{F}\big)^2}{4\mathcal{F}^2}\bigg)\,, \\
    \label{eq:2_der_term3}
    S_{2,3}&=\int d^{\llsp 3}x\sqrt{-g}\, \mathcal{F}^{-3/2}\big(\nabla^\mu\lnsp\mathcal{F}^\nu\,\nabla_{\!\mu} \mathcal{F}_\nu-\tfrac12\nabla^\mu\lnsp\mathcal{F}^\nu\,\nabla_{\!\nu}\mathcal{F}_\mu+\tfrac34\lsp\partial^\mu\lnsp\mathcal{F}\,\partial_\mu\mathcal{F}\big)\,,\\
    \label{eq:2_der_term4}
    S_{2,4}&=\int d^{\llsp 3}x\sqrt{-g}\, \mathcal{F}^{-3/2}\bigg(\frac{\mathcal{F}^\mu\mathcal{F}^\nu\,\nabla_{\!\mu}\mathcal{F}^\rho\,\nabla_{\!\nu}\mathcal{F}_{\!\rho}}{\mathcal{F}^2}+\frac{11\,\mathcal{F}^\mu\mathcal{F}^\nu\lsp\partial_\mu\mathcal{F}\,\partial_\nu\mathcal{F}}{4\mathcal{F}^2}-\frac{\mathcal{F}^\mu\mathcal{F}^\nu\,\nabla_{\!\mu}\partial_\nu\mathcal{F}}{\mathcal{F}}+\frac14\lsp\partial^\mu\lnsp\mathcal{F}\,\partial_\mu\mathcal{F}\bigg)\,.
\end{align}
We will not present here the expansion of the fourth order terms, which we have performed with the help of \href{http://www.xact.es}{\texttt{xAct}} in \emph{Mathematica}.

Comparing our connected functional \eqref{eq:curved_space_W} with \cite[Appendix A]{Boyack:2023uml} up to two-derivative order, we find that our coefficients $c_0, c_{2,1},c_{2,2},c_{2,3}$ are related to their $c_1,c_2,c_3,c_4$ via\footnote{Our term $S_{2,4}$ was missed by \cite{Boyack:2023uml}. Additionally, some contributions were missed in the $c_4$ terms of \cite{Boyack:2023uml} when going from hatted to unhatted variables.}
\begin{equation}\label{eq:map_EFT_coefficients}
    c_{0}\rightarrow c_1\,,\qquad
    c_{2,1}\to c_2+c_3\,,\qquad
    c_{2,2}\to c_3-\tfrac12c_4\,,\qquad
    c_{2,3}\to -c_4\,,
\end{equation}
for terms that appear in (\ref{eq:curved_space_W}).
\begin{table}
    \centering
    \renewcommand{\arraystretch}{1.05}
    \begin{tabular}{|c|c|c|c|}
        \hline
        \multirow{3}{*}{$n$} & \multirow{3}{*}{$\hat{\mathscr{L}}_{4,n}$} & \multirow{3}{*}{\makecell{Contributes to\\ two-point functions\\ for constant $\mathbb{F}_{\mu\nu}$}} & \multirow{3}{*}{\makecell{Contributes to\\ two-point functions\\ for constant $B$}} \\
        & & &\\
        & & &\\
        \hline
        1 & $\widehat{R}^2$ & \checkmark & \checkmark \\\hline
        2 & $\widehat{R}^{\mu\nu}\widehat{R}_{\mu\nu}$ & \checkmark & \checkmark \\\hline
        3 & $\widehat{R}^{\mu\nu}\widehat{R}_{\nu}{\lnsp}^\rho\widehat{\mathcal{F}}_\mu \widehat{\mathcal{F}}_\rho$ & \checkmark &\checkmark  \\\hline
        4 & $\widehat{R}^{\mu\nu}\widehat{R}^{\rho\sigma}\widehat{\mathcal{F}}_\mu \widehat{\mathcal{F}}_\nu \widehat{\mathcal{F}}_\rho \widehat{\mathcal{F}}_\sigma$ & \checkmark &\checkmark  \\\hline
        5 & $\widehat{R}\widehat{R}^{\mu\nu}\widehat{\mathcal{F}}_\mu\widehat{\mathcal{F}}_\nu$ & \checkmark & \checkmark \\\hline
        6 & $\widehat{R}^{\mu\nu}\widehat{\mathcal{F}}_\mu\widehat{\nabla}^2\lnsp\widehat{\mathcal{F}}_\nu$ & \checkmark & \checkmark \\\hline
        7 & $\widehat{R}^{\mu\nu}\widehat{\nabla}^\rho\lnsp\widehat{\mathcal{F}}_\mu \widehat{\nabla}_{\!\rho}\widehat{\mathcal{F}}_\nu$ &  & \\\hline
        8 & $\widehat{R}^{\mu\nu}\widehat{\nabla}_{\!\mu}\widehat{\mathcal{F}}^\rho \lsp\widehat{\nabla}_{\!\nu}\widehat{\mathcal{F}}_\rho$ &  & \\\hline
        9 & $\widehat{R}^{\mu\nu}\widehat{\mathcal{F}}^\rho \widehat{\nabla}_{\!\rho}\widehat{\nabla}_{\!\mu}\widehat{\mathcal{F}}_\nu$ & \checkmark & \checkmark \\\hline
        10 & $\widehat{R}\lsp\widehat{\nabla}^\mu\lnsp\widehat{\mathcal{F}}^\nu\lsp\widehat{\nabla}_{\!\nu}\widehat{\mathcal{F}}_\mu$ &  & \\\hline
        11 & $\widehat{R}\lsp\widehat{\mathcal{F}}^\mu\widehat{\nabla}^2\lnsp\widehat{\mathcal{F}}_\mu$ &  & \\\hline
        12 & $\widehat{R}^{\mu\nu}\lnsp\widehat{\mathcal{F}}_{\!\mu}\widehat{\mathcal{F}}^\rho\widehat{\mathcal{F}}^\sigma\widehat{\nabla}_{\!\sigma}\widehat{\nabla}_{\!\rho}\widehat{\mathcal{F}}_\nu$ & \checkmark & \checkmark \\\hline
        13 & $\widehat{R}\lsp\widehat{\mathcal{F}}^\mu\widehat{\mathcal{F}}^\nu\widehat{\mathcal{F}}^\rho\widehat{\nabla}_{\!\rho}\widehat{\nabla}_{\!\nu}\widehat{\mathcal{F}}_\mu$ & & \\\hline
        14 & $\widehat{\nabla}^2\lnsp\widehat{\mathcal{F}}^\mu\lsp\widehat{\nabla}^2\lnsp\widehat{\mathcal{F}}_\mu$ & \checkmark & \checkmark \\\hline
        15 & $(\widehat{\mathcal{F}}^\mu\widehat{\nabla}^2\lnsp\widehat{\mathcal{F}}_\mu)^2$ &  & \\\hline
        16 & $\widehat{\mathcal{F}}^\mu\widehat{\mathcal{F}}^\nu\lsp\widehat{\nabla}_{\!\mu}\widehat{\nabla}_{\!\nu}\widehat{\mathcal{F}}^\rho\lsp\widehat{\nabla}^2\lnsp\widehat{\mathcal{F}}_\rho$ & \checkmark & \checkmark \\\hline
        17 & $\widehat{\mathcal{F}}^\mu\widehat{\mathcal{F}}^\nu\lsp\widehat{\nabla}^{\rho}\widehat{\nabla}_{\!\nu}\widehat{\mathcal{F}}_\mu\lsp\widehat{\nabla}^2\lnsp\widehat{\mathcal{F}}_\rho$ &  & \\\hline
        18 & $\widehat{\mathcal{F}}^\mu\widehat{\mathcal{F}}^\nu\lsp\widehat{\nabla}^{\rho}\widehat{\nabla}_{\!\nu}\widehat{\mathcal{F}}^\sigma\lsp\widehat{\nabla}_{\!\sigma}\widehat{\nabla}_{\!\rho}\widehat{\mathcal{F}}_\mu$ &  & \\\hline
        19 & $\widehat{\mathcal{F}}^\mu\widehat{\mathcal{F}}^\nu\lsp\widehat{\nabla}^{\rho}\widehat{\nabla}^{\sigma}\widehat{\mathcal{F}}_\mu\lsp\widehat{\nabla}_{\!\rho}\widehat{\nabla}_{\!\sigma}\widehat{\mathcal{F}}_\nu$ &  & \\\hline
        20 & $\widehat{\mathcal{F}}^\mu\lsp\widehat{\nabla}^\nu\lnsp\widehat{\mathcal{F}}^\rho\lsp\widehat{\nabla}_{\!\rho}\widehat{\mathcal{F}}_\nu\lsp\widehat{\nabla}^2\widehat{\mathcal{F}}_\mu$ &  & \\\hline
        21 & $\widehat{\mathcal{F}}^\mu\lsp\widehat{\nabla}^\rho\lnsp\widehat{\mathcal{F}}^\nu\lsp\widehat{\nabla}^{\sigma}\widehat{\mathcal{F}}_\rho\lsp\widehat{\nabla}_{\!\sigma}\widehat{\nabla}_{\!\mu}\widehat{\mathcal{F}}_\nu$ &  & \\\hline
        22 & $\widehat{\mathcal{F}}^\mu \widehat{\mathcal{F}}^\nu \widehat{\mathcal{F}}^\rho \widehat{\mathcal{F}}^\sigma\lsp\widehat{\nabla}_{\!\mu}\widehat{\nabla}_{\!\nu}\widehat{\mathcal{F}}^\tau\lsp\widehat{\nabla}_{\!\rho}\widehat{\nabla}_{\!\sigma}\widehat{\mathcal{F}}_\tau$ & \checkmark & \checkmark  \\\hline
        23 & $\widehat{\mathcal{F}}^\mu \widehat{\mathcal{F}}^\nu \widehat{\mathcal{F}}^\rho \widehat{\mathcal{F}}^\sigma\lsp\widehat{\nabla}_{\!\rho}\widehat{\nabla}_{\!\nu}\widehat{\mathcal{F}}_\mu\lsp\widehat{\nabla}^2\widehat{\mathcal{F}}_\sigma$ &  & \\\hline
        24 & $\widehat{\mathcal{F}}^\mu \widehat{\mathcal{F}}^\nu \widehat{\mathcal{F}}^\rho \widehat{\mathcal{F}}^\sigma\lsp\widehat{\nabla}_{\!\sigma}\widehat{\nabla}_{\!\rho}\widehat{\mathcal{F}}^\tau\lsp\widehat{\nabla}_{\!\tau}\widehat{\nabla}_{\!\nu}\widehat{\mathcal{F}}_\mu$ &  & \\\hline
        25 & $\widehat{\mathcal{F}}^\mu \widehat{\mathcal{F}}^\nu \widehat{\mathcal{F}}^\rho \widehat{\mathcal{F}}^\sigma\lsp\widehat{\nabla}^\tau\widehat{\nabla}_{\!\sigma}\widehat{\mathcal{F}}_\rho\lsp\widehat{\nabla}_{\!\tau}\widehat{\nabla}_{\!\nu}\widehat{\mathcal{F}}_\mu$ & \checkmark & \\\hline
        26 & $\widehat{\mathcal{F}}^\mu \widehat{\mathcal{F}}^\nu \widehat{\mathcal{F}}^\rho\lsp \widehat{\nabla}_{\!\mu}\widehat{\mathcal{F}}^\sigma\lsp\widehat{\nabla}_{\!\sigma}\widehat{\mathcal{F}}^\tau\lsp\widehat{\nabla}_{\!\rho}\widehat{\nabla}_{\!\nu}\widehat{\mathcal{F}}_\tau$ &  & \\\hline
        27 & $\widehat{\mathcal{F}}^\mu \widehat{\mathcal{F}}^\nu \widehat{\mathcal{F}}^\rho\lsp \widehat{\nabla}_{\!\mu}\widehat{\mathcal{F}}^\sigma\lsp\widehat{\nabla}_{\!\nu}\widehat{\mathcal{F}}^\tau\lsp\widehat{\nabla}_{\!\tau}\widehat{\nabla}_{\!\rho}\widehat{\mathcal{F}}_\sigma$ &  & \\\hline
        28 & $\widehat{\mathcal{F}}^\mu \widehat{\mathcal{F}}^\nu \widehat{\mathcal{F}}^\rho\lsp \widehat{\nabla}^{\sigma}\widehat{\mathcal{F}}^\tau\lsp\widehat{\nabla}_{\!\tau}\widehat{\mathcal{F}}_\sigma\lsp\widehat{\nabla}_{\!\rho}\widehat{\nabla}_{\!\nu}\widehat{\mathcal{F}}_\mu$ &  & \\\hline
        29 & $\widehat{\mathcal{F}}^\mu \widehat{\mathcal{F}}^\nu \widehat{\mathcal{F}}^\rho \widehat{\mathcal{F}}^\sigma \widehat{\mathcal{F}}^\tau \widehat{\mathcal{F}}^\omega \lsp \widehat{\nabla}_{\!\rho}\widehat{\nabla}_{\!\nu}\widehat{\mathcal{F}}_\mu\lsp\widehat{\nabla}_{\!\omega}\widehat{\nabla}_{\!\tau}\widehat{\mathcal{F}}_\sigma$ &  & \\\hline
    \end{tabular}
    \caption{Terms in the four-derivative effective action. Non-trivial contributions to two-point functions of $J^\mu$ and $T^{\mu\nu}$ for constant $\mathbb{F}_{\mu\nu}$ and constant $B$ are indicated with a check mark.}
    \label{tab:four-der-effact-terms}
\end{table}

\section{The effective action in different backgrounds}\label{sec:EFT_backgrounds}

We pursue two complementary strategies to identify and study the Wilson coefficients of our large $B$ EFT.  The first is to compute the partition function explicitly in specific backgrounds with curvature and magnetic field.  The second is to examine two-point correlation functions of the current and stress-energy tensor in a flat background with constant magnetic field.  It is the first method that we consider in this section, looking first at a flat background with varying magnetic field, then moving to the monopole background ${\mathbb R} \times S^2$, and finally considering a squashed $S^3$ and a spinning $S^2$.  In later sections with examples, we compute Wilson coefficients for free theories from the $S^3$ background, while in the holographic example we analyse the spinning $S^2$ setup. The partition function for the flat background with varying magnetic field was given for free theories in \cite{Gusynin:1998bt}, while the monopole background for free theories has appeared in (or can be straightforwardly reconstructed from) \cite{Pufu:2013eda, PhysRevX.12.031012, Boyack:2023uml}.

\subsection{Expansion in magnetic field in flat background}
Terms in the effective action can be associated with inhomogeneities in the magnetic field \cite{Gusynin:1998bt}. To find the corresponding expression from our EFT, we expand \eqref{eq:curved_space_W} using the background\footnote{Note that the time-independence of $B$ below follows from a Bianchi identity.}
\begin{equation}
    g_{\mu\nu}=\eta_{\mu\nu}\, ,\qquad \mathcal{F}_{\mu\nu}=\mathbb{F}_{\mu\nu},\label{eq:special_background}
\end{equation}
where the only non-vanishing components of $\mathbb{F}_{\mu\nu}$ are given by $\mathbb{F}_{12}=-\mathbb{F}_{21}=B(\vec{x})$. The hatted objects defined in \eqref{eq:g_hat_definition} and \eqref{eq:hat_defns} are given by
\begin{equation}
    \hat g_{\mu\nu} = B\lsp \eta_{\mu\nu} \, , \qquad \widehat {\mathcal F}_\mu =  ( - \sqrt{B},0,0 ) \, .
\end{equation}

With this background choice and after integrations by parts, the EFT action \eqref{eq:curved_space_W} simplifies to
\begin{equation}
    \mathcal{W}=\int d^{\lsp 3}x\, \big[c_0\lsp B(\vec{x})^{\f{3}{2}} +\tfrac{1}{4}B(\vec{x})^{-\f{3}{2}}\left(\p_i B(\vec{x})\right)^2 \left(2c_{2,1}-c_{2,3}+c_{2,4}\right)\big]\,,\label{eq:EFT_prediction_magnetic_BG}
\end{equation}
so that matching to previously computed actions will determine the coefficient $c_0$ along with the combination $2c_{2,1}-c_{2,3}+c_{2,4}$. We omit the $c_{4,n}$ terms here because of their length, but they are straightforward to reconstruct.

\subsection{Free energy on sphere and monopole operator scaling dimension}
Consider the CFT in Euclidean signature with metric
\begin{equation}
    ds^2=dr^2 +r^2d\Omega^2\,,\hspace{1cm} d\Omega^2=d\theta^2+\sin^2\lnsp\theta\, d\phi^2\,.
\end{equation}
Under the coordinate transformation $r=L\lsp e^{\tau/L}$, the above line element on $\mathbb{R}^3$ maps to a Weyl factor times the metric on the cylinder $\mathbb{R}\times S^2$:
\begin{equation}\label{eq:Euclidean_metric_g}
    ds^2=e^{2\tau/L}(d\tau^2 +L^2 d\Omega^2)\,.
\end{equation}
The advantage in working with the coordinates $(\tau,\theta,\phi)$ is that the dilatation operator $D=r\lsp\partial/\partial r$ in $\mathbb{R}^3$ maps to $L\lsp\partial/\partial\tau$ on $\mathbb{R}\times S^2$, where $\partial/\partial\tau$ is the generator of Euclidean time translations. This implies that energies of states on $S^2$ are given by $\Delta/L$, where $\Delta$ is the scaling dimension of an operator at the origin of $S^2$.

Now let us consider the scenario with a constant background magnetic field and no fluctuations. The ground state energy on $S^2$ can be attributed to a single insertion of a monopole operator with magnetic charge $Q=2BL^2$ at its centre \cite{Pufu:2013eda}. The background field strength components due to the magnetic monopole are given by
\begin{equation}
    \bF_{\theta\phi}=-\bF_{\phi\theta}=\tfrac12 Q \sin\theta \,,\qquad \bF_{\tau\theta}=\bF_{\tau\phi}=0\,.
\end{equation}
The magnetic monopole background is described by
\begin{equation}
    \bF^\mu =(\bF^\tau ,\bF^\theta ,\bF^\phi)=(B,0,0)\,,\label{eq:monopole_background_F}
\end{equation}
and the hatted objects defined in \eqref{eq:g_hat_definition} and \eqref{eq:hat_defns} are given by
\begin{equation}
    \hat{g}_{\mu\nu}=\text{diag}(B,\tfrac12 Q, \tfrac12 Q \sin^2\lnsp\theta)\,,\qquad \widehat{\cF}^\mu=\widehat{\bF}^\mu= (1/\sqrt{B},0,0)\,,\qquad \widehat{\cF}_\mu =\widehat{\bF}_\mu=(\sqrt{B},0,0)\,.\label{eq:background_choice_hated}
\end{equation}
With the Euclidean time direction compactified via $\tau\sim\tau+\beta$, the free energy is defined as
\begin{equation}
    \mathcal{E}(B)= -\frac{1}{\beta} \log \mathcal{Z}_{S_\beta\times S^2}=\frac{1}{\beta}\lsp\mathcal{W}_E[\mathbb{A},g_{\mathbb{R}\times S^2}]\,,\label{eq:free_energy}
\end{equation}
evaluated with field configuration \eqref{eq:background_choice_hated}. Note that in \eqref{eq:free_energy}, we use the Euclidean functional $\mathcal{W}_E$ to compute the Euclidean free energy, which is related to the Lorentzian functional \eqref{eq:curved_space_W} by the relation\footnote{There is an additional sign associated with the analytic continuation of the electromagnetic field strength. Since such terms do not appear in our monopole effective action, we will postpone the discussion of this sign to Section  \ref{sec:squashedthreesphere}.}
\begin{equation}
    \mathcal{W}_E=-i\mathcal{W}\big{|}_{t=-i\tau}.   
\end{equation}

As discussed above, the monopole operator's scaling dimension, which corresponds to the ground state energy on $S^2$, is given by $\Delta= L\lsp \mathcal{E}(B)$. Most of the Wilson coefficients in \eqref{eq:curved_space_W} do not contribute to the partition function when evaluated in the background \eqref{eq:background_choice_hated}, in large part because $\widehat{\nabla}_\mu \widehat {\mathcal F}_\nu = 0$. Keeping terms up to order $Q^{-1/2}$, the scaling dimension of the magnetic monopole operator is
\begin{equation}
    \Delta = -\sqrt{2}\pi c_0  Q^{3/2} - 4\sqrt{2} \pi c_{2,1} Q^{1/2} - 8\sqrt{2} \pi (2c_{4,1} + c_{4,2}) Q^{-1/2} + \cdots.
    \label{eq:EFT_MonOpDim}
\end{equation}
where we have assumed $B>0$.\foot{Comparing with \cite[Eq.~(5)]{Boyack:2023uml} under the substitution rule \eqref{eq:map_EFT_coefficients}, we find that their $c_3$ there should be multiplied by an extra factor of $2$.} On physical grounds we expect $c_0\leq0$, for $\Delta$ should be positive when $Q\gg1$. As we will see below in Section \ref{sec:dispersion}, the bound $c_0\leq0$ follows from a dispersive argument. This means that the free energy $\mathcal{E}(B)$ is positive for CFTs in large magnetic field.

We may define a magnetic susceptibility as the dimensionless quantity
\begin{equation}
    \chi_m=-\frac{\sqrt{B}}{4\pi L^2}\frac{\partial^2 \mathcal{E}(B)}{\partial B^2}=\frac34c_0-\frac{1}{Q}\lsp c_{2,1}+\frac{6}{Q^2}(2c_{4,1}+c_{4,2})+\cdots.
\end{equation}
This is simply proportional to the EFT coefficient $c_0$ in the limit $Q\to\infty$. The bound $c_0\leq0$ then implies that unitary CFTs that become gapped in the presence of a magnetic field display diamagnetic behaviour at large field, as the free energy increases with $B$ at large $B$.

\subsection{The squashed three-sphere and the spinning two-sphere}
\label{sec:squashedthreesphere}

While the monopole above has the advantage of being well studied where results can be read off from the literature, it does not allow access to a broad set of Wilson coefficients.  The situation improves if the $S^2$ is swapped for a squashed $S^3$ (in a Euclidean setting) or spun (in a Lorentzian one).

In the free theory examples below, we will compute the partition function on an $S^3$.  For completeness, we give the result for a squashed $S^3$ here.  
Let the line element be 
\begin{equation}
    \label{eq:ds2squashedS3}
    \frac{ds^2}{L^2} = d \theta^2 + \cos^2\lnsp \theta  (1+ \epsilon \cos^2\lnsp \theta) \, d \psi^2 + \sin^2\lnsp \theta (1+ \epsilon \sin^2 \lnsp\theta)\, d \phi^2
    +2 \epsilon \sin^2 \lnsp\theta \cos^2 \lnsp\theta \, d \phi \, d \psi \, ,
\end{equation}
where the ranges of the coordinates are
$0 < \theta < \frac{\pi}{2}$, $0 < \psi < 2 \pi$, $0 < \phi < 2 \pi$ and $L$ is the radius of the sphere. In this coordinate presentation, the $\phi$ circle closes off smoothly at $\theta = 0$ while the $\psi$ circle closes off at $\theta = \frac{\pi}{2}$, giving overall a space with the topology of an $S^3$.  If the squashing parameter $\epsilon$ is set to zero, the line element becomes that of a round three-sphere. The magnetic field is included by turning on the components $F_{\theta \psi} = Q \sin 2 \theta$ and $F_{\theta \phi} = -Q \sin 2\theta$.

The hatted field strength and metric (in the unprimed coordinate system) are
\begin{equation}
    \hat g_{\mu\nu} = \frac{2 Q}{ L^2} g_{\mu\nu}  \, , \qquad \widehat {\mathcal F}^\mu =\sqrt{\frac{1}{2Q(1+\epsilon)}} ( 0 ,1,1) \, .
\end{equation}
The effective action in this background has the form
\begin{align}
    \label{effactionthreesphere}
    \smash{\frac{\log Z}{\sqrt{1+\epsilon}}} &= 4 \sqrt{2} \pi^2 c_0 Q^{3/2} \nonumber \\
    &\quad + 2 \sqrt{2} \pi^2 \big[2 c_{2,1}(3-\epsilon) + (2 c_{2,2} + 3 c_{2,3}) (1+\epsilon) \big] Q^{1/2} \nonumber \\
    &\quad + 2 \sqrt{2} \pi^2 \big[ 2 c_{4,1} (3-\epsilon)^2 + 2( c_{4,5}-c_{4,10} - c_{4,11} ) (3-\epsilon)(1+\epsilon) \nonumber \\
    &\quad + (2 c_{4,3} + 2 c_{4,4} - 2 c_{4,6} + 2 c_{4,14} + 2 c_{4,15} - c_{4,18} + c_{4,19} + 2 c_{4,20} - c_{4,21})(1+\epsilon)^2 \nonumber \\
    &\quad - 2 ( c_{4,7} + c_{4,8}) (1-\epsilon^2) + 2c_{4,2}(3 -2 \epsilon + 3 \epsilon^2) \big] Q^{-1/2} + \cdots \,.
\end{align}
While $\epsilon = 0$ is the round sphere, in the limit $\epsilon \to -1$, treating the three-sphere as a Hopf fibration over an $S^2$, the $U(1)$ fibre shrinks to zero size---a sort of high temperature limit of the monopole background.  In the case $\epsilon =1$, the Ricci tensor components in the Hopf fibre direction vanish, while when $\epsilon=3$ the Ricci scalar vanishes. As promised, this background gives access to an additional linear combination of the $c_{2,n}$ and two additional linear combinations of the $c_{4,n}$.  

In matching this Euclidean result to the Lorentzian effective action, there is a subtle sign that is produced from Wick rotation of the electromagnetic field strength.  The net effect is that each term in the Euclidean result above needs to be modified by a factor of $(-1)^{n/2}$ where $n$ is the number of $\widehat {\mathcal F}_\mu$ fields that appear in that term.  So for example, $c_{2,2}$ and $c_{2,3}$ pick up a sign in moving from Euclidean to Lorentzian signature, while $c_{2,1}$ does not. In more detail, each $\widehat {\mathcal F}_\lambda$ is defined from $\widehat {\mathcal F}^{\mu\nu}$ by contraction with an $\epsilon_{\lambda \mu\nu}$ while each  $\widehat {\mathcal F}^\lambda$ is defined from $\widehat {\mathcal F}_{\mu\nu}$ by contraction with an $\epsilon^{\lambda \mu\nu}$. While in Euclidean signature, both $\epsilon^{012}$ and $\epsilon_{012}$ have the same sign, in Lorentzian signature they have opposite sign.  Thus, contracting these objects pairwise introduces the extra factor of $(-1)^{n/2}$.  

In the holographic analysis in Section \ref{sec:holo}, a particular spinning metric appears, namely
\begin{equation}
    \label{eq:spinningmetric}
    g_{\mu\nu}  = r^2 \left(
    \begin{array}{ccc}
    -L^{-2}  & 0 & \frac{a \sin^2\lnsp \theta}{L^2-a^2} \\
    0 & \frac{L^2}{L^2 - a^2 \cos^2 \lnsp\theta}& 0 \\
    \frac{a \sin^2\lnsp \theta}{L^2-a^2}  & 0 & \frac{L^2 \sin^2\lnsp \theta}{L^2-a^2}
    \end{array}
    \right) , \qquad
    {\mathcal F}_{\theta \phi}
    = \tfrac12Q \sin \theta \, ,
\end{equation}
with all other components of ${\mathcal F}_{\mu\nu}$ set to zero and $Q = 2 B L^2$ as above. The rate of rotation is set by $a$ while $L$ sets the size of the $S^2$. We include an overall rescaling by $r$ in order to match what appears in the holographic example in Section \ref{sec:holo}, although it has no physical effect. The hatted metric and field strength are then
\begin{equation}
    \hat g_{\mu\nu}  =B\left(L^2-a^2\right) \left(
    \begin{array}{ccc}
    -L^{-2}  & 0 & \frac{a \sin^2\lnsp \theta}{L^2-a^2} \\
    0 & \frac{L^2}{L^2 - a^2 \cos^2\lnsp \theta}& 0 \\
    \frac{a \sin^2\lnsp \theta}{L^2-a^2}  & 0 & \frac{L^2 \sin^2\lnsp \theta}{L^2-a^2}
    \end{array}
    \right) , \qquad
    \widehat {\mathcal F}^t = -\frac{L}{\sqrt{B(L^2-a^2)}} \, ,
\end{equation}
with $\widehat {\mathcal F}^\theta = \widehat {\mathcal F}^\phi=0$. One finds
\begin{align}
    \label{eq:effectiveF}
    \frac{ \log Z}{\beta \sqrt{L^2-a^2}} &= 4 \pi c_0 B^{3/2} L 
    + 
    \frac{4 \pi B^{1/2}}{3L (L^2-a^2)}
    \left(a^2 (-4 c_{2,1} + 2 c_{2,2} + 3 c_{2,3}) + 6 c_{2,1} L^2 \right) \nonumber
    \\
    &\quad + \frac{8 \pi B^{-1/2}}{15 L^3 (L^2-a^2)^{2} }
    \big[a^4 \big(80 c_{4,1}  + 31 c_{4,2}-7 c_{4,3} + 6 c_{4,4} - 20 c_{4,5} \nonumber \\ 
    &\hspace{2cm}+ 7 (c_{4,6} -c_{4,7} - c_{4,8})+ 20 c_{4,10} + 20 c_{4,11}
    -7 c_{4,14} + 6 c_{4,15} - 3 c_{4,18} \nonumber \\
    &\hspace{2cm} + 3 c_{4,19} + 6 c_{4,20}-3 c_{4,21}\big)- 5 a^2 L^2 (8 c_{4,1}+ 4 c_{4,2} -c_{4,3}-2 c_{4,5}+ c_{4,6}   \nonumber  \\
    & \hspace{2cm}- c_{4,7}-c_{4,8}+2 c_{4,10} + 2 c_{4,11}-c_{4,14})  + 15 L^4 (2 c_{4,1} + c_{4,2}) \big]
    \,.
\end{align}
As this metric is explicitly Lorentzian, the discussion above concerning this factor of $(-1)^{n/2}$ does not apply in this case. Similar to the squashed three-sphere, this background gives an additional linear combination of $c_{2,n}$ coefficients compared to the monopole example and two extra linear combinations of the $c_{4,n}$.

\section{Two-point functions and Ward identities}\label{sec:2pt_fun_Ward_identities}
The EFT construction we have presented in Section \ref{S:EFT_construction} allows us to evaluate the time-ordered correlation functions of the current tensor $J^\mu$ and the energy momentum tensor $T^{\mu\nu}$ by means of appropriate functional derivatives of $\mathcal{W}[\mathcal{F},g]$ with respect to $A_\mu$ and $g_{\mu\nu}$. Let us emphasise here that our effective action, which is purely local once expanded around flat spacetime and constant background field strength, is valid in the deep IR, where no dynamical degrees of freedom remain as a consequence of our assumption that the phase of our magnetic-field perturbed CFT is gapped. As a result, the time-ordered correlation functions we will discuss are given purely by contact terms. These still contain physical information about the integrated-out UV physics, encoded in the coefficients of the contact terms, which relate to various response parameters of the system.

With the definitions \eqref{eq:JTdefns}, the two-point functions we will be analysing are given by
\begin{align}
    \label{eq:2pf_JJ}
    \langle J^\mu(x_1) J^\nu(x_2)\rangle&=-i\frac{\delta^2 \mathcal{W}[\mathcal{A},\eta]}{\delta A_{\mu}(x_1)\,\delta A_\nu(x_2)}\Big{|}_{A=0}\,,\\
    \label{eq:2pf_TJ}
    \langle T^{\mu\nu}(x_1) J^\rho(x_2)\rangle&=-2i\lsp\frac{\delta^2 \mathcal{W}[\mathcal{A},g]}{\delta g_{\mu\nu}(x_1)\,\delta A_\rho(x_2)}\Big{|}_{g=\eta,A=0}\,,\\
    \label{eq:2pf_TT}
    \langle T^{\mu\nu}(x_1) T^{\rho\sigma}(x_2)\rangle&=-4i\lsp\frac{\delta}{\delta g_{\mu\nu}(x_1)}\frac{1}{\sqrt{-g(x_2)}}\frac{\delta\mathcal{W}[\mathcal{A},g]}{\delta g_{\rho\sigma}(x_2)}\Big{|}_{g=\eta,A=0}\, .
\end{align}
It is a straightforward task to compute these two-point functions using \eqref{eq:curved_space_W}. The results will be given by contact terms with coefficients involving $c_0, c_{2,1},c_{2,2},c_{2,3},c_{2,4}$ and $c_{4,n}$ for $n=1,\ldots,29$. Let us give an example using $S_0$, which will not be specialised to the case of constant background magnetic field of \eqref{eq:B_background}, but will be expressed in terms of a general constant background $\mathbb{F}_{\mu\nu}$. First, the one-point functions are given by\footnote{There is a total derivative contribution to $\langle J^\mu \rangle$, often called a magnetisation current, that we discard here. That these types of currents are total derivatives is well known in the condensed matter literature \cite{PhysRevB.55.2344}.}
\begin{equation}\label{eq:1pfs}
    \langle J^\mu(x)\rangle=0\,,\qquad \langle T^{\mu\nu}(x)\rangle=c_0\Big(\mathbb{F}^{3/2}\eta^{\mu\nu}+\frac{3}{2\sqrt{\mathbb{F}}}\lsp\mathbb{F}^{\mu\rho}\lsp \mathbb{F}_\rho{\!}^\nu\Big)\,,
\end{equation}
where $\mathbb{F}^2=\frac12\mathbb{F}^{\mu\nu}\mathbb{F}_{\mu\nu}$. The two-point function of the current is given by
\begin{equation}
    \langle J^\mu(x) J^\nu(0)\rangle = -\frac{3\lsp i\lsp c_0}{2\sqrt{\mathbb{F}}}\Big(\partial^\mu\partial^\nu-\eta^{\mu\nu}\partial^2+\frac{1}{2\lsp\mathbb{F}^2}\lsp\mathbb{F}^{\mu\rho} \lsp\mathbb{F}^{\nu\sigma}\partial_\rho\partial_\sigma\Big)\delta^{(3)}(x)\,,
\end{equation}
while
\begin{equation}
    \langle T^{\mu\nu}(x) J^\rho(0)\rangle = -\frac{3\lsp i\lsp c_0}{2\sqrt{\mathbb{F}}}\Big(\mathbb{F}^{\mu\rho}\partial^\nu+\mathbb{F}^{\nu\rho}\partial^\mu-(\eta^{\mu\rho}\lsp\mathbb{F}^{\nu\sigma}+\eta^{\nu\rho}\lsp\mathbb{F}^{\mu\sigma}-\eta^{\mu\nu}\lsp\mathbb{F}^{\rho\sigma})\partial_\sigma -\frac{1}{2\lsp\mathbb{F}^2}\lsp \mathbb{F}^{\mu\sigma}\lsp \mathbb{F}_\sigma{\!}^\nu\lsp\mathbb{F}^{\rho\tau}\partial_\tau\Big)\delta^{(3)}(x)
\end{equation}
and
\begin{equation}\label{eq:TT2PFc0}
    \begin{aligned}
        \langle T^{\mu\nu}(x) T^{\rho\sigma}(0)\rangle &=i\lsp c_0\lsp \mathbb{F}^{3/2}(\eta^{\mu\rho}\eta^{\nu\sigma}+\eta^{\mu\sigma}\eta^{\nu\rho})\delta^{(3)}(x) - \frac{3\lsp i\lsp c_0}{2\sqrt{\mathbb{F}}}\Big(\mathbb{F}^{\mu\rho}\lsp\mathbb{F}^{\nu\sigma} + \mathbb{F}^{\mu\sigma}\lsp\mathbb{F}^{\nu\rho}\\
        &\quad-\eta^{\mu\rho}\lsp\mathbb{F}^{\nu\tau}\lsp\mathbb{F}_\tau{\!}^\sigma-\eta^{\mu\sigma}\lsp\mathbb{F}^{\nu\tau}\lsp\mathbb{F}_\tau{\!}^\rho-\eta^{\nu\rho}\lsp\mathbb{F}^{\mu\tau}\lsp\mathbb{F}_\tau{\!}^\sigma-\eta^{\nu\sigma}\lsp\mathbb{F}^{\mu\tau}\lsp\mathbb{F}_\tau{\!}^\rho+\eta^{\rho\sigma}\lsp\mathbb{F}^{\mu\tau}\lsp\mathbb{F}_\tau{\!}^\nu\\
        &\quad-\frac{1}{2\lsp\mathbb{F}^2}\lsp\mathbb{F}^{\mu\tau}\lsp\mathbb{F}_\tau{\!}^\nu\lsp \mathbb{F}^{\rho\omega}\lsp\mathbb{F}_{\omega}{\!}^\sigma\Big)\delta^{(3)}(x)\,.
    \end{aligned}
\end{equation}
These correlators obey the Ward identities we find below. Note that due to our definition \eqref{eq:2pf_TT}, the two-point function $\langle T^{\mu\nu}(x) T^{\rho\sigma}(0)\rangle$ is not symmetric under exchange of the two energy-momentum tensors:
\begin{equation}
    \langle T^{\mu\nu}(x) T^{\rho\sigma}(0)\rangle-\langle T^{\rho\sigma}(x) T^{\mu\nu}(0)\rangle=\frac{3\lsp i\lsp c_0}{2\sqrt{\mathbb{F}}}(\eta^{\mu\nu}\mathbb{F}^{\rho\tau}\mathbb{F}_\tau{\!}^\sigma-\eta^{\rho\sigma}\mathbb{F}^{\mu\tau}\mathbb{F}_\tau{\!}^\nu)\delta^{(3)}(x)\,.
\end{equation}
Although the one-point functions given in \eqref{eq:1pfs} do not receive contributions from higher-derivative terms in the effective action, this is not the case for the two-point functions, which do receive such contributions that actually get quite complicated.

Gauge invariance as well as Weyl and diffeomorphism invariance lead to Ward identities at the level of correlation functions involving the conserved current and stress-energy tensor. Starting with gauge invariance, we need to demand that the effective action is invariant under
\begin{eqnarray}
    \mathcal{A}_\mu(x)\to\mathcal{A}'_\mu(x)=\mathcal{A}_\mu(x)+\partial_\mu\alpha(x)\,,
\end{eqnarray}
for any function $\alpha(x)$. Following standard steps, this leads to
\begin{equation}\label{eq:curr_cons}
    \nabla_{\lnsp\mu}\langle J^\mu(x)\rangle=0
\end{equation}
for the one-point function of the conserved current. The two-point function of the conserved current satisfies $\nabla_{\lnsp\mu}\langle J^\mu(x)J^\nu(0)\rangle=0$, and this can be easily extended to higher-point functions. If the stress-energy tensor is involved, then the associated functional derivative with respect to the metric will act on the Christoffel symbol found in the covariant derivative in \eqref{eq:curr_cons} and lead to 
\begin{equation}
    \nabla^{(y)}_{\lnsp\rho}\langle T^{\mu\nu}(x)J^\rho(y)\rangle-i g^{\mu\nu}(x)\langle J^\rho(y)\rangle\lsp\partial^{(y)}_\rho\delta^{(3)}(x-y)=0\,.
\end{equation}

Weyl invariance is the assumption of invariance of the effective action under the transformation
\begin{equation}
    g_{\mu\nu}(x)\to g'_{\mu\nu}(x)=g_{\mu\nu}(x)+2\sigma(x)g_{\mu\nu}(x)\,,
\end{equation}
for any function $\sigma(x)$. This leads to the Ward identity
\begin{equation}
    g_{\mu\nu}(x)\langle T^{\mu\nu}(x)\rangle=0\,.
\end{equation}
Taking further functional derivatives with respect to the gauge field to generate insertions of the conserved current will lead to Ward identities like
\begin{equation}
    g_{\mu\nu}(x)\langle T^{\mu\nu}(x)J^\rho(0)\rangle=0\,,
\end{equation}
but taking further functional derivatives with respect to the metric will lead to more complicated Ward identities due to the Leibniz rule, e.g.\
\begin{equation}
    g_{\rho\sigma}(y)\langle T^{\mu\nu}(x)T^{\rho\sigma}(y)\rangle+2i\langle T^{\mu\nu}(x)\rangle \delta^{(3)}(x-y)=0\,.
\end{equation}

Finally, let us assume that our theory in curved space and with the added background gauge field is invariant under diffeomorphisms, which are given by
\begin{equation}
    \begin{aligned}
        x^\mu&\rightarrow x^{\prime\mu}=x^\mu +\xi^\mu(x)\,\\
        g_{\mu\nu}(x)&\rightarrow g'_{\mu\nu}(x')=\f{\p x^\rho}{\p x^{\prime\mu}}\f{\p x^\sigma}{\p x^{\prime\nu}}g_{\rho\sigma}(x)\,,\\
        \phi(x)&\rightarrow\phi'(x')=\phi(x)\,,\\
        \cA_\mu(x)&\rightarrow \cA'_\mu(x')=\f{\p x^\rho}{\p x^{\prime\mu}}\cA_\rho(x)\,.
    \end{aligned}
\end{equation}
Treating $\xi^\mu(x)$ as infinitesimal, we find the following infinitesimal transformation laws in terms of Lie derivatives:
\begin{equation}\label{eq:diffeo_lies}
    \begin{alignedat}{2}
        g'_{\mu\nu}(x)&=g_{\mu\nu}(x)-\pounds_{\lnsp\xi} g_{\mu\nu}(x)\,,\qquad &\pounds_{\lnsp\xi} g_{\mu\nu}&=\nabla_\mu\xi_\nu +\nabla_\nu \xi_\mu\,,\\
        \phi'(x)&=\phi(x)-\pounds_{\lnsp\xi} \phi(x)\,, &\pounds_{\lnsp\xi} \phi&=\xi^\mu \partial_\mu\phi\,,\\     
        \cA'_\mu(x)&=\cA_\mu(x)-\pounds_{\lnsp\xi} \cA_\mu(x)\,, &\pounds_{\lnsp\xi} \cA_\mu&=\xi^\nu \nabla_\nu \cA_\mu +\cA_\nu \nabla_\mu\xi^\nu\,.
    \end{alignedat}
\end{equation}
Invariance of the action $S$ of our theory under diffeomorphisms implies
\begin{equation}
    S[\phi',\cA'_\mu,g'_{\mu\nu}]=S[\phi,\cA_\mu,g_{\mu\nu}]\,,
\end{equation}
which for an infinitesimal diffeomorphism becomes
\begin{equation}
    S[\phi',\cA_\mu-\pounds_{\lnsp\xi} \cA_\mu, g_{\mu\nu}-\pounds_{\lnsp\xi} g_{\mu\nu}]=S[\phi,\cA_\mu,g_{\mu\nu}]
\end{equation}
or
\begin{equation}
    S[\phi',\cA_\mu,g_{\mu\nu}]=S[\phi,\cA_\mu+\pounds_{\lnsp\xi} \cA_\mu,g_{\mu\nu}+\pounds_{\lnsp\xi} g_{\mu\nu}]\,.\label{eq:invariance_L}
\end{equation}
If we assume that the path integral measure is also invariant under diffeomorphisms, then we get
\begin{align}
    \int[D\phi]\,e^{i[\phi,\cA_\mu,g_{\mu\nu}]}&= \int [D\phi']\,e^{iS[\phi', \cA_\mu,g_{\mu\nu}]}\nn\\
    \label{eq:diff_pi}&=\int [D\phi]\, e^{iS[\phi,\cA_\mu+\pounds_{\lnsp\xi} \cA_\mu,g_{\mu\nu}+\pounds_{\lnsp\xi} g_{\mu\nu}]}\,,
\end{align}
where in the first line we made a change of variable $ \phi\rightarrow\phi'$ in the path integral and in the second line we used \eqref{eq:invariance_L} and $[D\phi']=[D\phi]$. The relation \eqref{eq:diff_pi} can also be considered as a proof of the diffeomorphism invariance of the partition function $\mathcal{Z}[\cA_\mu,g_{\mu\nu}]= \mathcal{Z}[\cA_\mu+\pounds_{\lnsp\xi} \cA_\mu,g_{\mu\nu}+\pounds_{\lnsp\xi} g_{\mu\nu}]$, starting from the diffeomorphism invariance of the action. Now, expanding the right-hand side of \eqref{eq:diff_pi} and keeping terms up to linear order in $\xi$ we get
\begin{equation}
    \int [D\phi]\, e^{iS[\phi,\cA_\mu, g_{\mu\nu}]} \left( \int d^{\lsp d}x\, \pounds_{\lnsp\xi} \cA_\mu(x)\frac{\delta S}{\delta \cA_{\mu}(x)}+\int d^{\lsp d}x\, \pounds_{\lnsp\xi} g_{\mu\nu}(x)\frac{\delta S}{\delta g_{\mu\nu}(x)}\right)=0\,,
\end{equation}
and use of \eqref{eq:diffeo_lies} leads, after using suitable integrations by parts and gauge invariance, to
\begin{equation}
    \int d^{\lsp d}x\sqrt{-g}\,\xi_\nu\left(\mathcal{F}_\mu{\!}^\nu\frac{1}{\sqrt{-g}}\frac{\delta}{\delta A_\mu(x)}+\nabla_{\lnsp\mu}\frac{2}{\sqrt{-g}}\frac{\delta}{\delta g_{\mu\nu}(x)}\right)\mathcal{W}[\mathcal{F},g]=0\,,
\end{equation}
which, using the definitions \eqref{eq:JTdefns}, gives rise to the Ward identity
\begin{equation}
    \nabla_{\lnsp\mu}\langle T^{\mu\nu}(x)\rangle + \mathcal{F}_\mu{\!}^\nu\langle J^\mu(x)\rangle=0\,.
\end{equation}
At the level of two-point functions we find the diffeomorphism Ward identities\footnote{
See \cite{Herzog:2009xv} for a detailed discussion of these Ward identities in a related context.
}
\begin{equation}\label{eq:Ward_TJ_diff}
    \nabla_{\lnsp\mu}^{(x)}\langle T^{\mu\nu}(x)J^{\rho}(y)\rangle -\mathcal{F}^{\nu}{\!}_{\sigma}\langle J^{\sigma}(x)J^{\rho}(y)\rangle-ig^{\nu\rho}(x)\langle J^{\sigma}(x)\rangle\lsp \partial_{\sigma}^{(y)}\delta^{(3)}(x-y)=0\,,
\end{equation}
and
\begin{equation}\label{eq:Ward_TT_diff}
    \begin{aligned}
    \nabla_{\lnsp\rho}^{(y)}&\langle T^{\mu\nu}(x)T^{\rho\sigma}(y)\rangle+\mathcal{F}_{\rho}{\!}^{\sigma} \langle T^{\mu\nu}(x)J^{\rho}(y)\rangle -i g^{\mu\nu}(x)\langle T^{\rho\sigma}(y)\rangle\lsp\partial_\rho^{(y)}\delta^{(3)}(x-y)\\
    &-i\big(g^{\mu\sigma}(x)\langle T^{\nu\rho}(y) \rangle+g^{\nu\sigma}(x)\langle T^{\mu\rho}(y) \rangle\big)\partial_\rho^{(y)}\delta^{(3)}(x-y)+i\langle T^{\mu\nu}(y)\rangle\lsp\partial^\sigma_{(y)}\delta^{(3)}(x-y)\\
    &\hspace{5cm}+i\big(g^{\mu\sigma}(x)\mathcal{F}_\rho{\!}^\nu+g^{\nu\sigma}(x)\mathcal{F}_\rho{\!}^\mu\big)\langle J^\rho(y)\rangle\lsp \delta^{(3)}(x-y)=0\,.
    \end{aligned}
\end{equation}

Constraints implied by these Ward identities are most easily analysed in momentum space. In general there are two momentum scalars, $k^2$ and $\vec{k}^2$, where $\vec{k}$ is the spatial part of $k$ with zero in the temporal part. We define $\vec{\eta}^{\mu\nu}$ to be the spatial part of $\eta^{\mu\nu}$ with zero in parts involving temporal components. There are two form factors in $\langle\tilde{J}^{\mu}(k) \tilde{J}^{\nu}(-k)\rangle$ after Ward identities are solved,
\begin{equation}
    \label{eq:JJ}
    \langle  \tilde{J}^\mu(k) \tilde{J}^\nu(-k)\rangle=\frac{a_1}{\sqrt{B}} (k^\mu k^\nu-\eta^{\mu\nu}k^2)+\frac{a_2 }{\sqrt{B}}(\vec{k}^{\mu} \vec{k}^{\nu}-\vec{\eta}^{\lsp\mu\nu}\vec{k}^{2})\,,
\end{equation}
where $a_1$ and $a_2$ are dimensionless scalar functions of $k^2/B$ and $\vec{k}^2/B$. Starting from our EFT action \eqref{eq:curved_space_W} to four-derivative order, we find
\begin{align}
    \label{eq:a1_JJ}
    a_1&=\frac{3i}{2}\lsp c_0-2 i\lsp c_{2,3}\lsp\frac{k^2}{B}+2 i\lsp c_{2,4}\lsp\frac{k^2-\vec{k}^2}{B}-2 i\lsp c_{4,14}\lsp\frac{(k^2)^2}{B^2}+2 i\lsp c_{4,16}\lsp\frac{k^2(k^2-\vec{k}^2)}{B^2}-2 i\lsp c_{4,22}\lsp\frac{(k^2-\vec{k}^2)^2}{B^2}\,,\\
    a_2&=-\frac{3i}{4}\lsp c_0+i\lsp c_{2,1}\frac{k^2}{B}+i\lsp c_{2,2}\lsp\frac{k^2-\vec{k}^2}{B}+\frac{3 i}{2} c_{2,3} \frac{k^2}{B}+\frac{i}{2}\lsp c_{2,4}\lsp\frac{\vec{k}^2}{B}+8 i\lsp c_{4,1}\lsp \frac{(k^2)^2}{B^2}+3 i\lsp c_{4,2}\lsp \frac{(k^2)^2}{B^2}\nn\\
    &\quad-\frac{i}{2}\lsp c_{4,3} \frac{k^2 (4k^2-3\vec{k}^2)}{B^2}+\frac{i}{2} c_{4,4}\frac{(2 k^2-\vec{k}^2)^2}{B^2}-2 i \lsp c_{4,5} \frac{k^2 (2 k^2-\vec{k}^2)}{B^2}-i\lsp c_{4,6}\frac{k^2 (k^2-\vec{k}^2)}{B^2}\nn\\
    \label{eq:a2_JJ}
    &\quad-i\lsp c_{4,9}\lsp\frac{k^2(k^2-\vec{k}^2)}{B^2}+\frac{i}{2}\lsp c_{4,12}\lsp\frac{(k^2-\vec{k}^2)(2k^2-\vec{k}^2)}{B^2}+2 i\lsp c_{4,14}\lsp\frac{(k^2)^2}{B^2}-i\lsp c_{4,16}\lsp\frac{k^2 (k^2-\vec{k}^2)}{B^2}\nn\\
    &\quad-\frac{i}{2}\lsp c_{4,22}\lsp\frac{\vec{k}^2(k^2-\vec{k}^2)}{B^2}\,.
\end{align}

With $\tilde{k}^\mu=\epsilon^{0\mu\nu}k_\nu$, there are eight form factors in $ \langle\widetilde{T}^{\mu\nu}(k)\tilde{J}^{\rho}(-k)\rangle$ before Ward identities are solved, so that
\begin{equation}
    \begin{aligned}
    \langle\widetilde{T}^{\mu\nu}(k) \tilde{J}^{\rho}(-k)\rangle&=\frac{b_1}{\sqrt{B}} k^\mu k^\nu \tilde{k}^{\rho} + \frac{b_2}{\sqrt{B}}\vec{k}^{\lsp\mu}\vec{k}^{\lsp\nu}\tilde{k}^{\rho}+ \frac{b_3}{\sqrt{B}}(k^\mu k^\rho \tilde{k}^{\nu}+ k^\nu k^\rho \tilde{k}^{\mu})\\
    &\quad+\frac{b_4}{\sqrt{B}}(\mathbb{F}^{\mu\rho}k^\nu + \mathbb{F}^{\nu\rho}k^\mu)+\frac{b_5}{\sqrt{B}}(\mathbb{F}^{\mu\rho}\vec{k}^{\nu} + \mathbb{F}^{\nu\rho}\vec{k}^{\mu})\\
    &\quad+\sqrt{B}\lsp b_6\lsp\eta^{\mu\nu}\tilde{k}^\rho +\sqrt{B}\lsp b_7\lsp\vec{\eta}^{\lsp\mu\nu}\tilde{k}^\rho+\sqrt{B}\lsp b_8(\eta^{\mu\rho}\tilde{k}^{\nu}+\eta^{\nu\rho}\tilde{k}^{\mu})\,.
    \end{aligned}
\end{equation}
The solution of the Ward identities for $\langle\widetilde{T}^{\mu\nu}(k) \tilde{J}^{\rho}(-k)\rangle$ is
\begin{equation}
    \begin{aligned}
    b_3&=-b_1-\frac{b_2}{2}\frac{\vec{k}^{2}}{k^2}+\frac{i}{2}(a_1+2a_2)\frac{B}{k^2}\,,\qquad b_4=ia_1\,,\qquad b_5=0\,,\\
    b_6&=-ia_1-b_1 \frac{k^2}{B}\,,\qquad b_7=ia_2-b_2\frac{\vec{k}^{2}}{B}\,,\qquad
    b_8=\frac{i}{2}(a_1-2a_2)+b_1\frac{k^2}{B}+ \frac{b_2}{2}\frac{\vec{k}^{2}}{B}\,.
    \end{aligned}
\end{equation}
To be consistent with EFT principles in gapped phases we need to demand that $b_2$ is such that there are no $k^2$ poles in $b_3$, i.e.\ that $b_2\vec{k}^2-i(a_1+2a_2)B$ is proportional to at least one power of $k^2$. Expressions for $b_1,\ldots, b_8$ in terms of Wilson coefficients, akin to \eqref{eq:a1_JJ} and \eqref{eq:a2_JJ}, can be found in the ancillary \emph{Mathematica} file.

Finally, aside from the $c_0$ contribution in \eqref{eq:TT2PFc0}, there are 13 form factors in $ \langle\widetilde{T}^{\mu\nu}(k)\widetilde{T}^{\rho\sigma}(-k)\rangle$ before Ward identities are solved. These, along with constraints implied by the Ward identities and expressions in terms of Wilson coefficients can be found in the ancillary \emph{Mathematica} file.

\section{Dispersion relations and constraints}\label{sec:dispersion}
For Poincar\'e-invariant quantum field theories, constraints follow from the Kramers--Kronig dispersion relation of the time-ordered two-point function. In \cite{Creminelli:2022onn}, a generalised Kramers--Kronig representation for the current two-point function was developed to derive bounds on the coefficients of the superfluid EFT, where Poincar\'e symmetry is spontaneously broken.\footnote{For further progress on positivity bounds in theories that break Lorentz symmetry, we refer the reader to \cite{Hui:2023pxc,Serra:2024tmz,Hui:2025aja}.} In this section, we further generalise the setup of \cite{Creminelli:2022onn} to derive the sum rule for EFT coefficients appearing in various two-point functions involving current and stress-energy tensor operators in a theory where Poincar\'e symmetry is explicitly broken by turning on a constant background magnetic field. We start our discussion in general spacetime dimension $d$, but derive bounds on the EFT coefficients in $d=3$, as appeared in \eqref{eq:curved_space_W}, in the magnetic field background \eqref{eq:B_background} assuming the spectrum is gapped.

Let us define an operator $O(x)$ of mass dimension $d$ to be the linear combination of operators involving the current $J^\mu(x)$ and stress-energy tensor $T^{\mu\nu}(x)$ given by
\begin{equation}
    O(x)= \partial^0\lnsp J^\mu(x)\lsp V_\mu+T^{\mu\nu}(x)\lsp U_{\mu\nu}\,,\label{eq:O_definition}
\end{equation}
where $V_\mu$ and $U_{\mu\nu}$ are arbitrary constant coefficients. Note that $O(x)$ is not a Lorentz scalar, as $V_\mu$ and $U_{\mu\nu}$ do not transform under Lorentz transformations. Let us provide the definitions of time-ordered, retarded and advanced two-point correlation functions of the operator $O(x)$, respectively, 
\begin{align}
    G_T(x_{12})&= \langle O(x_1)O(x_2)\rangle=\theta(x_{12}^0)\langle 0| O(x_1)O(x_2)|0\rangle +\theta(-x_{12}^0)\langle 0| O(x_2)O(x_1)|0\rangle\,,\label{eq:time_ordered_propagator}\\
    G_R(x_{12})&= i\theta(x_{12}^0)\langle 0|\left[O(x_1),O(x_2)\right]|0\rangle\,,\label{eq:retarded_propagator}\\
    G_A(x_{12})&= -i\theta(-x_{12}^0)\langle 0|\left[O(x_1),O(x_2)\right]|0\rangle\,.\label{eq:advanced_propagator}
\end{align}
We assume that the theory does not break translation symmetry so that the correlation functions are only functions of $x_{12}= x_1-x_2$. Note that the retarded Green's function $G_R(x_{12})$ is supported in the forward light cone (FLC: $x_{12}^0>0, (x_{12})^2<0$), and the advanced Green's function $G_A(x_{12})$ is supported in the backward light cone (BLC: $x_{12}^0<0, (x_{12})^2<0$). Also, the reality of the operator $O(x)$ implies that the retarded and advanced correlators are real, i.e.\
\begin{equation}
    \left(G_{R/A}(x_{12})\right)^*=G_{R/A}(x_{12})\,.\label{eq:reality_condition_G}
\end{equation}
From the definitions of retarded and advanced Green's functions, under $x_{1}\leftrightarrow x_2$ exchange we have
\begin{equation}
    G_{R}(-x_{12})=G_{A}(x_{12})\,.\label{eq:x12_exchange_property}
\end{equation}
The momentum space Green's functions are defined as Fourier transforms of the position space Green's functions, i.e.\
\begin{equation}
    \widetilde{G}_{T/R/A}(k)= \int d^{\lsp d}x\, e^{-ik\cdot x} G_{T/R/A}(x)\,.\label{eq:G_TRA_def_mom_space}
\end{equation}

\paragraph{Analyticity and crossing} We would like to study the properties of retarded and advanced correlators for complex values of momentum $k^\mu$. Let us decompose the complex momenta into $k^\mu=k^\mu_{\Re}+ik^\mu_{\Im}$, where the momentum components $k^\mu_{\Re}$ and $k^\mu_{\Im}$ are real. For retarded and advanced Green's functions, let us rewrite \eqref{eq:G_TRA_def_mom_space} for complex momentum as
\begin{equation}
    \widetilde{G}_{R/A}(k_{\Re}+ik_{\Im})= \int d^{\lsp d}x_{12}\, e^{-ik_{\Re}\cdot x_{12}+k_{\Im}\cdot x_{12}} G_{R/A}(x_{12})\,.
\end{equation}
Note that for $k_{\Im}\cdot x_{12}< 0$ the corresponding momentum space Green's function will be analytic as the integrand decays exponentially in the limit $|x_{12}|\rightarrow\infty$. For simplicity, we would like to work with a single complex variable and study the analyticity properties of Green's functions. It is thus useful to work with the momentum configuration\footnote{In \cite{Creminelli:2022onn} it was shown that even if we choose the momentum configuration to be more generic, namely $\vec{k}=\vec{\zeta}+\omega\, \vec{\xi}$ with $\vec{\zeta},\vec{\xi}\in \mathbb{R}^{d-1}$, the bounds on EFT coefficients derived from the dispersion relation can not be improved by considering $\vec{\zeta}\neq 0$.}
\begin{equation}
    k=\omega(1,\vec{\xi})\,,\qquad \hbox{with } \vec{\xi}\in \mathbb{R}^{d-1}\,,\label{eq:momentum_config}
\end{equation}
so that for complex $k^\mu$ only $\omega$ becomes complex and we denote it by $\omega =\omega_{\Re}+i\omega_{\Im}$. With this momentum configuration, $\widetilde{G}_{R}$ is analytic in the upper-half-plane (UHP) of complex $\omega$, since $k_{\Im}\cdot x_{12}=-\omega_{\Im}(x^0_{12}-\vec{x}_{12}\cdot\vec{\xi})<0$ for $x^\mu_{12}\in \text{FLC}$ under the condition $|\vec{\xi}|<1$. Analogously, $\widetilde{G}_{A}$ is analytic in the lower-half-plane (LHP) of complex $\omega$ under the condition $|\vec{\xi}|<1$. 

Now, by complex conjugation of both sides of \eqref{eq:G_TRA_def_mom_space} for complex $k^\mu$ and using the reality of the position space Green's functions \eqref{eq:reality_condition_G}, we get
\begin{equation}
    \big(\widetilde{G}_{R/A}(k)\big)^* = \int d^{\lsp d}x_{12}\, e^{ik^*\cdot x_{12}} G_{R/A}(x_{12})=\widetilde{G}_{R/A}(-k^*)\,.\label{eq:G_prop_1}
\end{equation}
On the other hand, if we change the integration variable from $x_{12}\rightarrow -x_{12}$ on the right-hand side of the above expression and use the property \eqref{eq:x12_exchange_property}, we get 
\begin{equation}
    \big(\widetilde{G}_{R/A}(k)\big)^* = \int d^{\lsp d}x_{12}\, e^{-ik^*\cdot x_{12}} G_{R/A}(-x_{12})=\widetilde{G}_{A/R}(k^*)\,.\label{eq:G_prop_2}
\end{equation}
The properties \eqref{eq:G_prop_1} and \eqref{eq:G_prop_2} are summarised in
\begin{equation}
    \big(\wt{G}_{R/A}(k)\big)^*=\wt{G}_{R/A}(-k^*)=\wt{G}_{A/R}(k^*)\,.\label{eq:G_prop_combined}
\end{equation}

\paragraph{Dispersion relation}
For simplicity, from now on, we work with the momentum space Green's functions  \eqref{eq:G_TRA_def_mom_space} with the momentum configuration \eqref{eq:momentum_config} and restricting to $\xi<1$. We denote the corresponding Green's function by
\begin{equation} 
    \widetilde{G}_{R/A}\big(k=\omega(1,\vec{\xi})\big)=\mathbb{G}_{R/A}(\omega)\,,\qquad \text{with $|\vec{\xi}|<1$}\,.
\end{equation}
To derive the dispersion relation, we work with the retarded Green's function $\mathbb{G}_{R}(\omega)$, which is analytic in the upper half of the complex $\omega$ plane. We also assume that the QFT has a mass gap given by $m$, and any other possible non-analyticity of  $\mathbb{G}_{R}(\omega)$ present for the real values of $\omega$ in the region $\omega\leq -m$ and $\omega\geq m$, as denoted by zigzag lines in Fig.\ \ref{fig:analytic_structure_GR_omegaUHP}. However, since we never cross into the LHP in deriving the dispersion relation, this derivation also holds when the QFT has no mass gap unless the presence of a massless particle introduces some other non-analyticity in the UHP. Since $\mathbb{G}_{R}(\omega)$ is analytic in the UHP in Fig.\ \ref{fig:analytic_structure_GR_omegaUHP}, for any integer value of $\ell$ the integration of $\omega^{-\ell-1}\mathbb{G}_{R}(\omega)$ along the contour $C= C_1\cup C_0\cup C_2\cup C_{\text{arc}_\infty}$ vanishes, i.e.\
\begin{equation}
    \oint_{C}\frac{d\omega}{\omega^{\ell+1}} \mathbb{G}_R(\omega) =0\,.\label{eq:contour_integral_initial}
\end{equation}
\begin{figure}[ht]
    \centering
    \begin{tikzpicture}[scale=0.8]
        \draw[thick,->] (-6,0)-- (6,0);
        \draw[thick,->] (0,-0.5)--(0,6);
        \draw[thick] (5.7,5.5)--(6.3,5.5);
        \draw[thick] (5.7,5.5)--(5.7,6);
        \draw (6,5.8) node{$\omega$};
        \draw[Dark2-B, thick, decoration = {zigzag,segment length = 2.5mm, amplitude = 1mm}, decorate] (2.05,0) -- (5.9,0);
        \draw[Dark2-B] (2,0) node{$\bullet$} node[below] {$m$};
        \draw[Dark2-B, thick, decoration = {zigzag,segment length = 2.5mm, amplitude = 1mm}, decorate] (-2.05,0) -- (-6,0);
        \draw[Dark2-B] (-2.0,0) node{$\bullet$} node[below] {$-m$};
        \draw[thick,Dark2-A] (0.4,0.2) arc [start angle=0, end angle=90, radius=0.4];
        \draw[thick,Dark2-A] (0,0.6) arc [start angle=90, end angle=180, radius=0.4];
        \draw[thick,Dark2-A] (0.4,0.2)-- (2.4,0.2);
        \draw[thick,Dark2-A] (2.4,0.2)-- (6.0,0.2);
        \draw[thick,Dark2-A] (-6.0,0.2)-- (-2.4,0.2);
        \draw[thick,Dark2-A] (-0.4,0.2)-- (-2.4,0.2);
        \draw[thick,Dark2-A] (6,0.2) arc [start angle=0, end angle=90, radius=6];
        \draw[thick,->,Dark2-A] (0,6.2) arc [start angle=90, end angle=135, radius=6];
        \draw[thick,Dark2-A] (0,6.2) arc [start angle=90, end angle=180, radius=6];
        \draw[Dark2-A] (0.35, 0.8) node{$C_0$};
        \draw[Dark2-A] (-3, 0.5) node{$C_1$};
        \draw[Dark2-A] (3, 0.5) node{$C_2$};
        \draw[Dark2-A] (4.0, 4.0) node{$C_{\text{arc}_\infty}$};
    \end{tikzpicture}
    \caption{The analytic structure of $\mathbb{G}_R(\omega)$ in complex $\omega$ plane. The red zigzag lines and red dots represent the possible singularities of $\mathbb{G}_R(\omega)$, carrying information about single-particle poles or multi-particle branch-cuts.}
    \label{fig:analytic_structure_GR_omegaUHP}
\end{figure}

We would like to work with the value of $\ell\in \mathbb{Z}_+$ that satisfies $\lim_{|\omega|\rightarrow\infty}\omega^{-\ell}\mathbb{G}_R(\omega)=0$, so that the contribution from the arc at infinity $C_{\text{arc}_\infty}$ vanishes for the above integrand. From the CFT correlation functions involving the stress-energy tensor and current, we find that $\mathbb{G}_R(\omega)$ behaves like $\omega^d$ for large $\omega$. Hence, we choose $\ell>d$ and \eqref{eq:contour_integral_initial} can be rewritten as
\begin{equation}
    \int_{C_0} \f{d\omega}{\omega^{\ell+1}}  \mathbb{G}_R(\omega) + \lim_{\epsilon\rightarrow 0^+}\int _{R}^\infty \frac{ds}{(s+i\epsilon)^{\ell+1}} \mathbb{G}_R(s+i\epsilon)+  \lim_{\epsilon\rightarrow 0^+}\int _{-\infty}^{-R} \frac{ds}{(s+i\epsilon)^{\ell+1}} \mathbb{G}_R(s+i\epsilon)=0\,.
\end{equation}
In the last expression, the $C_0$ contour is parametrised by $\omega =-Re^{-i\phi},\ \phi\in [0,\pi]$ and $s$ is a real parameter. Finally, at the end of the computation, we would like to take the $R\rightarrow 0$ limit. Changing the integration variable from $s$ to $-s$ for the last integral above, we get
\begin{equation}
    \int_{C_0} \f{d\omega}{\omega^{\ell+1}}  \mathbb{G}_R(\omega) + \lim_{\epsilon\rightarrow 0^+}\int _{R}^\infty ds \left[\f{1}{(s+i\epsilon)^{\ell+1}} \mathbb{G}_R(s+i\epsilon)+   \f{(-1)^{\ell+1}}{(s-i\epsilon)^{\ell+1}} \mathbb{G}_R(-s+i\epsilon)\right]=0\,.
\end{equation}
Now using \eqref{eq:G_prop_combined} we can rewrite $\mathbb{G}_R(-s+i\epsilon)=\mathbb{G}_A(s-i\epsilon)$. Since the integration is supported for $s\in (-\infty , -R]\cup [R,\infty)$, the terms $\f{1}{(s\pm i\epsilon)^{\ell+1}}$ can be replaced by their principal values $s^{-\ell-1}$. Performing all these steps, the above expression becomes
\begin{equation}
    \int_{C_0} \frac{d\omega}{\omega^{\ell+1}}  \mathbb{G}_R(\omega) + \int _{R}^\infty ds\, s^{-\ell-1} \lim_{\epsilon\rightarrow 0^+}\left[ \mathbb{G}_R(s+i\epsilon)+   (-1)^{\ell+1} \mathbb{G}_A(s-i\epsilon)\right]=0\,.\label{eq:dispersion_intermediate}
\end{equation}
Using the definitions of retarded and advanced propagators from \eqref{eq:retarded_propagator} and \eqref{eq:advanced_propagator}, for $\ell\in 2\mathbb{Z}_+$ the second integral can be simplified using the identity
\begin{equation}
    \lim_{\epsilon\rightarrow 0^+}\left[\mathbb{G}_R(s+i\epsilon)- \mathbb{G}_A(s-i\epsilon)\right] = i\int d^{\lsp d}x\, e^{-i\mathbf{s}\cdot x}\, \langle 0|\left[O(x),O(0)\right]|0\rangle \,, 
\end{equation}
where $\mathbf{s}=s(1,\vec{\xi})$ and $ |\vec{\xi}|<1$. Now, inserting a complete set of basis states for the resolution of the identity between the two operators and using the property of the translation symmetry generator $\widehat{P}^\mu$, the above expression reduces to
\begin{align}
    \lim_{\epsilon\rightarrow 0^+}\left[\mathbb{G}_R(s+i\epsilon)- \mathbb{G}_A(s-i\epsilon)\right] &= i\int d^{\lsp d}x\, e^{-i\mathbf{s}\cdot x}\Bigg[\langle0|e^{-i\widehat{P}\cdot x}O(0)e^{i\widehat{P}\cdot x}\left(\SumInt\limits_{\mathbf{n}}|\mathbf{n}\rangle \langle \mathbf{n}|\right)O(0)|0\rangle \nn\\
    &\hspace{3cm} -\langle0|O(0)\left(\SumInt\limits_{\mathbf{n}}|\mathbf{n}\rangle \langle \mathbf{n}|\right)e^{-i\widehat{P}\cdot x}O(0)e^{i\widehat{P}\cdot x}|0\rangle \Bigg]\nn\\
    &= i(2\pi)^d \SumInt\limits_{\mathbf{n}}\left[\delta^{(d)}(\mathbf{s}-p_n) -\delta^{(d)}(\mathbf{s}+p_n)\right] \big{|}\langle 0|O(0)|\mathbf{n}\rangle\big{|}^2 .
\end{align}
Here $p_n$ represents the momentum of the state $|{\bf n} \rangle$. When we substitute the above expression in \eqref{eq:dispersion_intermediate}, the vacuum state does not contribute because $s>R$.  Moreover, regarding the excited states,
only the first delta function contributes because the integration range of $s$ is positive. Indeed, given our assumption of a mass gap,
the integral over $s$ only needs to be carried out in the range $s\in [m,\infty)$. Finally, taking the limit $R\rightarrow 0$ for $\ell\in 2\mathbb{Z}_+$, \eqref{eq:dispersion_intermediate} becomes
\begin{equation}
    \lim_{|\omega|\rightarrow 0} \int_{C_0} \f{d\omega}{\omega^{\ell+1}}  \mathbb{G}_R(\omega) + i (2\pi)^d \int _{m}^\infty ds\, s^{-\ell-1} \SumInt\limits_{\mathbf{n}}\delta^{(d)}(\mathbf{s}-p_n)  \big{|}\langle 0|O(0)|\mathbf{n}\rangle\big{|}^2=0.\label{eq:dispersion_temp}
\end{equation}
The first integral above can be performed on the contour $C_0$ using the parametrisation $\omega =-Re^{-i\phi},\ \phi\in [0,\pi]$ after Taylor expanding $\mathbb{G}_R(\omega)$ around $\omega=0$. Note that for a gapped QFT, $\mathbb{G}_R(\omega)$ has a regular Taylor expansion near $\omega=0$ given by
\begin{equation}
    \mathbb{G}_R(\omega)=\sum_{n=0}^\infty \f{\omega^n}{n!} \mathbb{G}^{(n)}_R(0),\hspace{1cm} \mathbb{G}^{(n)}_R(0)=\f{d^n}{d\omega^n}\mathbb{G}_R(\omega)\Big{|}_{\omega=0}\,.
\end{equation}
Finally, after substituting the above expression into the integrand of the first integral in \eqref{eq:dispersion_temp}, all terms with $n\neq\ell$ are real. However, the contribution of the second integral is purely imaginary. Hence, focusing on the imaginary contribution from \eqref{eq:dispersion_temp}, we obtain
\begin{equation}
    \mathbb{G}^{(\ell)}_R(0) = \f{\ell!}{\pi} (2\pi)^d \int _{m}^\infty ds\, s^{-\ell-1} \SumInt\limits_{\mathbf{n}}\delta^{(d)}(\mathbf{s}-p_n)  \big{|}\langle 0|O(0)|\mathbf{n}\rangle\big{|}^2\,.
\end{equation}
Now since the right-hand side of the above expression is non-negative, this implies
\begin{equation}
    \mathbb{G}^{(\ell)}_R(0) \geq 0\,, \qquad \text{for $\ell>d$ and $\ell\in 2\mathbb{Z}_+$}\,.\label{eq:positivity_GR}
\end{equation}

The above positivity condition is derived in terms of the retarded correlation function. However, the EFT action \eqref{eq:curved_space_W} contributes to the low-energy expansion of the time-ordered correlation function. From the definitions in \eqref{eq:time_ordered_propagator} and \eqref{eq:retarded_propagator}, we find
\begin{equation}
    G_R(x_{12})=i\llsp G_T(x_{12}) -i\langle 0|O(x_2)O(x_1)|0\rangle \,.
\end{equation}
Performing a Fourier transformation on both sides, using translation symmetry, and introducing the resolution of identity in terms of a complete set of basis states, the above expression in momentum space can be rewritten as
\begin{equation}
    \widetilde{G}_{R}(k)=i\llsp \widetilde{G}_{T}(k) -i(2\pi)^d \SumInt_{\mathbf{n}}\delta^{(d)}(k+p_n)\big{|}\langle 0|O(0)|\mathbf{n}\rangle \big{|}^2. 
\end{equation}
It turns out that, with the momentum configuration \eqref{eq:momentum_config}, the expansion of the retarded propagator around $\omega=0$ on the left-hand side receives contributions only from the contact terms present in the time-ordered Green's function on the right-hand side. Moreover, the second term on the right-hand side does not contribute to these terms, as rigorously argued in \cite{Creminelli:2022onn}. Hence, the positivity constraint in \eqref{eq:positivity_GR} translates to the following constraint on the low energy expansion of the time-ordered Green's function:
\begin{equation}
    i\lsp\mathbb{G}^{(\ell)}_T(0) \geq 0\,, \qquad \text{for $\ell>d$ and $\ell\in 2\mathbb{Z}_+$}\,,\label{eq:positivity_GT}
\end{equation}
where
\begin{equation}
    \mathbb{G}_T(\omega)=\widetilde{G}_T(\omega(1,\vec{\xi}))\,,\qquad \mathbb{G}^{(\ell)}_T(0)=\f{d^\ell}{d\omega^\ell}\mathbb{G}_T(\omega)\Big{|}_{\omega=0}\,,\qquad\text{with $|\vec{\xi}|<1$}\,.
\end{equation}

Using the definition \eqref{eq:O_definition} of the operator $O(x)$ in terms of current and stress-energy tensor and following the conventions of the time-ordered Green's function in \eqref{eq:time_ordered_propagator}, the above positivity constraint in $d=3$ can be re-written as
\begin{equation}
    \begin{pmatrix}
    V_\mu & U_{\rho\sigma}
    \end{pmatrix}  \left[\f{d^\ell}{d\omega^\ell}\begin{pmatrix}
    i\omega^2 \langle \widetilde{J}^\mu(k) \widetilde{J}^\nu(-k)\rangle & -\omega \langle \widetilde{J}^\mu(k) \widetilde{T}^{\alpha\beta}(-k)\rangle \\
    \omega  \langle \widetilde{T}^{\rho\sigma}(k) \widetilde{J}^\nu(-k)\rangle & i\langle \widetilde{T}^{\rho\sigma}(k) \widetilde{T}^{\alpha\beta}(-k)\rangle
    \end{pmatrix}\right]_{\omega=0}\begin{pmatrix}
    V_\nu \\ U_{\alpha\beta}
    \end{pmatrix}\geq 0\,,\label{eq:Semi_positive_condition}
\end{equation}
for the momentum configuration $k=\omega(1,\vec{\xi})$ with $|\vec{\xi}|<1$, for the values of $\ell\in 2\mathbb{Z}$ and for all $\ell\geq 4$, and for arbitrary choice of $V_\mu$ and $U_{\mu\nu}$. The above positivity constraint will be satisfied for an arbitrary choice of $V_\mu$ and $U_{\mu\nu}$ if the matrix
\begin{equation}
    \begin{pmatrix}
    i\omega^2 \langle \widetilde{J}^\mu(k) \widetilde{J}^\nu(-k)\rangle & -\omega \langle \widetilde{J}^\mu(k) \widetilde{T}^{\alpha\beta}(-k)\rangle \\
    \omega  \langle \widetilde{T}^{\rho\sigma}(k) \widetilde{J}^\nu(-k)\rangle & i\langle \widetilde{T}^{\rho\sigma}(k) \widetilde{T}^{\alpha\beta}(-k)\rangle
    \end{pmatrix},\label{eq:matrix_positive}
\end{equation}
yields a positive semi-definite matrix for each power of $\omega^\ell$ with $\ell=4,6,8,\ldots$ when the two-point functions are expanded in $\omega$.

\subsection{Bounds on EFT coefficients}
\label{sec:bounds}
In Section \ref{sec:2pt_fun_Ward_identities} we computed the low-energy expansion of the correlation functions appearing in the matrix \eqref{eq:matrix_positive}. From the Ward identities, it was observed that $T^{\mu\nu}$ is symmetric and traceless. Hence, in $d=3$, we select $T^{01}, T^{02}, T^{11}, T^{12}, T^{22}$ as the only independent components of $T^{\mu\nu}$, which makes \eqref{eq:matrix_positive} an $8 \times 8$ Hermitian matrix. Starting with this $8 \times 8$ matrix and imposing semi-positive definiteness of the coefficients of $\omega^4, \omega^6$, and $\omega^8$, we obtain the following inequalities, which arise from the diagonal entries at each order:
\begin{align}
    &c_0\leq0,\label{eq:bound_type_0_first}\\
    &c_{2,3}(1-\xi^2)-c_{2,4}\leq0\,,\label{eq:bound_type_1_first}\\
    &c_{4,14}-c_{4,16}+c_{4,22}+\xi^2(4c_{4,1}+2c_{4,2}-c_{4,3}+c_{4,4}-2c_{4,5})\geq0,\label{eq:bound_type_1_second}\\
    &c_{4,2}+c_{4,9}-\tfrac{1}{2}\xi^2(2c_{4,2}-c_{4,3}-c_{4,6}-c_{4,14})\geq0\,,\label{eq:bound_type_1_last}\\
    &\tfrac{1}{2}c_{2,3}(4-\xi^2)(1-\xi^2)+\xi^2\left(c_{2,2}+c_{2,1}(1-\xi^2)\right)-\tfrac{1}{2} c_{2,4}(2-\xi^2)^2\leq0\,, \label{eq:bound_type_2_first}\\
    &c_{4,22}-c_{4,16}(1-\xi^2)+c_{4,14}(1-\xi^2)^2\geq0\,,\label{eq:bound_type_2_second}\\
    &2(c_{4,14}-c_{4,16}+c_{4,22})+\xi^2(c_{4,9}-2c_{4,14}+c_{4,16}+c_{4,2}+c_{4,6}-c_{4,12})\nn\\
    &\hspace{5cm}-\tfrac{1}{2}(\xi^2)^2(2c_{4,2}-c_{4,3}+c_{4,6}-c_{4,14})\geq0\,,\label{eq:bound_type_2_third}\\
    &c_{4,2}+c_{4,9}+\tfrac{1}{2}\xi^2(c_{4,3}+2c_{4,6}+2c_{4,9}+4c_{4,14}-2c_{4,16}+c_{4,22}-c_{4,12})\nn\\
    &\hspace{5cm}+\tfrac{1}{2}(\xi^2)^2(16c_{4,1}+6c_{4,2}-c_{4,3}+c_{4,4}-4c_{4,5})\geq0\,,\label{eq:bound_type_2_fourth}\\
    &c_{4,2}+c_{4,9}-\tfrac{1}{2}\xi^2(4c_{4,2}-c_{4,3}+2c_{4,9}-c_{4,22}-c_{4,12})+\tfrac{1}{2}(\xi^2)^2(2c_{4,2}-c_{4,3}+c_{4,4})\geq0\,,\label{eq:bound_type_2_last}\\
    &4c_{4,14}(1-\xi^2)^2+c_{4,22}(2-\xi^2)^2-2(c_{4,6}+c_{4,9})\xi^2(1-\xi^2)-2c_{4,16}(2-\xi^2)(1-\xi^2)\nn\\
    &\hspace{5cm}+2(8c_{4,1}+3c_{4,2})\xi^2(1-\xi^2)^2+c_{4,4}\xi^2(2-\xi^2)^2+c_{4,12}\lsp\xi^2(2-\xi^2)\nn\\
    &\hspace{5cm}-c_{4,3}\xi^2(4-\xi^2)(1-\xi^2)-4c_{4,5}\xi^2(2-\xi^2)(1-\xi^2)\geq0\,.\label{eq:bound_type_3}
\end{align}
Here we have denoted $\xi^2=|\vec{\xi}|^2$.
There is a special locus of points where the coefficients of these polynomials in $\xi^2$ all separately vanish and the bounds are saturated: 
\begin{equation}
    \begin{aligned}
    c_0 &= 0 \ , \; \; \;
    c_{2,1} = c_{2,2}= c_{2,3} = c_{2,4} = 0 \ , \; \; \\
    c_{4,2} &= -c_{4,5} =- c_{4,9} = -2 c_{4,1} \, , \quad
    c_{4,3} = - 4 c_{4,1} \, , \quad
    c_{4,4}= c_{4,6} = c_{4,12} = c_{4,14} = c_{4,16} = c_{4,22} = 0 \, . \;
    \end{aligned}
\end{equation}
Comparing with (\ref{eq:EFT_MonOpDim}), we see for example that a theory in this corner of moduli space has zero magnetic monopole scaling dimension.

The first of these inequalities, \eqref{eq:bound_type_0_first}, is simple and requires no further analysis. After identifying $\xi^2=x$, the inequalities from \eqref{eq:bound_type_1_first} to \eqref{eq:bound_type_1_last} are of the form
\begin{equation}
    a+bx\geq 0\,,\qquad \forall\ x\in[0,1)\,,\label{eq:type_1_bound}
\end{equation}
which generates an allowed wedge-shaped region
\begin{equation}
    a\geq 0 \qquad\text{and}\qquad a+b\geq0\,.\label{eq:wedge}
\end{equation}
Applied to \eqref{eq:bound_type_1_first}, which after multiplication by $-1$ has $a=c_{2,4}-c_{2,3}$ and $b=c_{2,3}$, this yields the simple statements
\begin{equation}\label{eq:c24positivity}
    c_{2,4}\geq0\qquad\text{and}\qquad c_{2,3}\leq c_{2,4}\,.
\end{equation}

The inequalities from \eqref{eq:bound_type_2_first} to \eqref{eq:bound_type_2_last} are of the type
\begin{equation}
    a+bx+cx^2\geq 0\,,\qquad\forall\ x\in[0,1)\,.\label{eq:type_2_bound}
\end{equation}
Since this needs to be satisfied at $x=0$ we find again
\begin{equation}
    a\geq 0\,.
\end{equation}
In the boundary case $a=0$, \eqref{eq:type_2_bound} reduces to $x\lsp(b+c\lsp x)\geq0$, so that $b$ and $c$ are then subject to the wedge condition \eqref{eq:wedge} in place of $a$ and $b$. For $a>0$, dividing \eqref{eq:type_2_bound} by $a$ generates a condition on the two ratios $A=\frac{b}{a}$ and $B=\frac{c}{a}$, namely
\begin{equation}
    f(x)=Ax+Bx^2\geq -1\,,\qquad \forall\ x\in[0,1)\,.\label{eq:type_2_bound_simlified}
\end{equation}
First of all, the above condition has to be satisfied in the limit $x\to1$, i.e.\
\begin{equation}
    A+B\geq-1\,. \label{AB_condition}
\end{equation}
From the above expressions, the extremum is given by $f'(x_*)=0,\ x_*=-\f{A}{2B}$. When $B\leq 0$, $x_*$ corresponds to a maximum and this region is allowed once \eqref{AB_condition} is satisfied. When $B\geq 0$, we need to estimate whether $0\leq x_*<1$ or not. When $x_*<0$ and $x_*\geq 1$, the allowed region is $B\geq 0$ along with \eqref{AB_condition}. On the other hand, when $0\leq x_*<1$, the allowed region will be $f(x_*)\geq -1$. Following these steps of logic, from \eqref{eq:type_2_bound_simlified}, we find the allowed region shown in Fig.\ \ref{fig:genericquadbounds}. It is easiest to express this region in terms of its piecewise boundary. The bound $f(x_*)\geq-1$ is saturated when $B=A^2/4$, for which $x_*=-2/A$. As the bound $f(x_*)\geq-1$ is applicable only when $0\leq x_*< 1$, the boundary will only lie along the parabola $B=A^2/4$ as long as $A<-2$. We are thus led to the bounds
\begin{align}\label{eq:Bbranches}
    B & \geq 
    \begin{cases}
        -A - 1  & \text{if }  A \geq -2\,, \\
         \frac14 A^2 & \text{if } A< -2\,.
    \end{cases}
\end{align}
The two regions join smoothly at the point $(A,B) = (-2,1)$, which is shown as a blue dot in Fig.\ \ref{fig:genericquadbounds}.
\begin{figure}[ht]
    \centering
    \begin{tikzpicture}
        \begin{axis}[
            xmin=-10, xmax=10,
            ymin=-10, ymax=10,
            xlabel= $A$,
            ylabel= $B$,
            ylabel style={rotate=-90},
            title={Generic quadratic bound on Wilson coefficients}
        ]
        \addplot[name path=tr,domain=-10:-2,blue] {x^2/4};
        \addplot[name path=tl,domain=-2:10,blue] {-x-1};
        \path[name path=br] (axis cs:-10,10) -- (axis cs:-2,10);
        \path[name path=bl] (axis cs:-2,10) -- (axis cs:10,10);
        \addplot [
                thick,
                color=blue,
                fill=blue,
                fill opacity=0.1
            ]
            fill between[
                of=br and tr
            ];
        \addplot [
                thick,
                color=blue,
                fill=blue,
                fill opacity=0.1
            ]
            fill between[
                of=bl and tl
            ];
        \addplot +[
             only marks,
             mark=*,
             mark options={color=blue},
             mark size=2pt] coordinates {(-2,1)};
        \end{axis}
    \end{tikzpicture}
        \caption{Constraints of the type \eqref{eq:type_2_bound} lead, generically, to the allowed region shown here in blue for the ratios $A=\frac{b}{a}$ and $B=\frac{c}{a}$. The blue dot is the location where the two branches of \eqref{eq:Bbranches} meet.}
        \label{fig:genericquadbounds}
\end{figure}

\begin{figure}[ht]
\centering
\includegraphics[scale=0.55]{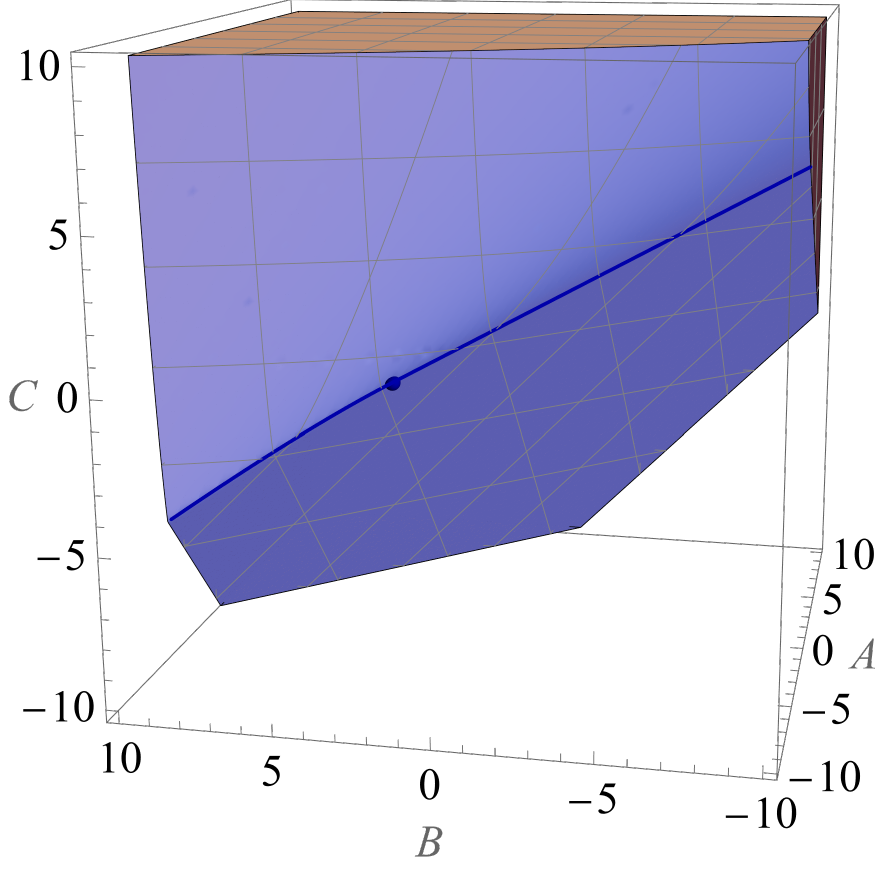}
\caption{Constraints of the type \eqref{eq:type_3_bound} lead, generically, to the allowed region shown here bounded by the orange surface for the ratios $A=\frac{b}{a}$, $B=\frac{c}{a}$, and $C=\frac{d}{a}$. The line indicates the piecewise curve along which the two bounds \eqref{eq:processedbounds} merge. The dot indicates the point $(-3,3,-1)$, at which the piecewise curve has a discontinuous second derivative.}
    \label{fig:genericcubicbounds}
\end{figure}

The final relation \eqref{eq:bound_type_3} is of the type
\begin{equation}
    a+bx+cx^2+dx^3\geq 0\,, \qquad \forall\ x\in[0,1)\,.\label{eq:type_3_bound}
\end{equation}
Since at $x=0$ the above inequality has to be satisfied, this implies
\begin{equation}
    a\geq 0\,.
\end{equation}
When $a=0$, the coefficients $b$, $c$ and $d$ are instead subject to the conditions derived above for $a$, $b$ and $c$ in \eqref{eq:type_2_bound}. For $a>0$, we can divide \eqref{eq:type_3_bound} by $a$ and find the two-variable condition
\begin{equation}
    g(x)= Ax+Bx^2+Cx^3\geq -1\,, \qquad \forall\ x\in[0,1)\,,\label{eq:type_3_bound_simlified}
\end{equation}
where $A=\frac{b}{a}$, $B=\frac{c}{a}$ and $C=\frac{d}{a}$. First of all, the above condition has to be satisfied in the limit $x\to1$, i.e.\
\begin{equation}
    A+B+C\geq-1\,. \label{ABC_condition}
\end{equation}
The extrema of $g(x)$ are given by
\begin{equation}
    g'(x_*^{\pm})=0\quad\Rightarrow\quad x_*^{\pm}=\f{-B\pm \sqrt{B^2-3AC}}{3C}\,,\qquad\text{with}\ g''(x^{\pm}_*)=\pm 2\sqrt{B^2-3AC}\,.
\end{equation}
First, note that when $B^2-3AC<0$ there is no extremum of $g(x)$ for real values of $x$; hence the condition \eqref{ABC_condition} is unchanged. Now for $B^2\geq 3AC$, once \eqref{ABC_condition} is obeyed we only need to focus on the minimum at $x^+_*$. If $x^+_*<0$ or $x^+_*\geq 1$,
the condition \eqref{ABC_condition} is unchanged. On the other hand, when $0\leq x^+_*<1$, there is the further restriction  $g(x^+_*)\geq -1$. Following these steps of logic, from \eqref{eq:type_3_bound_simlified}, we find the allowed region displayed in Fig.\ \ref{fig:genericcubicbounds}. Again, it is easiest to write down this region using its boundary, as
\begin{equation}
    \label{eq:processedbounds}
    C\geq \begin{cases}
        -1-A-B & \text{if } A\geq-3 \text{ and } B\geq -2A-3\,, \\
        -1-A-B & \text{if } A<-3 \text{ and } B\geq\tfrac{1}{4}(A^2-2A-3)\,,\\
       \frac{1}{27} \big(-2 A^3+2 (A^2-3 B)^{3/2}+9 A B\big) & \text{otherwise}\,.
    \end{cases}
\end{equation}

It is curious that our bounds involve curves and surfaces with discontinuous derivatives.
The simplest bound $a+bx\geq 0$ has a discontinuity at $a=0$ and $b=0$.  The
next simplest $a + bx + cx^2 \geq 0$ has a non-smooth point at $(A,B) = (-2,1)$.  The third type of bound $a + bx + cx^2 + dx^3 \geq 0$ has a non-smooth point at $(A,B,C) = (-3,3,-1)$.

For the examples in the following sections, it is useful to have the two-derivative bound expressed directly in terms of the $c_{2,n}$ instead of the linear combinations implicit in the general bound (\ref{eq:type_2_bound_simlified}). To that end, we start from (\ref{eq:bound_type_2_first}) and, setting $\xi^2=x$ as before, bring it to the form (\ref{eq:type_2_bound}) with
\begin{equation}
    a=2(c_{2,4}-c_{2,3})\,,\qquad b=\tfrac52 c_{2,3}-c_{2,1}-c_{2,2}-2c_{2,4}\,,\qquad c=c_{2,1}-\tfrac12 c_{2,3}+\tfrac12 c_{2,4}\,.
\end{equation}
The requirement $a\geq0$ is nothing but the $x=0$ case of \eqref{eq:bound_type_1_first}, and is guaranteed by \eqref{eq:c24positivity}. In the boundary case $a=0$, i.e.\ $c_{2,4}=c_{2,3}$, the bound reduces to $x\lsp(b+c\lsp x)\geq0$, and the wedge condition \eqref{eq:wedge} applied to $b+c\lsp x$ gives $c_{2,1}+c_{2,2}\leq\tfrac12 c_{2,3}$ and $c_{2,2}\leq\tfrac12 c_{2,3}$. For $a>0$ it is natural to introduce the rescaled coefficients $c'_{2,n}=c_{2,n}/(c_{2,4}-c_{2,3})$, which obey $c'_{2,4}-c'_{2,3}=1$ and in terms of which
\begin{equation}\label{eq:ABtwoderivative}
    A=\frac{b}{a}=\frac14\big(c'_{2,3}-2c'_{2,1}-2c'_{2,2}-4\big)\,,\qquad
    B=\frac{c}{a}=\frac14\big(2c'_{2,1}+1\big)\,.
\end{equation}
The allowed region is then precisely the one displayed in Fig.\ \ref{fig:genericquadbounds}, now shown again in Fig.\ \ref{fig:boundswithpoints} where we have included the values of the Wilson coefficients for the free complex scalar CFT and the free Dirac fermion CFT.
\begin{figure}[ht]
    \centering
    \begin{tikzpicture}
    \begin{axis}[
        xmin=-10, xmax=10,
        ymin=-5, ymax=5,
        xlabel= $A\,{=}\,\tfrac14\big(c'_{2,3}-2c'_{2,1}-2c'_{2,2}-4\big)$,
        ylabel= $B\,{=}\,\tfrac14\big(2c'_{2,1}+1\big)$,
        title={Quadratic bound on combinations of $c_{2,n}$'s},
        legend pos=outer north east
    ]
    \addplot +[
         only marks,
         mark=*,
         mark options={color=black},
         mark size=2pt] coordinates {(1.40918,-1.27279)};
    \addlegendentry{Free scalar}
    \addplot +[
         only marks,
         mark=asterisk,
         mark options={color=black},
         mark size=3pt] coordinates {(-0.6875,0.125)};
    \addlegendentry{Free fermion}
    \addplot[dashed, red!70!black, domain=-10:10] {0.0585215};
    \addlegendentry{Holographic example}
    \node[below right=-1pt, font=\footnotesize] at (axis cs:1.40918,-1.27279) {$(1.41,\, -1.27)$};
    \node[above right=-1.5pt, font=\footnotesize] at (axis cs:-0.6875,0.125) {$\left(-\frac{11}{16},\, \frac{1}{8}\right)$};
    \addplot[name path=tr,domain=-10:-2,blue] {x^2/4};
    \addplot[name path=tl,domain=-2:10,blue] {-x-1};
    \path[name path=br] (axis cs:-10,10) -- (axis cs:-2,10);
    \path[name path=bl] (axis cs:-2,10) -- (axis cs:10,10);
    \addplot [
            thick,
            color=blue,
            fill=blue,
            fill opacity=0.1
        ]
        fill between[
            of=br and tr
        ];
    \addplot [
            thick,
            color=blue,
            fill=blue,
            fill opacity=0.1
        ]
        fill between[
            of=bl and tl
        ];
    \end{axis}
    \end{tikzpicture}
    \caption{Bound (\ref{eq:bound_type_2_first}) with $A$ and $B$ given by (\ref{eq:ABtwoderivative}), including the free scalar and free fermion values for the $c_{2,n}$'s. One sees that the bound is obeyed in both cases, indicating that in the presence of a magnetic field the theories will be described by the EFT action (\ref{eq:curved_space_W}). (In the free fermion case there is also a zero mode. This needs to be added to the EFT, but it is not expected to affect substantially the local EFT we are discussing here.) The holographic example of Section \ref{sec:holo} determines $B$ but not $A$, and so corresponds to the dashed horizontal ray with $B\approx 0.0585$; see \eqref{eq:holoray} below.}
    \label{fig:boundswithpoints}
\end{figure}

\subsection{Complete constraints at order \texorpdfstring{$\omega^4$}{omega\^{}4}}
\label{sec:completeomega4}
The inequalities \eqref{eq:bound_type_0_first}--\eqref{eq:bound_type_3} follow from the diagonal entries of \eqref{eq:matrix_positive}. However, positive semi-definiteness is stronger than this. Expanding \eqref{eq:matrix_positive} in $\omega$, the Wilson coefficients appearing in each block are
\begin{equation}\label{eq:blockpattern}
    \begin{array}{|c|ccc|}
        \hline
         & \langle JJ\rangle & \langle JT\rangle & \langle TT\rangle\\
        \hline
        \omega^4 & c_0 & c_{2,n} & c_{4,n}\\
        \omega^6 & c_{2,n} & c_{4,n} & (c_{6,n})\\
        \omega^8 & c_{4,n} & (c_{6,n}) & (c_{8,n})\\
        \hline
    \end{array}
\end{equation}
At order $\omega^\ell$ the current-current block involves the coefficients of $(\ell-4)$-derivative terms, the mixed block those of $(\ell-2)$-derivative terms, and the stress-tensor block those of $\ell$-derivative terms. The entries shown in parentheses are not zero; they are simply not determined by \eqref{eq:curved_space_W}, which we have constructed only to four-derivative order, and would be generated by six- and eight-derivative operators. At orders $\omega^6$ and $\omega^8$ only the current-current block may consequently be used. In both cases that block has rank at most two, and its two potentially non-zero eigenvalues are precisely the inequalities already quoted, namely \eqref{eq:bound_type_1_first} and \eqref{eq:bound_type_2_first} at order $\omega^6$, and \eqref{eq:bound_type_2_second} and \eqref{eq:bound_type_3} at order $\omega^8$. Nothing further can be extracted there, except on loci where one of these eigenvalues vanishes: positivity of the full matrix then also requires the mixed entries along the corresponding null direction to vanish, since the undetermined stress-tensor block cannot compensate for them.

At order $\omega^4$, by contrast, the entire matrix is determined, and positivity does give more. Using invariance under rotations in the plane transverse to the magnetic field, we may set $\vec{\xi}=(\xi,0)$. The matrix then block-diagonalises into two pieces, spanned by $\{J^0,J^1,T^{02},T^{12}\}$ and $\{J^2,T^{01},T^{11},T^{22}\}$. Each piece has a null direction, $J^0-\xi\lsp J^1$ and $T^{11}+T^{22}-\xi\lsp T^{01}$ respectively, reflecting current conservation and the conservation and tracelessness of $T^{\mu\nu}$. Each therefore reduces to a $3\times3$ Hermitian matrix, which we take to be spanned by $\{J^1,T^{02},T^{12}\}$ and $\{J^2,T^{01},T^{11}\}$,
\begin{equation}\label{eq:omega4matrices}
    \mathcal{M}^{(a)}=\begin{pmatrix}
    \mu^{(a)} & i\lsp\nu^{(a)}_1 & i\lsp\nu^{(a)}_2\\[2pt]
    -i\lsp\nu^{(a)}_1 & \rho^{(a)}_1 & \rho^{(a)}_2\\[2pt]
    -i\lsp\nu^{(a)}_2 & \rho^{(a)}_2 & \rho^{(a)}_3
    \end{pmatrix},\qquad a=1,2\,,
\end{equation}
where, dropping an overall positive factor of $B^{-1/2}$,
\begin{align}
    \mu^{(1)} &= -\tfrac32\lsp c_0\,,\qquad
    \nu^{(1)}_1 = 2(c_{2,3}-c_{2,4})-\tfrac12(2c_{2,2}+3c_{2,3})\lsp\xi^2\,,\qquad
    \nu^{(1)}_2 = -\tfrac12(2c_{2,2}-c_{2,3})\lsp\xi\,,\nn\\
    \rho^{(1)}_1 &= 2(c_{4,14}-c_{4,16}+c_{4,22})+(c_{4,2}+c_{4,6}+c_{4,9}-c_{4,12}-2c_{4,14}+c_{4,16})\lsp\xi^2\nn\\
    &\hspace{5cm}+\tfrac12(c_{4,14}-2c_{4,2}+c_{4,3}-c_{4,6})\lsp\xi^4\,,\nn\\
    \rho^{(1)}_2 &= \big(c_{4,2}+c_{4,9}+c_{4,14}+\tfrac12 c_{4,6}-\tfrac12 c_{4,12}-\tfrac12 c_{4,16}\big)\lsp\xi-\tfrac12(c_{4,14}+2c_{4,2}-c_{4,3})\lsp\xi^3\,,\nn\\
    \rho^{(1)}_3 &= c_{4,2}+c_{4,9}+\tfrac12(c_{4,14}-2c_{4,2}+c_{4,3}+c_{4,6})\lsp\xi^2\,,
\end{align}
and
\begin{align}
    \mu^{(2)} &= -\tfrac34\lsp c_0(2-\xi^2)\,,\qquad
    \nu^{(2)}_1 = -2(c_{2,3}-c_{2,4})-(c_{2,1}-c_{2,3}+c_{2,4})\lsp\xi^2\,,\nn\\
    \nu^{(2)}_2 &= \tfrac12(2c_{2,2}-3c_{2,3}+2c_{2,4})\lsp\xi-\tfrac12(2c_{2,1}-c_{2,3}+c_{2,4})\lsp\xi^3\,,\nn\\
    \rho^{(2)}_1 &= 2(c_{4,14}-c_{4,16}+c_{4,22})+2(4c_{4,1}+2c_{4,2}-c_{4,3}+c_{4,4}-2c_{4,5})\lsp\xi^2\,,\nn\\
    \rho^{(2)}_2 &= \big(2c_{4,14}-\tfrac12 c_{4,12}-\tfrac32 c_{4,16}+c_{4,2}+c_{4,22}+\tfrac12 c_{4,6}+c_{4,9}\big)\lsp\xi\nn\\
    &\hspace{5cm}+(8c_{4,1}+3c_{4,2}-c_{4,3}+c_{4,4}-3c_{4,5})\lsp\xi^3\,,\nn\\
    \rho^{(2)}_3 &= c_{4,2}+c_{4,9}+\tfrac12(4c_{4,14}-c_{4,12}-2c_{4,16}+c_{4,22}+c_{4,3}+2c_{4,6}+2c_{4,9})\lsp\xi^2\nn\\
    &\hspace{5cm}+\tfrac12(16c_{4,1}+6c_{4,2}-c_{4,3}+c_{4,4}-4c_{4,5})\lsp\xi^4\,.
\end{align}
The complete content of \eqref{eq:Semi_positive_condition} at order $\omega^4$ is then the statement that both matrices are positive semi-definite,
\begin{equation}\label{eq:omega4complete}
    \mathcal{M}^{(1)}\succeq0\qquad\text{and}\qquad\mathcal{M}^{(2)}\succeq0\,,\qquad\forall\ \xi^2\in[0,1)\,.
\end{equation}

The diagonal entries of \eqref{eq:omega4matrices} recover what we already had: $\mu^{(1)},\mu^{(2)}\geq0$ are both \eqref{eq:bound_type_0_first}, while $\rho^{(2)}_1,\rho^{(1)}_3,\rho^{(1)}_1$ and $\rho^{(2)}_3$ are \eqref{eq:bound_type_1_second}, \eqref{eq:bound_type_1_last}, \eqref{eq:bound_type_2_third} and \eqref{eq:bound_type_2_fourth} respectively, and \eqref{eq:bound_type_2_last} is the quadratic form of $\mathcal{M}^{(2)}$ evaluated on $(0,\xi,-1)$, the direction corresponding to $T^{22}$. The off-diagonal entries, however, carry information that the diagonal analysis misses. Equivalently, all principal minors of $\mathcal{M}^{(1)}$ and $\mathcal{M}^{(2)}$ must be non-negative for all $\xi^2\in[0,1)$. It is straightforward to write down the resulting inequalities explicitly, but they are not very illuminating so we omit them. Let us only include the simplest instance, the $2\times2$ minor of either matrix at $\xi=0$, which gives
\begin{equation}\label{eq:cauchyschwarzbound}
    3\lsp|c_0|\lsp(c_{4,14}-c_{4,16}+c_{4,22})\geq4\lsp(c_{2,3}-c_{2,4})^2\,,
\end{equation}
a Cauchy--Schwarz-type relation bounding the two-derivative coefficients by the geometric mean of the zero- and four-derivative ones. This is not implied by \eqref{eq:bound_type_0_first} and \eqref{eq:bound_type_1_second}, which constrain the two factors on the left-hand side only separately. Indeed, the point $c_0=-1$, $c_{2,4}=1$, with all other Wilson coefficients set to zero, satisfies every one of \eqref{eq:bound_type_0_first}--\eqref{eq:bound_type_3} for all $\xi^2\in[0,1)$, while violating \eqref{eq:cauchyschwarzbound}.

\section{Free complex scalar field}\label{sec:complex_scalar}
Up to this point, the discussion concerning the EFT has been general, with minimal assumptions made about the underlying CFT. To be more concrete, we can choose a specific theory, compute the associated Wilson coefficients, and test the bounds (\ref{eq:bound_type_0_first})--(\ref{eq:bound_type_3}). Of considerable theoretical and phenomenological interest are free field theories in the presence of a strong magnetic field. These systems have been studied extensively in connection with  QED and scalar QED, and it is thus possible to obtain a number of Wilson coefficients simply by comparing quantities computed from (\ref{eq:curved_space_W}) with  results in the literature.

In this section, we consider the free complex scalar, relevant for the study of scalar QED in a strong magnetic field, and in the next we look at the free fermions.  By comparing with existing results \cite{Cangemi:1994by,Gusynin:1998bt} for systems in flat backgrounds with varying magnetic fields, the restricted effective action (\ref{eq:EFT_prediction_magnetic_BG}) determines the coefficient $c_0$ along with the combination $2c_{2,1}-c_{2,3}+c_{2,4}$.   Looking then at computations in monopole backgrounds \cite{Pufu:2013eda,PhysRevX.12.031012,Boyack:2023uml}, our (\ref{eq:EFT_MonOpDim}) provides a check of $c_0$ and a determination of $c_{2,1}$. To determine the remaining second order Wilson coefficients, we compute particular components of the current-current two-point function (\ref{eq:JJ}). Finally, as a check of the literature and our current-current computation, we compute the partition function on a three sphere, which comparing with (\ref{effactionthreesphere}) yields the combination $6 c_{2,1} - 2 c_{2,2} - 3 c_{2,3}$.  

\subsection{Current correlator for free scalar}\label{S:Scalar_propagator}
To set the stage, we begin with a brief literature review.  Comparing the scalar results \cite{Cangemi:1994by,Gusynin:1998bt}  in a flat background with varying magnetic field with (\ref{eq:EFT_prediction_magnetic_BG}), we can read off
\begin{equation}
    \label{eq:2c21mc23scalar}
    \begin{split}
    c_0&=\frac{(\sqrt{2}-1)\lsp\zeta\!\left(-\frac{1}{2}\right)}{2\pi} \ , \\
    2c_{2,1}-c_{2,3} + c_{2,4}&= \frac{1}{\sqrt{2}(2\pi)^2}\left[(\sqrt{2}-1)\frac{\pi}{4}\lsp\zeta\!\left(\f{1}{2}\right)+\left(1-\f{1}{2\sqrt{2}}\right) \frac{15}{16\pi}\lsp\zeta\!\left(\f{5}{2}\right)\right].
    \end{split}
\end{equation}
More precisely, $c_0$ and $2 c_{2,1} - c_{2,3}+c_{2,4}$ appear in \cite[Eq.\ (16)]{Cangemi:1994by} while $2 c_{2,1} - c_{2,3}+c_{2,4}$ can be extracted from \cite[Eq.\ (94)]{Gusynin:1998bt}.

Moving next to the monopole background with $Q$ units of magnetic flux, the eigenvalues and degeneracies of the Klein--Gordon operator with magnetic flux are well known. The free energy in the low temperature limit for a conformally coupled scalar can be computed from the sum \cite{Pufu:2013eda,PhysRevX.12.031012,Boyack:2023uml}
\begin{equation}
    E = \sum_{\ell=Q/2}^\infty (2 \ell+1) \sqrt{ \left( \ell+ \frac{1}{2}\right)^2 - \frac{Q^2}{4} } \ .
\end{equation}
Shifting the summation variable to $k = \ell-\frac{Q}{2}$ and expanding in the large $Q$ limit, this Casimir energy becomes
\begin{equation}
    E=\zeta \!\left(-\frac{1}{2}, \frac{1}{2} \right) Q^{3/2} + \frac{5}{2}   \zeta\! \left(-\frac{3}{2}, \frac{1}{2} \right) Q^{1/2}+ \frac{7}{8}\lsp\zeta\!\left(-\frac{5}{2}, \frac{1}{2} \right) Q^{-1/2} + \text{O}(Q^{-3/2})\,.
\end{equation}
Comparing this result with (\ref{eq:EFT_MonOpDim}), we can read off both $c_0$, which agrees with the previous result, and
\begin{equation}
    \label{eq:c0c21scalar}
    c_{2,1}=\frac{5 \left(2 \sqrt{2}-1\right) \zeta\! \left(-\frac{3}{2}\right)}{32 \pi }\,,
\end{equation}
along with a combination of $c_{4,1}$ and $c_{4,2}$ that we do not make precise here. Comparing (\ref{eq:2c21mc23scalar}) with (\ref{eq:c0c21scalar}), we can further isolate $c_{2,3}-c_{2,4}$:
\begin{equation}
    c_{2,3}-c_{2,4}=\frac{20 \left(2 \sqrt{2}-1\right) \zeta\! \left(-\frac{3}{2}\right)+\left(\sqrt{2}-2\right) \zeta\!\left(\frac{1}{2}\right)}{32 \pi }\,.
\end{equation}

To compute the remaining second order coefficients, we turn to $\langle \tilde{J}^\mu(k)\tilde{J}^\nu(-k)\rangle$. Following Schwinger's proper time formalism \cite{Schwinger:1951nm, Hattori:2023egw}, one can derive the full, non-perturbative scalar propagator, which takes a simpler form in momentum space
\begin{equation}
    \widetilde{G}(k)=\sum_{n=0}^\infty\frac{2e^{-\vec{k}^2/B}(-1)^n L_n(\frac{2\vec{k}^2}{B})}{k_3^2+m^2+(2n+1)B}\,.
    \label{eq:ScalarPropagator}
\end{equation}
Here $L_n(x)$ is the $n$'th Laguerre polynomial and $m$ is the mass of the scalar field which we later set to zero. Having completed the resummation of the background magnetic field, the correlation function $\langle \tilde{J}^\mu(k) \tilde{J}^\nu(-k)\rangle$  will receive contributions from four diagrams:
\begin{equation}
    \label{eq:JJdiagscalar}
    \begin{split}
    \langle \tilde{J}^\mu(k)\tilde{J}^\nu(-k)\rangle=&\begin{tikzpicture}[baseline=(vert_cent.base)]
    \node (vert_cent) at (0,0) {$\phantom{\cdot}$};
    \draw[ultra thick] (0,0) circle (1cm);
    \node[draw,circle,thick,black,cross,fill=white] at (-1,0) {};
    \node[draw,circle,thick,black,cross,fill=white] at (1,0) {};
    \node[inner sep=0pt] at (0,1) {$\blacktriangleright$};
    \node[inner sep=0pt] at (0,-1) {$\blacktriangleleft$};
    \draw (1,0) node[cross=4.5pt] {};
    \draw (-1,0) node[cross=4.5pt] {};
    \node[xshift=30pt] at (1,0) {$(2\ell+k)^\nu$};
    \node[xshift=-30pt] at (-1,0) {$(2\ell+k)^\mu$};
    \draw[-Stealth] (-0.3,1.2) arc[start angle=105,end angle=75,radius=1cm];
    \node[yshift=17pt] at (0,1) {$k+\ell$};
    \draw[-Stealth] (0.3,-1.2) arc[start angle=-75,end angle=-105,radius=1cm];
    \node[yshift=-17pt] at (0,-1) {$\ell$};
    \end{tikzpicture}+\begin{tikzpicture}[baseline=(vert_cent.base)]
    \node (vert_cent) at (0,0) {$\phantom{\cdot}$};
    \draw[ultra thick] (0,0) circle (1cm);
    \node[inner sep=0pt] at (0,1) {$\blacktriangleright$};
    \node[inner sep=0pt] at (0,-1) {$\blacktriangleleft$};
    \draw[decorate,decoration={coil,aspect=0}] (1,0) -- (2,1);
    \draw[decorate,decoration={coil,aspect=0}] (-1,0) -- (-2,1);
    \node[draw,circle,thick,black,cross,fill=white] at (2,1) {};
    \node[draw,circle,thick,black,cross,fill=white] at (-2,1) {};
    \node[xshift=13pt,yshift=2pt] at (2,1) {$\tilde{\mathbb{A}}^\nu$};
    \node[xshift=-13pt,yshift=2pt] at (-2,1) {$\tilde{\mathbb{A}}^\mu$};
    \draw[-Stealth] (-0.3,1.2) arc[start angle=105,end angle=75,radius=1cm];
    \node[yshift=17pt] at (0,1) {$k+\ell$};
    \draw[-Stealth] (0.3,-1.2) arc[start angle=-75,end angle=-105,radius=1cm];
    \node[yshift=-17pt] at (0,-1) {$\ell$};
    \end{tikzpicture}\\
    &+\left(\begin{tikzpicture}[baseline=(vert_cent.base)]
    \node (vert_cent) at (0,0) {$\phantom{\cdot}$};
    \draw[ultra thick] (0,0) circle (1cm);
    \node[inner sep=0pt] at (0,1) {$\blacktriangleright$};
    \node[inner sep=0pt] at (0,-1) {$\blacktriangleleft$};
    \draw[decorate,decoration={coil,aspect=0}] (1,0) -- (2,1);
    \node[draw,circle,thick,black,cross,fill=white] at (2,1) {};
    \node[draw,circle,thick,black,cross,fill=white] at (-1,0) {};
    \node[xshift=13pt,yshift=2pt] at (2,1) {$\tilde{\mathbb{A}}^\nu$};
    \node[xshift=-30pt] at (-1,0) {$(2\ell+k)^\mu$};
    \draw[-Stealth] (-0.3,1.2) arc[start angle=105,end angle=75,radius=1cm];
    \node[yshift=17pt] at (0,1) {$k+\ell$};
    \draw[-Stealth] (0.3,-1.2) arc[start angle=-75,end angle=-105,radius=1cm];
    \node[yshift=-17pt] at (0,-1) {$\ell$};
    \end{tikzpicture}+(\mu\leftrightarrow\nu)\right)-2\;\begin{tikzpicture}[baseline=(vert_cent.base)]
    \node (vert_cent) at (0,0.75) {$\phantom{\cdot}$};
    \node[inner sep=0pt] (c) at (0,0) {};
    \node[inner sep=0pt] (t) at (0,1.5) {$\blacktriangleright$};
    \draw[-Stealth] (-0.3,1.7) arc[start angle=105,end angle=75,radius=1cm];
    \draw[ultra thick] (0,0.75) circle (0.75cm);
    \node[yshift=-13pt] at (0,0) {$\eta^{\mu\nu}$};
    \node[yshift=17pt] at (0,1.5) {$\ell$};
    \node[draw,circle,thick,black,cross,fill=white] at (0,0) {};
    \end{tikzpicture}\;,
    \end{split}
\end{equation}
where $\tilde{\mathbb{A}}^\mu$ are the Fourier transforms of the background gauge fields responsible for the strong magnetic field. In the derivation of (\ref{eq:ScalarPropagator}), we have ignored the Schwinger phase. This is consistent only if we choose the background gauge field to be in the symmetric gauge:
\begin{equation}
    \mathbb{A}^1=-\frac{By}{2}\,,\qquad \mathbb{A}^2=\frac{Bx}{2}\,,\qquad \mathbb{A}^3=0\,.
\end{equation}
The Fourier transform of the background gauge field will produce derivatives with respect to the loop momentum acting on the scalar propagators, resulting in the correlation function
\begin{equation}
    \begin{split}
    \langle \tilde{J}^\mu(k)\tilde{J}^\nu(-k)\rangle=\int\frac{d^{\lsp 3}\ell}{(2\pi)^3}\bigg(&(2\ell+k)^\mu(2\ell+k)^\nu G(k+\ell)G(\ell)-2\eta^{\mu\nu}G(\ell)\\&-\frac{B}{2}\big((2\ell+k)^\mu(\delta^{\nu2}\partial_{\ell_1}-\delta^{\nu1}\partial_{\ell_2})G(k+\ell)G(\ell)+(\mu\leftrightarrow\nu)\big)\\ &-B^2(\delta^{\mu2}\partial_{\ell_1}-\delta^{\mu1}\partial_{\ell_2})G(k+\ell)(\delta^{\nu2}\partial_{\ell_1}-\delta^{\nu1}\partial_{\ell_2})G(\ell)\bigg)\,.
    \end{split}
\end{equation}
The second term may be computed fully, as it does not depend upon the external momentum, and is given by
\begin{equation}
    -2\eta^{\mu\nu}\int\frac{d^{\lsp 3}\ell}{(2\pi)^3}\sum_{n=0}^\infty\frac{2e^{-\vec{k}^2/B}(-1)^n L_n(\frac{2\vec{k}^2}{B})}{k_3^2+(2n+1)B}\,.
\end{equation}
Switching the order of the summation and integration, we can perform the integration over $\ell_3$ and $\vec{\ell}$ independently. Using the Laguerre polynomial identity
\begin{equation}
    \int_0^\infty dx\,e^{-x/2}L_n(x)=2(-1)^n\,,
\end{equation}
one finds that
\begin{equation}
    -2\;\begin{tikzpicture}[baseline=(vert_cent.base)]
    \node (vert_cent) at (0,0.75) {$\phantom{\cdot}$};
    \node[inner sep=0pt] (c) at (0,0) {};
    \node[inner sep=0pt] (t) at (0,1.5) {$\blacktriangleright$};
    \draw[-Stealth] (-0.3,1.7) arc[start angle=105,end angle=75,radius=1cm];
    \draw[ultra thick] (0,0.75) circle (0.75cm);
    \node[yshift=-13pt] at (0,0) {$\eta^{\mu\nu}$};
    \node[yshift=17pt] at (0,1.5) {$\ell$};
    \node[draw,circle,thick,black,cross,fill=white] at (0,0) {};
    \end{tikzpicture}=-\frac{\sqrt{B}}{2\pi}\left(1-\frac{1}{\sqrt{2}}\right)\zeta\left(\frac{1}{2}\right)\eta^{\mu\nu}\,,
    \label{eq:ffdiag}
\end{equation}
where we have also used the fact that
\begin{equation}
    \sum_{n=0}^\infty\frac{1}{(2n+1)^s}=(1-2^{-s})\lsp\zeta(s)
\end{equation}
when using zeta-function regularisation. 

The remaining diagrams must be evaluated in a small-$k$ expansion in order to match with the explicit expression (\ref{eq:JJ}). First, turning on only $k_3$, one finds that the contribution from the diagram
$$\begin{tikzpicture}[baseline=(vert_cent.base)]
    \node (vert_cent) at (0,0) {$\phantom{\cdot}$};
    \draw[ultra thick] (0,0) circle (1cm);
    \node[inner sep=0pt] at (0,1) {$\blacktriangleright$};
    \node[inner sep=0pt] at (0,-1) {$\blacktriangleleft$};
\draw[decorate,decoration={coil,aspect=0}] (1,0) -- (2,1);
\node[draw,circle,thick,black,cross,fill=white] at (2,1) {};
\node[draw,circle,thick,black,cross,fill=white] at (-1,0) {};
\node[xshift=13pt,yshift=2pt] at (2,1) {$\tilde{\mathbb{A}}^\nu$};
\node[xshift=-30pt] at (-1,0) {$(2\ell+k)^\mu$};
\draw[-Stealth] (-0.3,1.2) arc[start angle=105,end angle=75,radius=1cm];
\node[yshift=17pt] at (0,1) {$k+\ell$};
\draw[-Stealth] (0.3,-1.2) arc[start angle=-75,end angle=-105,radius=1cm];
\node[yshift=-17pt] at (0,-1) {$\ell$};
\end{tikzpicture}$$
in fact vanishes identically, as it will either be proportional to $\ell_1\ell_2$ or $\ell_1^2-\ell_2^2$, both of which vanish by the rotational invariance of the $\vec{\ell}$ integral. Using the Laguerre polynomial identities
\begin{equation}
    xL_n'(x)=nL_n(x)-nL_{n-1}(x)\,,\qquad L_n'(x)=-L_{n-1}^{(1)}(x)\,,
\end{equation}
to evaluate the derivatives on the Laguerre polynomials, one finds after another involved computation that
\begin{equation}
    \begin{split}
    \langle \tilde{J}^\mu(k)\tilde{J}^\nu(-k)\rangle\supset&(\delta^{\mu 1}\delta^{\nu 1}+\delta^{\mu 2}\delta^{\nu 2})\bigg( \frac{3(\sqrt{2}-1)\lsp\zeta\!\left(-\frac{1}{2}\right)}{4\pi \sqrt{|B|}}k_3^2\\&-\frac{640(2\sqrt{2}-1)\zeta\left(-\frac{3}{2}\right)+32(\sqrt{2}-2)\lsp\zeta\!\left(\frac{1}{2}\right)}{512\pi |B|^{3/2}}(k_3^2)^2\bigg)\,.
    \end{split}
\end{equation}
Matching with the expected form of the correlator yields an independent re-derivation of $c_0$ and $c_{2,3}-c_{2,4}$, consistent with results already obtained.

In order to obtain all of the two-derivative coefficients, we must consider also turning on spatial momenta. However, one notices from (\ref{eq:ScalarPropagator}) that this will result in terms with derivatives acting on the Laguerre polynomials, which drastically increases the difficulty of the integrals one must evaluate. Fortunately, a complete evaluation of $\langle \tilde{J}^{\mu}(k)\tilde{J}^\nu(-k)\rangle$ is not necessary, and it is possible to evaluate $c_{2,2}$ and disentangle $c_{2,3}$ from $c_{2,4}$ by considering only specific combinations of components and powers of the external momentum. Turning on only $k_3$ and $k_2$, one finds from (\ref{eq:JJ}) that at order $k_3^2k_2^2$ we have
\begin{equation}
    \langle \tilde{J}^1(k)\tilde{J}^1(-k)\rangle\supset k_2^2k_3^2\frac{2c_{2,1}+2c_{2,2}-5c_{2,3}
     + 4 c_{2,4}}{2 |B|^{3/2}}\,, \qquad \langle \tilde{J}^2(k)\tilde{J}^2(-k)\rangle\supset -k_2^2k_3^2\frac{2c_{2,3}
     }{|B|^{3/2}}\,.
\end{equation}
By evaluating the right-hand side of (\ref{eq:JJdiagscalar}) we can determine enough distinct combinations of coefficients to isolate them individually. As before, one finds that the diagram 
$$\begin{tikzpicture}[baseline=(vert_cent.base)]
    \node (vert_cent) at (0,0) {$\phantom{\cdot}$};
    \draw[ultra thick] (0,0) circle (1cm);
    \node[inner sep=0pt] at (0,1) {$\blacktriangleright$};
    \node[inner sep=0pt] at (0,-1) {$\blacktriangleleft$};
\draw[decorate,decoration={coil,aspect=0}] (1,0) -- (2,1);
\node[draw,circle,thick,black,cross,fill=white] at (2,1) {};
\node[draw,circle,thick,black,cross,fill=white] at (-1,0) {};
\node[xshift=13pt,yshift=2pt] at (2,1) {$\tilde{\mathbb{A}}^\nu$};
\node[xshift=-30pt] at (-1,0) {$(2\ell+k)^\mu$};
\draw[-Stealth] (-0.3,1.2) arc[start angle=105,end angle=75,radius=1cm];
\node[yshift=17pt] at (0,1) {$k+\ell$};
\draw[-Stealth] (0.3,-1.2) arc[start angle=-75,end angle=-105,radius=1cm];
\node[yshift=-17pt] at (0,-1) {$\ell$};
\end{tikzpicture}$$
leads only to trivially vanishing contributions for both components. Evaluating the remaining two diagrams one finds that
\begin{equation}
\begin{gathered}
    2c_{2,1}+2c_{2,2}-5c_{2,3}+4 c_{2,4}=-\frac{110\left(2 \sqrt{2}-1\right) \zeta\!\left(-\frac{3}{2}\right)+7\left(\sqrt{2}-2\right)\zeta\! \left(\frac{1}{2}\right)}{64\pi }\,,\\
    c_{2,3}=\frac{30\left(2 \sqrt{2}-1\right) \zeta\!\left(-\frac{3}{2}\right)+ \left(\sqrt{2}-2\right)\zeta\! \left(\frac{1}{2}\right)}{256 \pi }\,,
\end{gathered}
\end{equation}
which, combined with the previous expressions for $c_{2,1}$ and $c_{2,3}-c_{2,4}$, yields
\begin{equation}
\begin{gathered}
    c_{2,1}=\frac{5(-1+2\sqrt2)\lsp\zeta\left(-\frac{3}{2}\right)}{32\pi}\,,\qquad c_{2,2}=\frac{150(-1+2\sqrt{2})\lsp\zeta\left(-\frac{3}{2}\right)+5(-2+\sqrt{2})\lsp\zeta\left(\frac{1}{2}\right)}{512\pi}\,,\\
    c_{2,3}=\frac{30(-1+2\sqrt{2})\lsp\zeta\left(-\frac{3}{2}\right)+(-2+\sqrt{2})\lsp\zeta\left(\frac{1}{2}\right)}{256\pi}\,,\\ c_{2,4}=-\frac{130(-1+2\sqrt{2})\lsp\zeta\left(-\frac{3}{2}\right)+7(-2+\sqrt{2})\lsp\zeta\left(\frac{1}{2}\right)}{256\pi}\,.
\end{gathered}
\end{equation}
Given that $c_{2,3}\approx-0.000675$ and $c_{2,4}\approx0.000086$, one sees that the free complex scalar obeys the bounds in (\ref{eq:c24positivity}). Furthermore, computing the specific combination of coefficients $A$ and $B$ in (\ref{eq:ABtwoderivative}) one finds that the free complex scalar field obeys the bounds shown in Fig.\ \ref{fig:boundswithpoints}, as is to be expected because the theory is fully gapped by the presence of a magnetic field.

\subsection{Three-sphere partition function for free scalar}
We compute the partition function for a complex scalar field on a round three-sphere in the presence of $Q$ units of magnetic
flux.  In Appendix \ref{app:scalarS3}, we derive the eigenvalues and degeneracies in a slightly more general case, the squashed $S^3$. For the round case, the eigenvalues reduce to
\begin{equation}
    \lambda = (\ell+1)^2 - 2 Q (2p-\ell) + Q^2 - \tfrac{1}{4} \, ,
\end{equation}
where $\ell \in {\mathbb N} \cup \{0 \}$, $p = 0, 1, 2, \ldots, \ell$ and each $\lambda$ appears with degeneracy $\ell+1$. From this data, it is easy to write a formal expression for the partition function,
\begin{equation}
    \log Z = \sum_{\ell=0}^\infty \sum_{p = 0}^\ell (\ell+1) \log \left[ (\ell+1)^2 - 2 Q (2p-\ell) + Q^2 - \tfrac{1}{4} \right] \, ,
\end{equation}
which is also clearly divergent.  To regularise the answer, we consider the related sum
\begin{equation}
    I(s) =  \sum_{\ell=0}^\infty \sum_{p = 0}^\ell (\ell+1)  \left[  (\ell+1)^2 - 2 Q (2p-\ell) + Q^2 - \tfrac{1}{4} \right]^{-s} \, ,
\end{equation}
and define
\begin{equation}
    \log Z = - \partial_s I(s) |_{s=0} \, .
\end{equation}
Following this redefinition with a Mellin transform
\begin{equation}
    X^{-s} = \frac{1}{\Gamma(s)} \int_0^\infty dt \, t^{s-1} e^{-t X} \, ,
\end{equation}
we at last have a path toward an answer:
\begin{align}
    I(s) &= -\sum_{\ell=1}^\infty \frac{1}{\Gamma(s)} \int_0^\infty dt \, t^{s-1} \ell e^{-t (\ell^2+Q^2) + \frac{t}{4}}  \frac{\sinh 2 Q \ell t}{\sinh 2 Q t} \nonumber\\
    &= -\frac{1}{\Gamma(s)} \int_0^\infty dt \, \frac{t^{s-2} e^{-t(Q^2-1/4)}}{2 \sinh 2 Q t} \frac{d}{dQ} \sum_{\ell=1}^\infty e^{-t \ell^2} \cosh 2 Q \ell t\,,
\end{align}
where in the first line we shifted the summation index $\ell$ by one.

The next step in the strategy is to relate this sum to an elliptic theta function.
Note that
\begin{equation}
    \theta_3(z,q) = 1 + 2 \sum_{n=1}^\infty q^{n^2} \cos (2 \pi n z) \, .
\end{equation}
Thus,
\begin{equation}
    I(s) = - \frac{1}{\Gamma(s)} \int_0^\infty dt \, \frac{t^{s-2} e^{-t(Q^2+1/4)}}{4 \sinh 2 Q t} \frac{d}{dQ} \left[  \theta_3(iQt/\pi, e^{-t})- 1 \right] .
\end{equation}
We take advantage of the periodicity of this elliptic theta function,
\begin{equation}
    \theta_3(z+\tau; \tau)= e^{-i \pi (\tau+2z)} \theta_3(z;\tau) \, ,
\end{equation}
where $\theta_3 (z ; \tau) = \theta_3(z, e^{i \pi \tau})$. Periodicity and the integrality of $Q$ allow us to remove the $Q$ dependence from the theta function:
\begin{equation}
    \theta_3(n\tau; \tau) =
    \exp \left(-i \pi n^2 \tau \right) \theta_3(0 ; \tau) \ .
\end{equation}
Thus,
\begin{align}
    I(s) &= - \frac{1}{\Gamma(s)} \int_0^\infty dt \, \frac{t^{s-2} e^{-t(Q^2-1/4)}}{4 \sinh 2 Q t} \frac{d}{dQ} \left[ e^{Q^2 t} \lsp \theta_3\!\left(0; \frac{it}{ \pi} \right)  \right]\nonumber 
    \\
    &= - \frac{1}{\Gamma(s)} \int_0^\infty dt \, \frac{Q t^{s-1} e^{t/4}}{2 \sinh 2 Q t}\lsp
    \theta_3\!\left(0; \frac{it}{ \pi} \right)  .
    %   \sum_{n} e^{-n^2 t}  
\end{align}
Finally, we use the behaviour of the theta function under $\tau \to -1 / \tau$,
\begin{equation}
    \theta_3( 0; -1/\tau) = (-i \tau)^{1/2}  \theta_3(0; \tau)\,, 
\end{equation}
to find
\begin{align}
    I(s) &= - \frac{1}{\Gamma(s)} \int_0^\infty dt \, \frac{Q t^{s-2} e^{t/4}}{2 \sinh 2 Q t}  \left(\pi t \right)^{1/2} \theta_3\left(  0 ; \frac{i \pi}{t} \right) \nonumber \\
    &=  -\frac{1}{\Gamma(s)} \int_0^\infty dt \, \frac{\sqrt{\pi} Q t^{s-\frac{3}{2}}}{2 \sinh 2 Q t} \sum_n e^{- \frac{\pi^2 n^2}{t} + \frac{t}{4}}
    \, .
\end{align}

Let's define a new variable $Qt = t'$:
\begin{align}
    I &= - \frac{Q^{\frac{3}{2}-s} }{\Gamma(s)} \int_0^\infty dt' \, \frac{\sqrt{\pi}  t'^{s-\frac{3}{2}}}{2 \sinh 2 t'} \sum_n e^{- \frac{Q \pi^2 n^2}{t'}  + \frac{t'}{4Q}}  \, .
\end{align}
In this form, we at last see the leading $Q^{3/2}$ behaviour that should emerge in the $s\to 0$ limit at large $Q$. The $n=0$ and $n\neq 0$ terms in the sum need to be treated differently. We find
\begin{align}
    -\partial_s I |_{s=0} &= - 4\pi Q^{3/2} \sum_{j=0}^\infty  \sqrt{j + \frac{1}{2} - \frac{1}{16 Q}} + \sum_{j=0}^\infty \sum_{n=1}^\infty \frac{2Q}{n} e^{-\pi n \sqrt{-1 + 8Q + 16jQ}} \nonumber\\
    &= -4 \pi Q^{3/2} \zeta\!\left(-\frac{1}{2}, \frac{1}{2} - \frac{1}{16Q} \right) -2 Q  \sum_{j=0}^\infty \log \left(1 - e^{-\pi \sqrt{-1+8Q + 16 j Q}} \right).
\end{align}
The second sum is exponentially suppressed at large $Q$ so we focus on the first sum:
\begin{align}
    -\partial_s I |_{s=0} &= -4\pi  Q^{3/2}  \sum_{k=0}^\infty \left(
    \left(k + \frac{1}{2}\right)^{\frac{1}{2}} - \frac{1}{32  Q} \left(k + \frac{1}{2} \right)^{-\frac{1}{2}} - \frac{1}{2048 Q^2} \left( k + \frac{1}{2} \right)^{-3/2} + \ldots 
    \right),
\end{align}
where the ellipses denote higher terms in the $1/Q$ expansion as well as terms exponentially suppressed in $Q$.  Using zeta-function regularisation, the final answer is
\begin{align}
    -\partial_s I |_{s=0}&\approx -2  \pi ( \sqrt{2}-2)\lsp\zeta\! \left( -\frac{1}{2} \right) Q^{3/2} + \frac{\pi}{8} (\sqrt{2}-1) \lsp\zeta\! \left(\frac{1}{2} \right) Q^{1/2} 
    \nonumber \\
    &\quad +\frac{\pi}{512} (2 \sqrt{2}-1) \lsp\zeta\! \left(\frac{3}{2} \right) Q^{-1/2} + \cdots\,.
\end{align}
Comparing with the effective action (\ref{effactionthreesphere}), we find
\begin{align}
    c_0 &= \frac{1}{2\pi} (\sqrt{2}-1) \lsp\zeta\! \left(-\frac{1}{2} \right) , \\
    6 c_{2,1} + 2 c_{2,2} + 3 c_{2,3} &= -\frac{1}{32 \pi} (\sqrt{2}-2) \lsp\zeta \!\left( \frac{1}{2} \right) . 
\end{align}
We get a constraint on the $c_{4,n}$ coefficients as well. Making the adjustment described in Section \ref{sec:squashedthreesphere} that accompanies the transition from Euclidean to Lorentzian signature, $6 c_{2,1} + 2 c_{2,2} + 3 c_{2,3} \to 6 c_{2,1} - 2 c_{2,2} - 3 c_{2,3}$, we find agreement with the $c_{2,n}$ computed earlier!

\section{Free Dirac fermion}\label{sec:fermion}
\subsection{Current correlator for free fermion}
\label{S:Fermion_propagator}
First, consider placing a single, free, four-component fermion in a strong magnetic field. The flat space EFT in a varying magnetic field (\ref{eq:EFT_prediction_magnetic_BG}) at two derivative order has been computed in \cite{Gusynin:1998bt,Cangemi:1994by},
which lets us extract
\begin{align}
    \label{eq:c0_Cangemi_spinor}
    c_0&=\frac{2}{\sqrt{2}\pi}\lsp\zeta\!\left(-\frac{1}{2}\right)\,, \\
    2c_{2,1}-c_{2,3} +c_{2,4}&=\frac{15}{32\sqrt{2} \pi^3}\lsp\zeta\!\left(\frac{5}{2}\right).\label{eq:c2_Gusynin}
\end{align}
More precisely, \cite[Eq.\ (16)]{Cangemi:1994by} provides a result for $c_0$ and $2 c_{2,1} - c_{2,3}+c_{2,4}$ for two-component fermions while  \cite[Eq.\ (64)]{Gusynin:1998bt} gives $2 c_{2,1} - c_{2,3}+c_{2,4}$ in the four component case. The monopole dimension, or equivalently free energy on $S^1_\beta\times S^2$, has been computed in \cite{Pufu:2013eda,PhysRevX.12.031012,Boyack:2023uml}.  Comparing with our (\ref{eq:EFT_MonOpDim}) yields both $c_0$ (which agrees with (\ref{eq:c0_Cangemi_spinor})) and
\begin{equation}
    c_{2,1}=-\frac{15}{64\sqrt{2} \pi^3}\lsp\zeta\!\left(\f{5}{2}\right).\label{eq:c2_Boyack}
\end{equation}
Combining this expression with (\ref{eq:c2_Gusynin}) for $2c_{2,1}-c_{2,3}+c_{2,4}$, one finds 
\begin{equation}
    c_{2,3}-c_{2,4}=-\frac{15}{16\sqrt{2}\pi^3}\lsp\zeta\!\left(\frac{5}{2}\right)\,.
\end{equation}

The monopole analysis is straightforward once the spectrum and its degeneracies are known.  The dimension comes from a large magnetic field expansion of the sum \cite{Pufu:2013eda,PhysRevX.12.031012,Boyack:2023uml}
\begin{align}
    \Delta &=-4\sum_{\ell=\frac{Q}{2}+1}^\infty \ell\sqrt{\ell^2-\frac{Q^2}{4}}\nn\\
    &=-4\sum_{j=1}^\infty \left(j+\frac{Q}{2} \right)\sqrt{\left(j+\frac{Q}{2} \right)^2-\frac{Q^2}{4}} \nn\\
    &= -2Q^{3/2}\zeta\!\left(-\f{1}{2}\right) -5Q^{1/2}\zeta\!\left(-\frac{3}{2}\right)-\frac{7}{4} Q^{-1/2}\zeta\!\left(-\frac{5}{2}\right)+\cdots\,,
\end{align}
from which we can read off additionally a combination of $c_{4,1}$ and $c_{4,2}$ that we do not make precise. (We are assuming $Q$ is an integer and the initial sum over $\ell$ is either over values in ${\mathbb Z}$ or ${\mathbb Z} + \frac{1}{2}$, depending on the parity of $Q$.)

To determine the remaining second order coefficients, and as an independent check on these expressions, we must now turn to the two-point function $\langle \tilde{J}^\mu(k)\tilde{J}^\nu(-k)\rangle$. Using Schwinger's proper time technique \cite{Schwinger:1951nm}, it is possible to resum the contributions of the magnetic field and derive the non-perturbative fermion propagator, which, following \cite{Miransky:2015ava}, is most naturally expressed in terms of a sum over Landau Levels,\footnote{For the free fermion we are able to ignore the Schwinger phase, which will cancel when computing correlation functions.}
\begin{equation}
    \widetilde{G}(k)=-ie^{-\vec{k}^2/|B|}\sum_{n=0}^\infty(-1)^n\frac{D_n(B,\vec{k})}{k_3^2+2|B|n}\,,
    \label{eq:Greens_fun_momentum_space}
\end{equation}
where we have Wick rotated to Euclidean space with $k_0\rightarrow ik_3$. The functions $D_n(B,\vec{k})$ are given by
\begin{equation}
    \begin{split}
    D_n(B,\vec{k})=&-k_3\gamma_3\left[\bigg(1+i\gamma_1\gamma_2\,\text{sign}(B)\bigg) L_n^0\left(\frac{2\vec{k}^2}{|B|}\right)-\bigg(1-i\gamma_1\gamma_2\,\text{sign}(B)\bigg) L_{n-1}^0\left(\frac{2\vec{k}^2}{|B|}\right)\right]\\ &+4\vec{k}_i\gamma_i\, L_{n-1}^1\left(\frac{2\vec{k}^2}{|B|}\right)\,,
    \end{split}
    \label{eq:Dndef}
\end{equation}
where $L_n^k(x)$ are associated Laguerre polynomials.
Once the effect of the background magnetic field has been resummed into the full electron propagator, the correlation function $\langle \tilde{J}^\mu(k)\tilde{J}^\nu(-k)\rangle$ receives a contribution only from the diagram
$$\begin{tikzpicture}[baseline=(vert_cent.base)]
    \node (vert_cent) at (0,0) {$\phantom{\cdot}$};
    \draw[ultra thick] (0,0) circle (1cm);
    \node[draw,circle,thick,black,cross,fill=white] at (-1,0) {};
    \node[draw,circle,thick,black,cross,fill=white] at (1,0) {};
    \node[inner sep=0pt] at (0,1) {$\blacktriangleright$};
    \node[inner sep=0pt] at (0,-1) {$\blacktriangleleft$};
\draw (1,0) node[cross=4.5pt] {};
\draw (-1,0) node[cross=4.5pt] {};
\node[xshift=15pt] at (1,0) {$\tilde{J}^\nu$};
\node[xshift=-15pt] at (-1,0) {$\tilde{J}^\mu$};
\draw[-Stealth] (-0.3,1.2) arc[start angle=105,end angle=75,radius=1cm];
\node[yshift=17pt] at (0,1) {$k+\ell$};
\draw[-Stealth] (0.3,-1.2) arc[start angle=-75,end angle=-105,radius=1cm];
\node[yshift=-17pt] at (0,-1) {$\ell$};
\end{tikzpicture}$$
leading to the expression
\begin{equation}
    \langle \tilde{J}^\mu(k)\tilde{J}^\nu(-k)\rangle=\int \frac{d^{\lsp 3}\ell}{(2\pi)^3}\tr\left[\gamma^\mu \widetilde{G}(k+\ell)\gamma^\nu\widetilde{G}(\ell)\right]\,.
    \label{eq:JJfermion}
\end{equation}
To match with the form of the correlator given in (\ref{eq:JJ}) and to compute the form factors $a_1(k^2,\vec{k}^2)$ and $a_2(k^2,\vec{k}^2)$ explicitly, we will expand this expression about small $k$. First, let us consider turning on only the $k_3$ momentum, that is to say let us consider only those terms in the small-$k$ expansion which contain no factors of $k_1$ and $k_2$. Using the orthogonality properties
\begin{equation}
    \int_0^\infty dx\, x^ke^{-x}L_n^k(x)L_m^k(x)=\frac{(n+k)!}{n!}\delta_{nm}
\end{equation}
of the Laguerre polynomials, after an involved computation one finds the expression
\begin{equation}
    \langle \tilde{J}^\mu(k)\tilde{J}^\nu(-k)\rangle\supset \bigg(\frac{3}{\sqrt{2}\pi}\frac{1}{\sqrt{|B|}}\lsp\zeta\!\left(-\frac{1}{2}\right)k_3^2+\frac{15}{8\sqrt{2}\pi^3}\frac{1}{|B|^{3/2}}\lsp\zeta\!\left(\frac{5}{2}\right)\big(k_3^2\big)^2\bigg)(\delta^{\mu 1}\delta^{\nu 1}+\delta^{\mu 2}\delta^{\nu 2})\,.
\end{equation}
In fact, this is not quite correct. Because the sum over $n$ begins at $n=0$ in (\ref{eq:Greens_fun_momentum_space}), we have had to discard an explicitly divergent factor to arrive at the zeta function. This infinity comes from attempting to integrate out the lowest Landau level zero mode, which remains massless even upon the addition of the magnetic field. Ignoring this zero mode is equivalent to fixing the vacuum state to be a particular filling of the lowest Landau level \cite{Miransky:2015ava}. Comparing with the expected form of the propagator (\ref{eq:JJ}), we correctly find that only $\langle \tilde{J}^1(k)\tilde{J}^1(-k)\rangle$ and $\langle \tilde{J}^2(k)\tilde{J}^2(-k)\rangle$ include purely $k_3$ terms. Recalling that $k_0^2=-k_3^2$ and that the Euclidean correlation function is defined without the factor of $-i$, this fixes $c_0$ and $c_{2,3}-c_{2,4}$ in agreement with the prior results from matching with the literature.

As before, we compute the remaining two-derivative coefficients by considering the correlators $\langle \tilde{J}^1\tilde{J}^1\rangle$ and $\langle \tilde{J}^2 \tilde{J}^2\rangle$ at $\text{O}(k_2^2 k_3^2)$. The expression (\ref{eq:JJfermion}) hides the sum over the six terms obtained by taking the product of two terms in (\ref{eq:Dndef}). Each term will be associated with a different trace over the spinor indices, and thus a different tensor structure in $\mu$ and $\nu$. These traces are universal, and after some consideration one finds that 
\begin{equation}
\begin{gathered}
    2c_{2,1}+2c_{2,2}-5c_{2,3}+4c_{2,4}=\frac{165\,\zeta\!\left(\frac{5}{2}\right)}{64\sqrt{2}\pi^{3}}\,,\\
    c_{2,3}=\frac{8\pi^{2}\zeta\!\left(\frac{1}{2}\right)-45\,\zeta\!\left(\frac{5}{2}\right)}{256\sqrt{2}\pi^{3}}\,,
\end{gathered}
\end{equation}
which, combined with the previous expressions for $c_{2,1}$ and $c_{2,3}-c_{2,4}$, yields
\begin{equation}
\begin{gathered}
    c_{2,1}=-\frac{15\,\zeta\!\left(\frac{5}{2}\right)}{64\sqrt{2}\pi^{3}}\,,\qquad
    c_{2,2}=\frac{8\pi^{2}\zeta\!\left(\frac{1}{2}\right)-225\,\zeta\!\left(\frac{5}{2}\right)}{512\sqrt{2}\pi^{3}}\,,\\
    c_{2,3}=\frac{8\pi^{2}\zeta\!\left(\frac{1}{2}\right)-45\,\zeta\!\left(\frac{5}{2}\right)}{256\sqrt{2}\pi^{3}}\,,\qquad
    c_{2,4}=\frac{8\pi^{2}\zeta\!\left(\frac{1}{2}\right)+195\,\zeta\!\left(\frac{5}{2}\right)}{256\sqrt{2}\pi^{3}}\,.
\end{gathered}
\end{equation}
As $c_{2,3}\approx -0.0156$ and $c_{2,4}\approx0.0130$, the free fermion will obey (\ref{eq:c24positivity}), and computing the combinations $A$ and $B$ in (\ref{eq:ABtwoderivative}) one finds that the free fermion also obeys the bounds shown in Fig.\ \ref{fig:boundswithpoints}. Thus, fixing the vacuum state to remove the divergence associated with the massless lowest Landau level is sufficient for the free fermion to satisfy the two-derivative positivity bounds. Whether the gapless lowest Landau level nevertheless obstructs a local effective field theory description is a separate question, which the bounds alone cannot settle. Interestingly, the values of $A$ and $B$ for the free fermion are completely rational
\begin{equation}
    A=\frac14\Big(c'_{2,3}-2c'_{2,1}-2c'_{2,2}-4\Big)=-\frac{11}{16}\,,\qquad
    B=\frac14\Big(2c'_{2,1}+1\Big)=\frac18\,,
\end{equation}
with the zeta values cancelling between numerator and denominator. 

\subsection{Three-sphere partition function for free fermion}
We consider the partition function for four component fermions, adding together $\log \det (\slashed D)$ for two different choices of the gamma matrices. Using the shorthand for the eigenvalues derived in Appendix \ref{app:fermionS3},
\begin{equation}
    \lambda_{\ell,p \pm} = - \tfrac{1}{2} \pm \left((\ell+1)^2 - 2 Q (2p - \ell-1) + Q^2 \right)^{1/2} \, ,
\end{equation}
which have degeneracy $\ell+1$,
we find the partition function
\begin{align}
    \log Z &= \sum_{\ell=0}^\infty \biggl\{ \sum_{p=0}^{\ell+1} (\ell+1) \left[ \log( \lambda_{\ell,p -} )
    + \log (- \lambda_{\ell,p -} ) \right]
    + \sum_{p=1}^{\ell} (\ell+1) \left[ \log (\lambda_{\ell,p+} )  + \log (-\lambda_{\ell,p+}) \right] \biggr\} \nonumber \\
    &=   \sum_{\ell=0}^\infty \left[ 2\sum_{p=0}^{\ell} (\ell+1) \log \left((\ell+1)^2 - 2 Q (2p - \ell-1) + Q^2 -  \tfrac{1}{4} \right) +  (\ell+1) \log 
    \left(\frac{\ell + \frac{3}{2} - Q}{\ell + \frac{1}{2} + Q} \right)^2 \right]
    \nonumber  \\
    &= \log Z_1 +  \log Z_2 \,.
\end{align}
(Note we are making a smoothness assumption about the behaviour of the eigenvalues as $Q$ passes through $\ell + 1$ and do not always take the positive branch of the square root.)
Let's start with $\log Z_1$,
\begin{align}
    \log Z_1 &=  2\sum_{\ell=0}^\infty \sum_{p=0}^{\ell} (\ell+1) \log \left( (\ell+1)^2 - 2 Q (2p - \ell-1) + Q^2  - \tfrac{1}{4} \right) \ ,
\end{align}
which, similar to the scalar case, we convert to 
\begin{align}
    I_1(s) &=  2\sum_{\ell=0}^\infty \sum_{p=0}^{\ell} (\ell+1) \left( (\ell+1)^2 - 2 Q (2p - \ell-1) + Q^2  -\tfrac{1}{4} \right)^{-s}\nonumber \\
    %&=2\sum_{\ell=0}^\infty  \frac{1}{\Gamma(s)} \int_0^\infty dt \, t^{s-1}  (\ell+1) e^{-t ( (\ell+1)^2 +2 Q + Q^2 - \frac{1}{4})} \frac{ \sinh 2 (\ell+1) Q t}{\sinh 2 Q t}  \nonumber \\ 
    &=  \frac{1}{\Gamma(s)}  \int_0^\infty dt \, t^{s-2} \frac{e^{-t (2 Q + Q^2 - \frac{1}{4})}}{ \sinh 2 Q t} \frac{d}{dQ} \sum_{\ell=1}^\infty  e^{-t \ell^2} \cosh 2 Q \ell t \, ,
\end{align}
where in the third line we have re-indexed the sum and also introduced a derivative to trade the sum for an elliptic theta function, just as we did in the scalar case. Indeed, in close analogy to the calculation for the scalar, this sum reduces to
\begin{align}
    I_1(s) &= \frac{ Q^{\frac{3}{2}-s}}{\Gamma(s)} \int_0^\infty dt' \frac{\sqrt{\pi} t'^{s-\frac{3}{2}}}{ \sinh 2 t'} \sum_n e^{- \frac{Q \pi^2 n^2}{t'} + \frac{t'}{4 Q} -2 t'} \ , 
\end{align}
where again we have introduced a new variable $t' = Qt$ to make the scaling with $Q$ explicit. The $n=0$ term in the sum gives the contributions we are looking for while the $n\neq 0$ contributions are again exponentially suppressed in $Q$:
\begin{align}
    -\partial_s I_1|_{s=0}
    &=8 \pi Q^{3/2} \zeta \left( - \frac{1}{2}, 1 - \frac{1}{16 Q} \right) 
    - 4 Q \sum_{j=0}^\infty \log \left( 1 - e^{-\pi  \sqrt{16 Q(1+j) - 1}} \right) .
\end{align}
However, we also need to look at $\log Z_2$. With a similar regularisation as above we find
\begin{align}
    I_2 &= 2\sum_{\ell=0}^\infty (\ell+1)\left[ \left( 1 - \frac{\ell + \frac{3}{2}}{Q}  \right)^{-s} - \left( 1 + \frac{\ell + \frac{1}{2}}{Q}  \right)^{-s}  \right] 
    \\
    &= 2 \sum_{\ell=0}^\infty  \frac{1}{\Gamma(s)} \int_0^\infty dt \, t^{s-1} (\ell+1) \left[ e^{-t(1 -( \ell+\frac{3}{2})/Q)}
    -e^{-t(1 +( \ell+\frac{1}{2})/Q)}  \right] \nonumber \\
    &= 0 \, ,
\end{align}
where the sum on $\ell$ vanishes.

The final answer is then given by $I_1(s)$ alone:
\begin{eqnarray}
    \log Z = - \partial_s I_1(s)  |_{s=0} \, .
\end{eqnarray}
Expanding out $\partial_s I_1(s)$ at large $Q$, we get
\begin{align}
    -\partial_s I_1 |_{s=0} 
    &= 8\pi Q^{3/2} \sum_{n=0}^\infty \left[ (n+1)^{1/2} - \frac{1}{32 (n+1)^{1/2}}\frac{1}{ Q} - \frac{1}{2048 (1+n)^{3/2}} \frac{1}{ Q^2} + \cdots \right] \\
    =& 8 \pi Q^{3/2} \left[ \zeta\! \left(-\frac{1}{2} \right) - \frac{1}{32Q} \zeta\! \left(\frac{1}{2} \right) - \frac{1}{2048 Q^2} \zeta\!\left(\frac{3}{2} \right) + \cdots \right].
\end{align}
From comparing $I_1$ with (\ref{effactionthreesphere}) in the round case $\epsilon=0$, we can read off
\begin{align}
    c_0 &=  \frac{2}{\sqrt{2} \pi}\lsp \zeta\!\left( -\frac{1}{2} \right) , \\
    6 c_{2,1} + 2 c_{2,2} + 3 c_{2,3} &= -\frac{1}{8 \sqrt{2} \pi}\lsp \zeta \!\left(\frac{1}{2} \right) .
\end{align}
Taking into account the shift in sign of $c_{2,2}$  and $c_{2,3}$ in moving from Euclidean to Lorentzian signature, these results
match exactly the results from the previous subsection.

\section{A holographic example}\label{sec:holo}
Our holographic model is the Einstein--Hilbert--Maxwell action with a negative cosmological constant, $\Lambda = - 3/L^2$.  We follow the notation of
\cite{Herzog:2009xv}, where the action is
\begin{equation}
    S_{\rm bulk} = \frac{1}{2 \kappa^2} \int d^{\lsp 4}x \sqrt{-g} \left(R -2 \Lambda\right) - \frac{1}{4 g^2} \int d^{\lsp 4} x \sqrt{-g}\, F_{\mu\nu} F^{\mu\nu} \, .
\end{equation}
$L$ sets the curvature scale, $\kappa$ is the gravitational interaction strength, and $g$ the gauge coupling.
To regulate divergences and have a well posed variational principle, we require the standard boundary terms
\begin{equation}
    S_{\rm bry} = \frac{1}{\kappa^2} \int d^{\lsp 3} x \sqrt{-\gamma} \left( K - \frac{2}{L} - \frac{L}{2} \tilde R  \right),
\end{equation}
where $K$ is the trace of the extrinsic curvature and $\tilde R$ is the boundary Ricci scalar computed from the boundary metric $\gamma_{\mu\nu}$ which is the induced
metric on a slice near the conformal boundary.

This type of action is a consistent truncation of many known AdS/CFT correspondences, allowing one access to correlation functions involving the stress tensor and a $U(1)$ R-symmetry current. In typical AdS/CFT set-ups \cite{Aharony:1999ti}, the factor $\frac{L^{2}}{\kappa^2}$ is assumed to be very large, to ensure a classical analysis is valid.  On the field theory side, the classical limit typically corresponds to taking some kind of large $N$ limit. For example, for ${\mathcal N}=4$ $SU(N)$ super Yang-Mills, where $d=5$, the ratio would be of order $N^2$.   Perhaps more appropriate to the current set-up where $d=4$, for ABJM theory \cite{Aharony:2008ug}, it would be of order $N^{3/2}$. Note that we could rescale $F_{\mu\nu} \to \sqrt{2} \frac{ L g}{\kappa} F_{\mu\nu}$ to have a universal factor of $1 / 2 \kappa^2$ in front of the action.  We find it convenient to allow $g$ to be an additional free parameter in the problem.

We have two calculations in mind.  First, we look at small fluctuations around a magnetically charged black brane solution in the extremal limit. From these fluctuations, we can deduce current two-point functions that should match those computed from the magnetic effective action.  To keep things as simple as possible, we will look only at fluctuations in the long wavelength $\vec k \to 0$ limit. Next, we look at the on-shell action for a magnetically charged and spinning black hole, from which we constrain further combinations of the Wilson coefficients in the effective action.  These two computations together will allow us to deduce four combinations of the Wilson coefficients, namely $c_0$, $c_{2,1}$, $c_{2,2}  + \frac{3}{2} c_{2,4}$, and $c_{2,3}  - c_{2,4}$, discussed previously, as well as three linear combinations of the $c_{4,n}$.

If we were to turn on $\vec k \neq 0$ in order to fix the remaining $c_{2,4}$, existing holographic analyses show that the proposed effective field theory matching would face a variety of problems. The authors of \cite{Buchbinder:2008nf} demonstrated how the current-current two point functions at non-zero temperature separate into shear and sound channels.  They found that the sound channel supports a diffusive mode $\omega = -i D k^2$ while the shear channel supports a sub-diffusive one $\omega = - i D' k^4$.  The zero temperature limits of these modes are regular,
\begin{equation}
    D = \frac{z_h}{2} \ , \; \; \; D' = \frac{z_h^3}{12} \ .
\end{equation}
(The horizon radius $z_h^2 = \sqrt{6} g L / B \kappa$ can be re-expressed in terms of the magnetic field $B$.) These values follow from Eqs.\ (2.30) and (3.19) of \cite{Buchbinder:2008nf}, $\mathfrak{w}=-\tfrac{3i}{16}\lsp\mathfrak{q}^2(3-h^2)(1+h^2)/h^2$ and $\mathfrak{w}=-i\lsp\mathfrak{q}^4(3-h^2)^3/32h^2$, where $\mathfrak{w}=\omega/2\pi T$ and $\mathfrak{q}=k/2\pi T$, with temperature $T=\alpha(3-h^2)/4\pi$, magnetic field $h\lsp\alpha^2$ and horizon radius $z_h=1/\alpha$ in the conventions of that paper. The factors of $(3-h^2)$ cancel against those in $T$, giving $D=3(1+h^2)/8h^2\alpha$ and $D'=1/4h^2\alpha^3$, which remain finite in the extremal limit $h^2\to3$. These modes, which show up as poles in the current-current two-point functions, violate the form of our effective field theory expansion and introduce non-uniformities in the $\vec k \to 0$ and $\omega \to 0$ limits.

There is another issue that will likely arise even with $\vec k = 0$, but at higher order in the $\omega$ expansion.  At zero temperature the geometry has an $AdS_2$ throat, which is known to produce non-analytic $\log \omega$ corrections to the two point functions \cite{Moitra:2020dal, Faulkner:2009wj}.

\subsection{Two-point functions}
We consider fluctuations around the magnetically charged black brane solution in asymptotically anti-de Sitter space.
The line element for a black brane with both electric and magnetic charge is
\begin{equation}
    \frac{1}{L^2} ds^2 = \frac{1}{z^2} \left[ - f(z) dt^2 + dx^2 + dy^2 \right] + \frac{1}{z^2} \frac{dz^2}{f(z)} \ ,
\end{equation}
with gauge field and warp factor
\begin{equation}
    A = \frac{hx}{z_h} dy - q \left( 1- \frac{z}{z_h} \right) dt \, , \qquad f(z) = 1 + (h^2+q^2) \alpha \frac{z^4}{z_h^4}  - \left(1+(h^2 + q^2) \alpha \right) \frac{z^3}{z_h^3} \, ,
\end{equation}
where
\begin{equation}
    \alpha = \frac{ \kappa^2 z_h^2}{2 g^2 L^2} \, .
\end{equation}
The AdS boundary is at $z=0$ and there is a horizon at $z=z_h$ with Hawking temperature
\begin{equation}
    T = \frac{3 - (h^2 +q^2) \alpha}{4 \pi z_h} \, .
\end{equation}
The magnetic field is $B = h / z_h$.
We will be particularly interested in the limit $T=0$ and $q=0$, in which case $h=\sqrt{3/\alpha}$ and the magnetic field strength is $B = \sqrt{3/\alpha} / z_h $.

To compute current and stress-energy tensor two-point functions, we look at fluctuations around the background metric $g_{\mu\nu}$ and gauge potential $A_\mu$.  In particular, we consider fluctuations of the form
\begin{equation}
    \delta A_\mu = a_\mu(z) e^{-i \omega t + i \vec k \cdot \vec x} \, , \qquad \delta g_{\mu\nu} = h_{\mu\nu}(z) e^{-i \omega t+ i \vec k \cdot \vec x} \, ,
\end{equation}
where we further make the gauge choice that $a_z = 0$ and $h_{\mu z} = 0$.

The idea behind the calculation is to read off the current two-point function from the boundary behaviour of the fluctuations $a_\mu(z)$.  If we vary the action with respect to the gauge potential, we find the term
\begin{equation}
    \delta S = \frac{1}{g^2} \int_{z=0} d^3 x \, \sqrt{-g} \, \delta A_\mu F^{z \mu} + \cdots\,,
\end{equation}
from which we can read off the current one-point function in the dual CFT, as per the usual AdS/CFT dictionary:
\begin{equation}
    \langle J_\mu \rangle = \frac{\delta S}{\delta A^{\rm (bry)}_\mu} = \left. \frac{1}{g^2} F_{z \mu} \right|_{\rm bry} \ .
    \label{eq:Jdef}
\end{equation}
We then want to vary $\langle J_x \rangle$ with respect to the boundary values of $a_x$ or $a_y$ to get a corresponding two-point function. From a differential equations point of view, having fixed the leading behaviour of $a_\mu$ at the boundary $z=0$, a suitable boundary condition in the interior at the black hole horizon then lets us deduce the subleading $\text{O}(z)$ behaviour of $a_\mu$, from which we can read off the two-point function.

Following reference \cite{Hartnoll:2007ip}, we make the further restriction  $\vec k =0$ and consider only the fluctuations $a_x$, $a_y$, $h_{xt}$ and $h_{yt}$.  In this case, it is convenient to look at the combinations
\begin{align}
    E_a &= - (-i L^2 z_h \omega a_a + h z^2 \epsilon_{ab} h_{bt}) \, , \\
    B_a &= -L^2 z_h f(z) \epsilon_{ab} a'_b(z) \, ,
\end{align}
where $a,b = x,y$ and then further introduce the complex combinations
\begin{equation}
    {\mathcal E}_\pm = E_x \pm i E_y \, ,\qquad {\mathcal B}_\pm = B_x \pm i B_y \, .
\end{equation}
The combination of Einstein's and Maxwell's equations then imply
\begin{align}
    f(q {\mathcal E}_+ + h {\mathcal B}_+)' + \omega(h {\mathcal E}_+ - q {\mathcal B}_+) &= 0 \, , \\
    \frac{\omega z_h^4}{4  \alpha z^2} \left( {\mathcal E}_+' - \frac{\omega}{f} {\mathcal B}_+ \right) +  h^2 {\mathcal B}_+ +  q h {\mathcal E}_+ &= 0 \, .
\end{align}
The system for ${\mathcal E}_-$ and ${\mathcal B}_-$ is the same but with the swap $q \to -q$ and ${\mathcal E}_+(z) \to - {\mathcal E}_-(z)$.

We momentarily postpone a solution of this system of differential equations and instead simply quote the result, a boundary and small frequency expansion from which we can read off the desired current two-point function:
\begin{align}
    \label{eq:Ebry}
    {\mathcal E}_\pm &=   \pm {\mathcal A}_\pm \left( 1 - \frac{\omega^2 z_h^2\cot^{-1} \sqrt{2}}{\sqrt{2}} +  \omega^2 z_h z + \text{O}(z^2, \omega^4) \right) \, , \\
    \label{eq:Bbry}
    {\mathcal B}_\pm &= \omega z_h {\mathcal A}_\pm \left( 1-\frac{z}{z_h}  + \frac{\omega^2 z_h^2 \cot^{-1} \sqrt{2}}{\sqrt{2}}z + \text{O}(z^2, \omega^4) \right) \, ,
\end{align}
where ${\mathcal A}_\pm$ are integration constants. From the definition of the current one-point function (\ref{eq:Jdef}), we have
\begin{equation}
    \langle J_a \rangle = \frac{1}{g^2} A'_a(0)  =  \left. \frac{1}{L^2 z_h g^2}  \epsilon_{ab} B_b \right|_{\rm bry} \ .
\end{equation}
We also have
\begin{align}
    B_y(0) &= \left. \frac{1}{2i}({\mathcal B}_+ - {\mathcal B}_- ) \right|_{\rm bry} = \frac{\omega z_h}{2i} ({\mathcal A}_+  - {\mathcal A}_-) \, , \\
    \frac{C}{2} ({\mathcal A}_+ - {\mathcal A}_-) = E_x(0) &=i L^2 z_h \omega a_x(0) \, ,
\end{align}
where $C = 1 - 2^{-1/2} \omega^2 z_h^2 \cot^{-1} \sqrt{2}$ can be read off from the boundary expansion (\ref{eq:Ebry}). Assembling this chain of three equalities, we find that
\begin{equation}
    \langle J_x \rangle= \frac{z_h}{g^2} \frac{\omega^2}{C}  a_x(0) \, ,
\end{equation}
from which we can identify $z_h \omega^2 / g^2 C$ as the leading small-frequency contribution to the (Fourier transform of the) $\langle J_x J_x \rangle$ two-point function.

Comparing with (\ref{eq:JJ}), we can read off $c_0$ and $c_{2,3}-c_{2,4}$ from this expression. In particular, their ratio is
\begin{equation}
    \label{eq:c23ratio}
    \frac{c_{2,3}- c_{2,4}}{c_0} = \frac{3 B}{4}  \frac{ \cot^{-1} \sqrt{2}}{\sqrt{2}} z_h^2=  \frac{3 \sqrt{3}}{4} \cot^{-1} \left(\sqrt{2} \right)\, \frac{g L}{ \kappa}
    \, .
\end{equation}

\paragraph{Derivation} Returning now to a derivation of (\ref{eq:Ebry}) and (\ref{eq:Bbry}), let us first discuss the boundary conditions at the horizon.  Historically, these calculations were done with $T\neq 0$ and in-going boundary conditions at the horizon. The exponential form of the electric and magnetic fields is taken to be
\begin{equation}
    {\mathcal E}_\pm \sim {\mathcal B}_\pm \sim \left(1 - \frac{z}{z_h} \right)^{-i \omega/ 4 \pi T } \, .
\end{equation}
Balanced against an $e^{-i \omega t}$ factor, as $z$ increases toward the horizon, so does $t$ to maintain the same phase factor in the exponential.  We are interested instead in a zero-temperature limit, in which case the point $z=z_h$ becomes an essential singularity of the differential equation.  In this case, the near horizon behaviour is different.  For the particular case $q=0$, one finds
\begin{equation}
    \label{Bplushorizon}
    {\mathcal B}_+ \sim \left(1 - \frac{z}{z_h} \right)^{-\frac{2 i \omega z_h}{9}} \exp \left( \frac{i \omega z_h}{6 \left(1-\frac{z}{z_h} \right)} \right)
    \left( 1 + \text{O}(1-z) \right) \ .
\end{equation}

For the problem at $T \neq 0$, it is usually a straightforward affair to match the near-horizon expansion onto a small-frequency expansion.  In this extremal $T=0$ case, the matching is more delicate.  We have a boundary layer problem where an approximate solution can be found using the method of matched asymptotic expansions.  This strategy was used successfully in \cite{Davison:2013bxa}, in the closely related case of an electrically charged black hole.  Because the near horizon behaviour in the extremal limit is $AdS_2$, the calculations are similar in spirit to calculations in the early days of AdS/CFT where the deep throat $AdS_d$ region of a D-brane is matched onto the asymptotically flat region far away from the brane  (see e.g.\ \cite{Gubser:1996xe}).

To find the solution near the horizon,  we introduce a variable $\zeta = (z_h - z)/\omega z_h^2$ and then re-expand the differential equation in small $\omega$:
\begin{equation}
    {\mathcal B}_+'' + \frac{2}{\zeta} {\mathcal B}_+' + \frac{1}{36 \zeta^4}(1 - 72 \zeta^2) {\mathcal B}_+ + \text{O}(\omega)= 0 \, ,
\end{equation}
which has the pair of solutions
\begin{equation}
    \label{Bpluszeta}
    {\mathcal B}_+ = a_+ (1 + 6 i \zeta) e^{i / 6 \zeta} + a_- (1 - 6 i \zeta) e^{-i / 6 \zeta} \, .
\end{equation}
The ingoing solution corresponds to setting $a_-= 0$.  The strategy is then to re-expand this near-horizon solution in the limit of large $\zeta$ and match it to the small-frequency solution we will derive below that is valid far from the horizon. Expanding (\ref{Bpluszeta}) in the limit of large $\zeta$ yields
\begin{equation}
    \label{eq:Bbrylayer}
    {\mathcal B}_+ = a_+ \left(6 i \zeta + \frac{i}{12 \zeta}  + \text{O}(\zeta^{-2})\right) + \text{O}(\omega) \ .
\end{equation}

We next solve this system in a small $\omega$ expansion in the limit $T=0$ and $q=0$ but now using the original radial variable $z$. We look for solutions of the form,
\begin{align}
    {\mathcal E}_+ &=  f(z)  {\mathcal A}_+ + \omega^2 z_h^2 {\mathcal E}_2(z) + \text{O}(\omega^4) \, , \\
    {\mathcal B}_+ &=  \omega z_h {\mathcal A}_+ \left(1-\frac{z}{z_h} \right) + \omega^3 z_h^3{\mathcal B}_3(z) + \text{O}(\omega^5) \, .
\end{align}
The expansion involves every other power of $\omega$. The expressions for ${\mathcal E}_2$ and ${\mathcal B}_3$ are horrendous, involving sums of logs and polylogs, but analytic.  Matching to the large $\zeta$ expansion, the integration constants can be fixed.  We give here just the expansions of the final solutions at the horizon: 
\begin{align}
    {\mathcal B}_+ &={\mathcal A}_+\omega^3 z_h^3 \left( - \frac{1}{72\left(\frac{z}{z_h}-1\right) }+ \frac{1}{18} + \text{O}(z/z_h-1) \right)  , \\
    {\mathcal E}_+ &= -\frac{{\mathcal A}_+ \omega^2 z_h^2}{12} + \text{O}(z/z_h-1) \, .
\end{align}
One can check the expression for ${\mathcal B}_+$ has the right form to match (\ref{eq:Bbrylayer}) after the redefinition $\zeta = (z_h - z)/\omega z_h^2$.  The boundary expansion was given already as (\ref{eq:Ebry}) and (\ref{eq:Bbry}).

\subsection{Partition function on spinning \texorpdfstring{$S^2 \times {\mathbb R}$}{S2xR}}

Using holography, by looking at a charged and rotating black hole in $AdS_4$  \cite{Carter:1968ks,Plebanski:1976gy}, we can identify more of the Wilson coefficients in the magnetic effective action than by looking at just a charged black hole. This computation is largely based off the notation, solution, and analysis of Ref.\ \cite{Caldarelli:1999xj} although the matching of the Wilson coefficients is new.

The line element is
\begin{equation}
    ds^2 = - \frac{\Delta_r}{\rho^2} \left[ dt - \frac{a \sin^2 \!\theta}{\Xi} d \phi \right]^2 + \frac{\rho^2}{\Delta_r} dr^2 + \frac{\rho^2}{\Delta_\theta} d \theta^2 + \frac{\Delta_\theta \sin^2\! \theta}{\rho^2} \left[ a \, dt - \frac{r^2+a^2}{\Xi} d \phi \right]^2 \, ,
\end{equation}
where
\begin{equation}
    \rho^2 = r^2 + a^2 \cos^2\! \theta \,,\qquad \Xi = 1 - \frac{a^2}{L^2} \, ,
\end{equation}
\begin{equation}
    \Delta_r = (r^2 + a^2) \left( 1 + \frac{r^2}{L^2} \right) - 2 M r + \frac{P^2}{2} \, , \qquad \Delta_\theta = 1 - \frac{a^2}{L^2} \cos^2 \!\theta \, .
\end{equation}
The metric is supported by electromagnetic flux
\begin{equation}
    A = - \frac{q_e r}{\rho^2} \left(dt - \frac{a \sin^2 \!\theta}{\Xi} d \phi \right) - \frac{q_m \cos \theta}{\rho^2} \left(a \, dt - \frac{r^2 + a^2}{\Xi} d \phi \right) .
\end{equation}
We further have the relation
\begin{equation}
    P^2 = \frac{\kappa^2}{g^2}(q_e^2 + q_m^2) \, .
\end{equation}

The on-shell action boils down to
\begin{equation}
    S = S_{\rm bulk} + S_{\rm bry} = -\frac{2 \pi \beta}{1 - \frac{a^2}{L^2}} \left( \frac{(r_h^2+a^2)(r_h^2-L^2)}{r_h L^2 \kappa^2}
    - \frac{q_e^2+q_m^2}{2 g^2 r_h} + \frac{(q_e^2-q_m^2)r_h}{g^2 (a^2 + r_h^2)}  \right)  .
\end{equation}
Here $r_h$ is the horizon, which is the largest root of $\Delta_r$, and $\beta$ is the inverse Hawking temperature:
\begin{equation}
    \beta = \frac{4\pi (r_h^2 + a^2)}{r_h \left(1 + \frac{a^2}{L^2} + 3 \frac{r_h^2}{L^2}  - \frac{2a^2 + P^2}{2r_h^2} \right)}   .
\end{equation}
The magnetic field can be read off from the large $r$ limit of $dA \sim -\frac{L^2 q_m}{L^2-a^2} \sin \theta \, d \theta \wedge d\phi$.

We now expand the log of the partition function  $S = - \log Z$ in the limit $q_e=0$, $T=0$, and large $B$. The result from the gravity side is
\begin{align}
    \label{eq:gravityF}
    \frac{\kappa^{2} \log Z}{\beta \sqrt{L^2-a^2}} &= -16\pi \left( \frac{\kappa L B}{\sqrt{6} g} \right)^{3/2}  - 4 \pi \frac{ L^2 - 3 a^2}{L^2-a^2} \left( \frac{ \kappa L B}{\sqrt{6} g} \right)^{1/2} \nn \\
    & + \frac{\pi}{6} \frac{-55 a^4 + 18 a^2 L^2 + L^4}{(L^2-a^2)^{2}} \left( \frac{\kappa L B}{ \sqrt{6} g}\right)^{-1/2} + \text{O}(B^{-3/2}) \ ,
\end{align}
defining $B=- q_m / (L^2-a^2)$.

We compare this result against the large $B$ effective action computed for the boundary metric along a large-$r$ slice and field strength (\ref{eq:spinningmetric}). Comparing the gravity result (\ref{eq:gravityF}) with the effective field theory result (\ref{eq:effectiveF}), and including the ratio $(c_{2,3}-c_{2,4})/c_{0}$ (\ref{eq:c23ratio}) that we determined in the previous subsection, we can specify the following combinations of  Wilson coefficients for this theory:
\begin{align}
    c_0 &= -\frac{2 \cdot 2^{1/4}}{3^{3/4}} \sqrt{\frac{L}{g^3 \kappa}} \ , \\
    c_{2,1} &= -\frac{1}{2 \cdot 6^{1/4}} \sqrt{\frac{L^3}{g \kappa^3}}
    \ , \\
    c_{2,2}+\frac{3}{2} c_{2,4} %&= \frac{7}{2 \cdot 6^{1/4}} \sqrt{\frac{L^3}{g \kappa^3}} - \frac{3}{2} c_{2,3}
    %\nonumber \\
    &= \left( \frac{7}{2 \cdot 6^{1/4}} + \left(\frac{3}{2} \right)^{7/4} \cot^{-1} \sqrt{2}\right) \sqrt{\frac{L^3}{g \kappa^3}} \ ,
    \\
    c_{2,3}-c_{2,4} &= -\left( \frac{3}{2} \right)^{3/4} \cot^{-1} \sqrt{2} \sqrt{\frac{L^3}{g \kappa^3}}
    \ .
\end{align}
We can also identify three linear combinations of the $c_{4,n}$.

Because $c_{2,2}$ and $c_{2,4}$ are fixed only in the combination $c_{2,2}+\tfrac32 c_{2,4}$, testing the two-derivative bound \eqref{eq:bound_type_2_first} against this theory requires a little care. Written in the form \eqref{eq:type_2_bound}, its coefficients are
\begin{equation}
    a=2(c_{2,4}-c_{2,3})\,,\qquad c=c_{2,1}-\tfrac12(c_{2,3}-c_{2,4})\,,\qquad b=\tfrac52 c_{2,3}-c_{2,1}-c_{2,2}-2c_{2,4}\,,
\end{equation}
so that $a$ and $c$ involve $c_{2,2}$ and $c_{2,4}$ only through the determined quantities $c_{2,3}-c_{2,4}$ and $c_{2,1}$, while $b$ grows linearly in $c_{2,4}$. The holographic theory therefore does not correspond to a point of Fig.\ \ref{fig:boundswithpoints} but to a horizontal ray, traced out as $c_{2,4}$ is varied,
\begin{equation}\label{eq:holoray}
    B=\frac14-\frac{\sqrt{2}}{12\cot^{-1}\sqrt{2}}\approx0.0585\,,\qquad
    A=\bigg(\frac{2}{3}\bigg)^{\!3/4}\frac{1}{\cot^{-1}\sqrt{2}}\lsp c_{2,4}\sqrt{\frac{g\kappa^3}{L^3}}-2-\frac{\sqrt{2}}{2\cot^{-1}\sqrt{2}}\,.
\end{equation}
Since $B$ is positive but small, the ray can never meet the parabolic branch of \eqref{eq:Bbranches}, which for $A<-2$ would require $B\geq1$; the bound therefore reduces to the linear condition $A+B\geq-1$, which the ray crosses at $A=-(1+B)\approx-1.0585$. Equivalently, and more simply, $a+b+c=\tfrac12 c_{2,4}-c_{2,2}$, so that the $\xi^2\to1$ endpoint of \eqref{eq:bound_type_2_first} reads
\begin{equation}\label{eq:c24geq2c22}
    c_{2,4}\geq2\lsp c_{2,2}\,,
\end{equation}
and for the case at hand this endpoint is the entire bound. In terms of the combination we have actually determined, \eqref{eq:c24geq2c22} states that $c_{2,4}$ must be at least half of $c_{2,2}+\tfrac32 c_{2,4}$, that is
\begin{equation}\label{eq:holoc24bound}
    c_{2,4}\geq\bigg(\frac{7}{4\cdot6^{1/4}}+\frac34\bigg(\frac32\bigg)^{\!3/4}\cot^{-1}\sqrt{2}\bigg)\sqrt{\frac{L^3}{g\kappa^3}}\approx1.74\lsp\sqrt{\frac{L^3}{g\kappa^3}}\,.
\end{equation}
This single inequality exhausts the two-derivative constraints for this model: it implies $c_{2,4}\geq0$, while $c_{2,3}\leq c_{2,4}$ holds identically here, so that both statements in \eqref{eq:c24positivity} are subsumed. The computations of this section cannot decide it. Setting $c_{2,4}$ to zero by hand would give $A\approx-3.15$, deep inside the excluded region of Fig.\ \ref{fig:boundswithpoints}, so whether this theory respects our dispersive bounds turns entirely on the size of the one coefficient we have been unable to determine.

Upon reflection, it is not entirely surprising that this holographic system does not have an effective field theory expansion consistent with a gapped theory.  The extremal black hole we find at zero temperature has a non-zero entropy, which implies that the ground state of this system is not gapped.  There must be many zero-energy ground states to maintain a non-zero entropy. There is a longer discussion to be had here about whether this non-zero entropy is physical or merely an artifact of the large-$N$ limit in these holographic systems.  Recent work on quantum effects on black hole entropy suggest that in the non-supersymmetric cases like the one studied here, loop effects will remove the ground state degeneracy \cite{Iliesiu:2022onk}.

\section{Conclusion}\label{sec:conclusion}
This work has four principal results: the computation up to fourth order in derivatives of an EFT for a parity- and charge-conjugation-invariant 3d CFT gapped by a large magnetic field; a demonstration that the resultant EFT satisfies the Ward identities associated with diffeomorphism and Weyl invariance; the derivation of bounds on the corresponding Wilson coefficients from positivity of a spectral density; and the computation of the Wilson coefficients up to second order in derivatives for three simple theories---a free scalar, a free four-component fermion, and a holographic example. That the leading and subleading terms in the EFT involve only the five Wilson coefficients $c_0$, $c_{2,1}$, $c_{2,2}$, $c_{2,3}$ and $c_{2,4}$ implies a certain universality to the long distance behaviour of CFTs in a magnetic field \cite{Boyack:2023uml}. In contrast, the explosion to 29 terms we find at third order suggests we need a better organising principle before continuing the expansion to yet higher order. 

The dispersive bounds that we found might provide just such a principle.  Already at leading order, the result $c_0 \leq 0$ shows the physical result that 3d CFTs at large magnetic field are diamagnetic (and that the scaling dimension of external magnetic monopole operators is positive). Indeed, in section \ref{sec:bounds} we do find a number of bounds on the $c_{4,n}$ coefficients, but we hope more stringent conditions on the coefficients could be derived by considering correlation functions involving three or more currents.  Perhaps the S-matrix bootstrap program in Lorentz-symmetry breaking backgrounds \cite{Hui:2023pxc} shows a way to proceed. Ultimately, we would like to find small islands in which the $c_{2a,n}$ are allowed to take values.

One of the most interesting issues that arose with our examples is that the dispersive bounds we derived were satisfied by the free scalar and the free fermion, while for the holographic example we could not tell, because the answer hinges on the one two-derivative coefficient that our holographic computations leave undetermined.  We did identify a problem, that neither the free fermion nor the holographic example has a mass gap at strong magnetic field, and it would be useful to understand exactly what the implications of a failure of the bounds would be. Presumably there are terms in the EFT for these theories that are not encapsulated by our derivative expansion.  As discussed above, calculations involving $\langle J^\mu(x) J^\nu(x) \rangle$ in the holographic example suggest the appearance of diffusive and sub-diffusive poles, which are forbidden in our initial setup by the assumption of a mass gap.  Is there a way to incorporate these zero energy modes as extra degrees of freedom in the EFT that would then allow us to describe the holographic example?

A less pressing issue is that our scheme for computing the $c_{2a,n}$ in examples was piecemeal, involving putting together several different overlapping strategies.  There are easy ways to improve our results slightly, for example in all cases by computing the full $\langle J^\mu(x) J^\nu(0) \rangle$ correlator instead of special components, or in the case of the free theories by including squashing in the computation of $\mathcal{W}[g,A]$ on a three-sphere. While we were able to deduce all the $c_{2,n}$ in the free theories, our strategies were insufficient for computing all 29 of the $c_{4,n}$.  Is there a strategy for computing these Wilson coefficients systematically?

One future direction is introducing a chemical potential ($\mu$) along with a constant magnetic field background ($B$) in our CFT in three dimensions. Introducing a chemical potential corresponds to turning on a background temporal component of the gauge field in the EFT, i.e.\ $\mathbb{A}_0 = \mu$. One might be tempted to gauge away the chemical potential by choosing a gauge parameter $\Lambda = \mu t$, which leads to the gauge transformation $\mathcal{A}_\mu \rightarrow \mathcal{A}_\mu + \partial_\mu \Lambda$. However, since $\Lambda$ does not vanish as $|t| \rightarrow \infty$, such a gauge transformation is not allowed.

In general, we expect the addition of chemical potential to lead to additional massless degrees of freedom that would then need to be incorporated into the EFT.  In the free scalar theory for example, the chemical potential acts as a negative mass term and should destabilise the vacuum once the chemical potential is of the order of the magnetic field.  More generally, with interaction where the run-away behaviour of the scalar can be stabilised by a $\phi^4$ term in the action, introducing a chemical potential leads to spontaneous symmetry breaking and the formation of a superfluid phase. These fluid phases are expected to have gapless shear and sound modes that would need to be included in the EFT. 

For fermions, the behaviour can be rich and subtle.  In the absence of a magnetic field, the 2d electron gas at large chemical potential supports zero sound, due to coherent oscillations of the Fermi surface. With attractive interactions, there can in addition be a superconducting instability leading to yet more gapless degrees of freedom, in particular second sound.  Adding a magnetic field in general suppresses zero sound and also the superconducting instability, but how exactly and where in the $(\mu, B)$ phase diagram is sensitive to theory-specific details.

For theories with a large magnetic field and small enough chemical potential that the gap assumption remains true, the general structure of the EFT action \eqref{eq:curved_space_W} should remain valid because the effective action depends on the background gauge field only through the field strength. However, the specific value of the chemical potential $\mu$ will modify the coefficients $c_{2a,n}$ in the EFT, as the spectrum of occupied modes changes. As an example, an effective action for a free Dirac fermion in the presence of both a background magnetic field and chemical potential has been derived in \cite{Nguyen:2016itg}. The key difference between our analysis in subsection \ref{S:Fermion_propagator} and that of \cite{Nguyen:2016itg} is that the chemical potential alters which Landau levels are filled. To modify our result in section \ref{S:Fermion_propagator} to incorporate $\mu$, when integrating out the Dirac fermion modes, we need to include contributions only from the unfilled Landau levels. Changing which Landau levels are full and empty also provides a natural interpretation of the computation performed in subsection \ref{S:Fermion_propagator}, where the LLL was removed by hand. This procedure is effectively equivalent to working with a chemical potential in the range $0 < \mu < \sqrt{2|B|}$, between the lowest and the next-to-lowest Landau level.

Finally, one may wish to generalise the construction of the EFT action \eqref{eq:curved_space_W} to conformal field theories in four spacetime dimensions in the presence of a strong background magnetic field. However, as discussed below \eqref{eq:g_hat_definition}, in four dimensions it is possible to construct two independent Weyl-invariant effective metrics: $\hat{g}^{(1)}_{\mu\nu}=g_{\mu\nu}\left(\mathcal{F}^{\mu\nu}\mathcal{F}_{\mu\nu}\right)^{1/2}$ and $\hat{g}^{(2)}_{\mu\nu}=g_{\mu\nu}\left(\mathcal{F}^{\mu\nu}\mathcal{F}_{\nu\rho}\mathcal{F}^{\rho\sigma}\mathcal{F}_{\sigma\mu}\right)^{1/4}$. The presence of two inequivalent Weyl-invariant metrics complicates the construction of the EFT action.

\ack{We thank Ritam Sinha for collaboration in the initial stages of this project. We thank D.~Anninos, F.~Gautason, and F.D.M.~Haldane for discussions.  For many of our calculations we have used \emph{Mathematica} with the package \href{http://www.xact.es}{\texttt{xAct}}~\cite{xact}. 
This work was supported by STFC under grant ST/X000753/1.
AS is also supported by the Royal Society under grant URF\textbackslash{}R1\textbackslash211417.}

\begin{appendix}

\section{Free theories and squashed spheres}\label{app:squashed}

The goal is to find the spectrum of a massless scalar and a massless free fermion on a squashed three-sphere in the presence of magnetic flux.  It will be convenient to use two different coordinate systems on the squashed three-sphere $S^3_\epsilon$. The first task is then to  review some details about these two different presentations of $S^3_\epsilon$.

The first metric makes the structure of the manifold  as a Hopf fibration over $S^2$ manifest:
\begin{equation}
    \frac{d s'^2}{L^2}=  \frac{1}{4} \left[ (1+\epsilon) (d  \psi' + \cos  \theta' \, d \phi')^2 + d  \theta'^2 + \sin^2 \! \theta' \, d  \phi'^2  \right] ,
\end{equation}
where $\epsilon$ (not necessarily small) is the squashing parameter.  When $\epsilon=0$, we recover the round metric on a sphere of radius $L$, volume $2 \pi^2 L^2$ and Ricci tensor $ R'_{\mu\nu} = \frac{2}{L^2}  g'_{\mu\nu}$.  
The variables have the range
$0< \psi' < 4 \pi$, $0 <  \theta' < \pi$, and $0 <  \phi' < 2 \pi$.  

The second coordinate system, which will be more convenient for understanding the quantum numbers of the scalar and fermion
wavefunctions, has the line element (\ref{eq:ds2squashedS3}) and unprimed coordinates $(\psi, \theta, \phi)$. The coordinate transformation that takes us from one metric to the other is
\begin{equation}
    \psi = \tfrac{1}{2} (\psi' + \phi' )\, , \qquad \phi = \tfrac{1}{2}( \psi' -  \phi' ) \, , \qquad  \theta = \tfrac{1}{2} \theta' \, .
\end{equation}
The ranges of the coordinates have to be adjusted such that $0 < \theta < \frac{\pi}{2}$, $0 < \psi < 2 \pi$, $0 < \phi < 2 \pi$. The way in which the two coordinate systems is related is illustrated in Fig.\ \ref{fig:coordinateplot}.

\begin{figure}[ht]
    \begin{center}
        \begin{tikzpicture}[scale=0.75]
            \filldraw[draw=black, fill=red!20] (-2,-2) -- (0,-4) -- (2,-2) -- cycle;
            \filldraw[draw=black, fill=red!20] (-2,2) -- (0,0) -- (2,2) -- cycle;
            \filldraw[draw=black, fill=blue!20] (-2,-2) -- (2,-2) -- (2,2) -- cycle;
            \filldraw[draw=black, fill=green!20] (-2,-2) -- (-2,2) -- (0,0) -- cycle;
            \filldraw[draw=black, fill=green!20] (2,-2) -- (2,2) -- (4,0) -- cycle;
            \draw[->] (-2,-2) to (4,-2);
            \draw[->] (-2,-2) to (-2,4);
            \draw[->] (-2,-2) to (3,3);
            \draw[->] (-2,-2) to (1,-5);
            \draw[thick,blue] (-2,2) to (2,2);
            \draw[thick,blue] (2,-2) to (2,2);
            \draw[thick,blue] (-2,-2) to (-2,2);
            \draw[thick,blue] (-2,-2) to (2,-2);
            \draw[thick,red] (-2,-2) to (2,2);
            \draw[thick,red] (-2,-2) to (0,-4);
            \draw[thick,red] (0,-4) to (4,0);
            \draw[thick,red] (4,0) to (2,2);
            \draw[ultra thick] (-1.8,2) to (-2.2,2);
            \draw[ultra thick] (-1.8,0) to (-2.2,0);
            \draw[ultra thick] (2,-2.2) to (2,-1.8);
            \draw[ultra thick] (0,-2.2) to (0,-1.8);
            \draw[ultra thick] (1.9,2.1) to (2.1,1.9);
            \draw[ultra thick] (-0.1,0.1) to (0.1,-0.1);
            \draw[ultra thick] (-0.1,-4.1) to (0.1,-3.9);
            \node[xshift=10pt] at (4,-2) {$\psi$};
            \node[yshift=10pt] at (-2,4) {$\phi$};
            \node[xshift=5pt,yshift=5pt,rotate=45] at (3,3) {$\psi'$};
            \node[xshift=5pt,yshift=-5pt,rotate=-45] at (1,-5) {$\phi'$};
            \node[xshift=-12pt] at (-2,2) {$2\pi$};
            \node[xshift=-12pt] at (-2,0) {$\pi$};
            \node[yshift=-11pt] at (0,-2) {$\pi$};
            \node[yshift=-11pt] at (2,-2) {$2\pi$};
            \node[xshift=-7pt,yshift=7pt,rotate=45,below=12pt] at (0,0) {$2\pi$};
            \node[xshift=-7pt,yshift=7pt,rotate=45] at (2,2) {$4\pi$};
            \node[xshift=-7pt,yshift=-7pt,rotate=-45] at (0,-4) {$2\pi$};
        \end{tikzpicture}
    \end{center}
    \caption{The way in which the $(\psi, \phi)$ and $(\psi', \phi')$ tori are related.  The $(\psi, \phi)$ coordinates are taken to lie inside a square of width $2\pi$ while the $(\psi', \phi')$ coordinates lie in the tilted rectangle of length $4\pi$ and width $2 \pi$ (lengths measured in the primed coordinate system).}
    \label{fig:coordinateplot}
\end{figure}

We will pursue a separation of variables approach in identifying the eigenfunctions of the Laplacian and Dirac operator on this space. In the unprimed coordinate system, it is more straightforward to assign standard periodicity properties to the $e^{i m \phi}$ and $e^{i m' \psi}$ pieces of the wavefunction, choosing $m$ and $m'$ to be integers for the scalar and in fact integers for the fermion as well (antiperiodicity is taken care of by the two-component nature of the wavefunction). In the Hopf fibre metric, $\phi'$ and $\psi'$ are periodic under shifts by $4 \pi$, and the quantisation conditions have to be adjusted accordingly.

While for the Laplacian we can discard the Hopf fibre metric and work just in the unprimed coordinate system, for the fermion it is easier to set up a simple orthonormal frame using the primed metric.  Thus, we keep both metrics. In particular, for the fermion, we introduce the vielbeins
\begin{align}
    e^1 &=\frac{L}{2} (\cos \psi' \, d \theta'+ \sin \psi' \sin \theta' \, d \phi')\,,
    \\
    e^2 &= \frac{L}{2} (\sin \psi' \, d \theta' - \cos \psi' \sin \theta' \, d \phi')\,,
    \\
    e^3 &= \frac{L}{2} \tau (d \psi'+ \cos \theta' \, d \phi')\,,
\end{align}
where $\tau = \sqrt{ 1 + \epsilon}$ is a different way of writing the squashing parameter. The associated spin connection $\omega^{ab}$, which satisfies the vanishing torsion condition
$d e^a + {\omega^a}_b \wedge e^b = 0$, has the nonvanishing components
\begin{align}
    \omega^{12} = \frac{1}{L} \left(\frac{2}{\tau} - \tau \right)e^3 
    \, , \qquad
    \omega^{13} = - \frac{\tau}{L} e^2
    \, , \qquad
    \omega^{23} = \frac{\tau}{L} e^1
    \, . 
\end{align} 

In addition to the squashing, we include a magnetic field through the $S^2$ base of the Hopf fibration,
\begin{equation}
    F_{\theta'  \phi'} = \tfrac12 Q \sin \theta' \, .
\end{equation}
There is a standard quantisation condition on this flux, that $Q$ must be an integer. (More generally, the electric charge of the fermion or scalar $q$ multiplied by $Q$ should be an integer. Here we will set $q=1$ and suppress it.) A magnetic field of this specific form does not break any further symmetries, which gives us a better chance of being able to solve for the eigenfunctions of the Laplacian and Dirac operator. In the unprimed coordinate system, the non-zero components of the Maxwell field strength are instead
$F_{\theta \psi} =Q \sin 2 \theta$ and $F_{\theta \phi} = -Q \sin 2 \theta$.

\subsection{The free scalar and the squashed sphere}
\label{app:scalarS3}
We solve for the Laplacian eigenfunctions and eigenvalues on the squashed $S^3$ in the presence of a magnetic field:
\begin{equation}
    (\nabla^\mu + i q A^\mu) (\nabla_\mu + i q A_\mu) \Phi - \frac{\mu}{8} R \Phi = - \frac{\lambda}{L^2} \Phi \, .
\end{equation}
The conformally coupled scalar has $\mu=1$ where $R$ is the Ricci scalar on the squashed $S^3$. 
We pick a gauge where
\begin{equation}
    A_\mu = \tfrac12 Q (1-\cos 2 \theta) (0,1,-1) \, .
\end{equation}

Separating variables, we find
\begin{align}
    \Phi &= e^{i (m_2 - Q) \psi} e^{i m_1 \phi} (1-x)^{\frac{|m_1|}{2}} (1+x)^{\frac{|m_2|}{2}} \times \nonumber\\
    & {}_2 F_1 \left(-n ,1+|m_1| + |m_2| + n, 1+|m_1|,  \frac{1-x}{2} \right)  ,
\end{align}
where $n$ is a non-negative integer, $m_1$ and $m_2$ are any integers, and $x = \cos 2\theta$. These mode functions have not been normalised. The hypergeometric can also be written as a Jacobi polynomial
\begin{equation}
    P^{(m_1, m_2)}_n(x) = \frac{(m_1 + 1)_n}{n!} {}_2 F_1 (-n, 1+m_1 + m_2 + n, m_1+1, \tfrac{1}{2}(1-x) ) \, .
\end{equation}

Let us explain the restrictions on $n$, $m_1$ and $m_2$. The scalar should be periodic about the $\phi$ and $\psi$ circles, restricting the allowed values of $m_1, m_2 \in {\mathbb Z}$.  At the poles $x = \pm 1$, these circles shrink, and thus the wavefunction must vanish as well, forcing us to choose positive exponents for the $(1 \pm x)$ prefactors. The hypergeometric function must be well-behaved not only at $x=1$ but also $x=-1$ forcing us to restrict it to have a polynomial form, i.e.\ $n \in {\mathbb N} \cup \{0 \}$ must be a non-negative integer.  This restriction on $n$ leads to the following quantisation condition for the eigenvalues:
\begin{eqnarray}
    \label{spectrumoriginal}
    \lambda = (2n + |m_1| + |m_2| + 1)^2  - \left( \frac{Q}{\epsilon} + m_1 + m_2 \right)^2 \frac{\epsilon}{1+\epsilon}  + \frac{1}{\epsilon} Q^2  + \frac{\mu}{4} (3-\epsilon)-1 \, .
\end{eqnarray}

To make evident the transformation properties of these wavefunctions under the residual rotational symmetries, we substitute $\ell = 2n + |m_1| + |m_2|$ and write the eigenvalues instead as
\begin{equation}\label{spectrumred}
    \lambda = (\ell+1)^2 - \left( \frac{Q}{\epsilon} + 2p - \ell \right)^2 \frac{\epsilon}{1+\epsilon} + \frac{Q^2}{\epsilon} + \frac{\mu}{4}(3-\epsilon) - 1 \, ,
\end{equation}
where $p = 0, 1, \ldots, \ell$ and $\ell = 0, 1, 2, \ldots$ and the degeneracy is $\ell+1$.
The smallest eigenvalue occurs near $\ell = |Q| - 1 - \epsilon$ and is bounded below by $2 |Q| - \epsilon +\frac{\mu}{4}(3-\epsilon) - 1 $. Thus, there is an $\text{O}(|Q|)$ gap in the spectrum when the magnetic field is large.

The result (\ref{spectrumred}) matches a calculation in an appendix of \cite{Egidi_2024}.\footnote{More precisely, \cite[Eq.\ (A.9)]{Egidi_2024},
\begin{equation}
    k(k+2)+  \left(\frac{1}{\varepsilon^2} - 1\right) (2p-k)^2 + 2 (2p-k) t + \varepsilon^2 t^2\,,
\end{equation}
 becomes our result with the identifications
\begin{equation}
    t  = -Q/(1+\epsilon) \, , \qquad k = \ell \, , \qquad \varepsilon = \sqrt{1+\epsilon} \, ,
\end{equation}
and setting the conformal coupling $\mu=0$.}
We can check several other limits as well.  If we turn off the squashing, magnetic field, and the conformal coupling, we recover the familiar eigenvalues of the Laplacian on a round $S^3$, namely $\ell(\ell+2)$ with degeneracy $(\ell+1)^2$, giving us confidence we have found all of the states.  If we turn off only the magnetic field, then we recover the eigenvalues on the squashed $S^3_\epsilon$ \cite{critchley1981vacuum,shen1987higher}, 
\begin{equation}
    \lambda |_{Q=0} = (\ell+1)^2 - (2p-\ell)^2 \frac{\epsilon}{1+\epsilon} + \frac{\mu}{4}(3-\epsilon) - 1 \, .
\end{equation}

Below, we will be interested in computing the partition function in the limit the squashing disappears $\epsilon \to 0$:
\begin{eqnarray}
    \label{lapeigenvalues_nosquashing}
    \lim_{\epsilon \to 0} \lambda = (\ell+1)^2 - 2 Q \left(  2p - \ell \right) + Q^2 + \frac{3 \mu}{4} - 1 \, .
\end{eqnarray}

\subsection{The free fermion and the squashed sphere}
\label{app:fermionS3}
It is more convenient to solve for the spinor eigenfunctions $\slashed{D} \psi = i \frac{\lambda}{L} \psi$ in the primed coordinate system.
In our conventions, the Dirac operator on curved spacetime is 
\begin{equation}
    \slashed{D} \Psi =  \gamma^a e_a^\mu (\partial_\mu + \tfrac{1}{4} \omega_\mu^{bc} \gamma_b \gamma_c + i q A_\mu) \Psi \,.
\end{equation}
We choose $A_{\phi'} = \frac{Q}{2}(1- \cos  \theta')$ with the other components set to zero.   As before, we set $q=1$.  

We use the explicit basis $\gamma^a = \{\sigma_1, \sigma_2, \sigma_3 \}$ and explore a separation of variables ansatz where $\Psi(\theta', \psi', \phi') = e^{i(n_1 - \frac{Q}{2}) \phi'} e^{i (n_2 -  \frac{Q}{2}) \psi'} \Psi(\theta')$. Furthermore, we look for solutions where
\begin{equation}
    \Psi = \left(
    \begin{array}{c}
        f_1(\theta') \\
        e^{i \psi'}  \frac{1}{\sqrt{\sin\theta'}} \, f_2(\theta')
    \end{array}
    \right) .
\end{equation}  
A common problem with solving the Dirac equation in 2d or 3d is that while it is relatively easy to get some second order differential equation for one of the components, it can often be tricky to find a simple second order differential equation even if one exists.  The prefactors above benefited from some trial and error exploration before landing on a suitable choice. With the further variable substitution $x = \cos 2 \theta'$, a relatively simple second order differential equation for either $f$ can be derived, which has hypergeometric solutions.  

The relations between $f_1$ and $f_2$ are
\begin{align}
    f_1 &= - \frac{4 i \tau}{1-x^2} \frac{  (n_1 - n_2 x) f_2 +  (1-x^2) f_2' }{2 + 4 n_2 - 2 Q + \tau^2 -2 \tau \lambda}\, , \\
    \label{f1tof2}
    f_2 &= - 4 i \tau \frac{-(n_1 - n_2 x ) f_1 + (1-x^2) f_1'}{-2 - 4 n_2 + 2 Q + \tau^2 - 2 \tau \lambda} \,  .
\end{align}
Looking more closely at the maps from $f_1$ to $f_2$, we see there may in general be some problems when the denominator and numerator vanish.  Examining these cases, we find the eigenfunctions
\begin{align}
    \label{lambdaplusbranch}
    \begin{pmatrix}
    f_1 \\
    f_2
    \end{pmatrix} &= (1-x)^{(n_2-n_1)/2} (1+x)^{(n_1+n_2)/2} \begin{pmatrix}
    1 \\
    0 
    \end{pmatrix}, \text{ with }
    \lambda = \frac{1}{2 \tau}(2 + 4 n_2 - 2 Q + \tau^2) \, , \\
    \label{specialsols}
    \begin{pmatrix}
    f_1 \\
    f_2 
    \end{pmatrix} &=   (1-x)^{-(n_2-n_1)/2} (1+x)^{-(n_1+n_2)/2} \begin{pmatrix}
    0 \\
    1 
    \end{pmatrix},
    \text{ with }
    \lambda = \frac{1}{2\tau}(-2 - 4 n_2 + 2 Q + \tau^2) \, .
\end{align}

To choose the proper quantum numbers for the wavefunction on the $(\psi, \phi)$ torus, let us take a moment to look at the relation between the primed and unprimed coordinate systems.  We have
\begin{eqnarray}
    e^{i(n_1 - \frac{Q}{2}) \phi'} e^{i (n_2 -  \frac{Q}{2}) \psi'} = e^{i (n_1 + n_2 - Q) \psi} e^{i(n_2-n_1) \phi} \ .
\end{eqnarray}
In correspondence to the variables in the Laplacian case, we define  $m_2 = n_1 + n_2 $
and $m_1 = n_2 - n_1 $.  By allowing $m_1, m_2 \in {\mathbb Z}$, we arrange for the correct anti-periodic behaviour around the cycles of the torus.  The anti-periodicity is guaranteed by the two-component nature of the wavefunction, which has eigenvalues $\pm \frac{1}{2}$ under the spin generators $\frac{\sigma_a}{2}$.  

The second order differential equation  for $f_1$ has characteristic exponents $\pm \frac{m_i}{2}$ at the poles $x = \pm 1$, with $m_1$ associated with the pole at $x=1$ and $m_2$ the pole at $x=-1$.  Note the $\psi$ circle, which is associated with $m_2$, closes off at $\theta' = \pi$ or  equivalently $x = -1$.  The $\phi$ circle is associated with $m_1$ and closes off at $\theta' = 0$ or equivalently $x=1$.   To have a well-behaved wavefunction at the poles, we want to associate a positive power of $(1+x)$ and $(1-x)$ to the wavefunction.

The solutions for $f_1$ are very similar to the Laplacian case
\begin{equation}
    f_1 = (1-x)^{\frac{|m_1|}{2}} (1+x)^{\frac{|m_2|}{2}}
    {}_2 F_1 (-n , 1 + |m_1| + |m_2| + n , 1+|m_1|, \tfrac{1}{2}(1-x) ) \, ,
\end{equation}
where the polynomial condition on the hypergeometric gives the eigenvalues
\begin{eqnarray}
    \label{fermionlambda}
    \lambda_{\pm} = \frac{\sqrt{1+\epsilon}}{2} \pm \sqrt{
    (2n + |m_1| + |m_2| +1)^2 -\left( \frac{Q}{\epsilon} + m_1 + m_2 + 1 \right)^2 \frac{\epsilon}{1+\epsilon} 
    + \frac{Q^2}{\epsilon} 
    }\,,
\end{eqnarray}
and  we have replaced $\tau^2$ with $1+\epsilon$. The integers $n \in {\mathbb N} \cup \{ 0 \}$ are non-negative while $m_1, m_2 \in {\mathbb Z}$.  

Naively, we can proceed as before and set $\ell = 2n + |m_1| + |m_2|$, finding $(\ell+1)^2$ states for each allowed value of $\ell$.  However, comparing with the round sphere limit where $\epsilon = Q = 0$, we find either an over- or an under-counting of the degeneracy.  In this limit, the eigenvalues are $\ell + \frac{3}{2}$ and $- \ell - \frac{1}{2}$. For eigenvalues $\pm (\ell + \frac{3}{2})$, the  degeneracy should be $(\ell+1)(\ell+2)$ (see e.g.\  \cite{Camporesi:1995fb}).  Thus, we have over-counted the $\lambda = -\ell - \frac{1}{2}$ eigenvalues by $(\ell+1)$ and undercounted the $\lambda = \ell + \frac{3}{2}$ eigenvalues by $(\ell+1)$.

The undercounting is easily resolved.  Looking at (\ref{specialsols}), we see that we missed $\ell+1$ states where only $f_2$ is turned on. These are states with $m_1 + m_2 = -\ell-2$ and $m_1$ and $m_2$ both negative and in fact strictly less than $-1/2$ because of the factor of $(\sin \theta)^{-1/2}$ invoked in the definition of $f_2$. The over-counting has a similar resolution.  The map (\ref{f1tof2}) from $f_1$ to $f_2$ has a singularity when $\lambda = \frac{\tau}{2} -  \frac{1}{\tau}(1 + m_1 +  m_2 -  Q)$.  This value is obtained for $n=0$, $m_1$ and $m_2$ both positive, and $1+m_1 + m_2 - Q \geq 0$.  For small values of $Q$, these $\ell+1$ states should be discarded. Indeed, by the smooth dependence of the eigenvalues on $Q$, increasing $Q$ through the branch point, these states should be discarded for general $Q$.  

Using the $\ell$ quantum number,  we find for each integer $p \in \{ 0, 1, \ldots, \ell+1 \}$, $\ell+1$ states of the form
\begin{equation}
    \lambda_+ = \frac{\sqrt{1+\epsilon}}{2} + \sqrt{
    (\ell +1)^2 -\left( \frac{Q}{\epsilon} + 2p -\ell - 1 \right)^2 \frac{\epsilon}{1+\epsilon} 
    + \frac{Q^2}{\epsilon} 
    } \, .
\end{equation}
Note that the square root becomes zero when $\ell = |Q|-1$ and $p=0$ or $\ell+1$, depending on the sign of $Q$.  For these eigenvalues that pass through the branch cut, we interpret the square root in the formula such that 
\begin{equation}
    \lambda_+ = \frac{\sqrt{1+\epsilon}}{2} + \frac{1+\ell-|Q|}{\sqrt{1+\epsilon}} \, ,
\end{equation}
in agreement with the eigenvalues of the special solutions (\ref{lambdaplusbranch}). Note there is no $\text{O}(Q)$ gap in this spectrum.  Instead, there is a zero crossing at $\ell = |Q|-\frac{3}{2} - \frac{\epsilon}{2}$. For each $p \in \{ 1, 2, \ldots, \ell \} $, we find $\ell+1$ states of the form
\begin{equation}
    \lambda_- = \frac{\sqrt{1+\epsilon}}{2} - \sqrt{
    (\ell +1)^2 -\left( \frac{Q}{\epsilon} + 2p -\ell -1 \right)^2 \frac{\epsilon}{1+\epsilon} 
    + \frac{Q^2}{\epsilon}  
    } \ .
\end{equation}

While (\ref{fermionlambda}) is a new result to our knowledge, the  $Q=0$ limit has been well known for decades \cite{hitchin1974parallel,dowker1999effective}.  We can write our result in the form
\begin{equation}
    \lambda_+ = \frac{\tau}{2} + \frac{1}{\tau} \sqrt{ (\ell+1)^2 + 4(\tau^2-1) p (\ell +1- p) }\, ,
\end{equation}
from which it is relatively straightforward to match \cite[Eq.\ (2)]{dowker1999effective}.

Our interest will be the $\epsilon \to 0$ limit, where we obtain\footnote{The quantum numbers in this limit are implicit in the discussion 
below \cite[Eq.\ (15)]{Gautason:2025per}.  We thank F.~Gautason for discussion
about his recent work.}
\begin{align}
    \lambda_+ |_{\epsilon=0} &=  \frac{1}{2} + \sqrt{
    (\ell +1)^2 -2 Q( 2p -\ell - 1 )
    +  Q^2
    } \, , \qquad p \in \{0, 1, \ldots, \ell+1 \}\,, \\
    \lambda_- |_{\epsilon=0} &=  \frac{1}{2} - \sqrt{
    (\ell +1)^2 -2 Q( 2p -\ell -1)
    +  Q^2
    } \, , \qquad p \in \{1, 2, \ldots, \ell \}\,,
\end{align}
each of which has degeneracy $\ell+1$. In calculating the partition function, it will be convenient to add a second set of spinors with  a parity-reversed set of gamma matrices $\{ -\sigma_1, -\sigma_2, -\sigma_3 \}$. From the structure of the Dirac equation, it is easy to see we will get a second set of eigenspinors with eigenvalue minus what is given above.  

\end{appendix}

\bibliography{main}

\end{document}